\documentclass[universe,review,accept,pdftex]{Definitions/mdpi} 

\usepackage{multirow,tabularx}
\usepackage{natbib}
\usepackage{amssymb}
\usepackage[normalem]{ulem}
\setcitestyle{authoryear}

\newcommand{\hmpc}{h^{-1}{\rm Mpc}}

\newcommand{\kms}{{\rm km}\,{\rm s}^{-1}}

\newcommand{\Zsolar}{\;{\rm Z}_{\odot}}
\newcommand{\msolar}{{\rm M}_{\odot}}
\newcommand{\hmsolar}{h^{-1}{\rm M}_{\odot}}
\newcommand{\msolaryr}{{\rm M}_{\odot} {\rm yr}^{-1}}

\newcommand{\vw}{{v_{\rm wind}}}

\newcommand{\msol}{{\rm M}_{\odot}}

\newcommand{\gadtwo}{{\sc Gadget-2}}
\newcommand{\gadget}{{\sc Gadget-3}}
\newcommand{\arepo}{{\sc Arepo}}
\newcommand{\changa}{{\sc CHANGA}}
\newcommand{\ramses}{{\sc RAMSES}}
\newcommand{\gizmo}{{\sc GIZMO}}
\newcommand{\CIV}{\hbox{C\,{\sc iv}}}
\newcommand{\OV}{\hbox{O\,{\sc v}}}
\newcommand{\OVI}{\hbox{O\,{\sc vi}}}
\newcommand{\OVII}{\hbox{O\,{\sc vii}}}
\newcommand{\OVIII}{\hbox{O\,{\sc viii}}}
\newcommand{\OIX}{\hbox{O\,{\sc ix}}}
\newcommand{\HI}{{\hbox{H\,{\sc i}}}}
\newcommand{\HII}{{\hbox{H\,{\sc ii}}}}
\newcommand{\SiIII}{\hbox{Si\,{\sc iii}}}
\newcommand{\FeXVII}{\hbox{Fe\,{\sc xvii}}}
\newcommand{\MgXII}{\hbox{Mg\,{\sc xii}}}

\newcommand{\fb}{{f_{\rm b}}}

\newcommand{\Chandra}{\textit{Chandra}}
\newcommand{\Athena}{\textit{Athena}}
\newcommand{\Lynx}{\textit{Lynx}}
\newcommand{\XMM}{XMM-\textit{Newton}}

\newcommand{\XRISM}{\textit{XRISM}}
\newcommand{\eROSITA}{\textit{eROSITA}}

\DeclareMathSymbol{\la}{3}{AMSa}{46}
\DeclareMathSymbol{\ga}{3}{AMSa}{38}

\newcommand{\dataaccess}[1]{\vspace{12pt}\noindent{\fontsize{9}{9}\selectfont\textbf{Data Access:} {#1}\par}}

\firstpage{1} 
\pubvolume{7}
\issuenum{7}
\articlenumber{209}
\pubyear{2021}
\copyrightyear{2021}

\history{Received: 9 April 2021; Accepted: 2 June 2021; Published: 24 June 2021}

\Title{Simulating Groups and the IntraGroup Medium: The Surprisingly Complex and Rich Middle Ground Between Clusters and Galaxies }

\Author{Benjamin D. Oppenheimer$^{1,2}$\orcidA{}, Arif Babul$^{3}$\orcidB{}, Yannick Bah\'e$^{4}$\orcidC{}, Iryna S. Butsky$^{5}$\orcidD{}, and Ian G. McCarthy $^{6}$\orcidE{}}

\AuthorNames{Benjamin D. Oppenheimer, Arif Babul, Yannick Bah\'e, Iryna S. Butsky, Ian G. McCarthy}

\address{%
$^{1}$ \quad CASA, Department of Astrophysical and Planetary Sciences, University of Colorado, 389 UCB, Boulder, CO 80309, USA;\\
$^{2}$ \quad Harvard Smithsonian Center for Astrophysics, 60 Garden Street, Cambridge, MA 02138, USA;\\
$^{3}$ \quad Department of Physics and Astronomy, University of Victoria, Victoria, BC V8W 2Y2, Canada;\\
$^{4}$ \quad Leiden Observatory, Leiden University, P.O. Box 9513, 2300 RA Leiden, The Netherlands;\\
$^{5}$ \quad Astronomy Department, University of Washington, Seattle, WA 98195, USA\\
$^{6}$ \quad Astrophysics Research Institute, Liverpool John Moores University, 146 Brownlow Hill, Liverpool L53RF, UK
}

\corres{Correspondence: benjamin.oppenheimer@colorado.edu}

\abstract{
Galaxy groups are more than an intermediate scale between clusters and halos hosting individual galaxies, they are crucial laboratories capable of testing a range of astrophysics from how galaxies form and evolve to large scale structure (LSS) statistics for cosmology.  Cosmological hydrodynamic simulations of groups on various scales offer an unparalleled testing ground for astrophysical theories.  Widely used cosmological simulations with $\sim$(100 Mpc)$^3$ volumes contain statistical samples of groups that provide important tests of galaxy evolution influenced by environmental processes.  Larger volumes capable of reproducing LSS while following the redistribution of baryons by cooling and feedback are essential tools necessary to constrain cosmological parameters.  Higher resolution simulations can currently model satellite interactions, the processing of cool ($T\sim 10^4$ K) multi-phase gas, and non-thermal physics including turbulence, magnetic fields, and cosmic ray transport.  We review simulation results regarding the gas and stellar contents of groups, cooling flows and the relation to the central galaxy, the formation and processing of multi-phase gas, satellite interactions with the intragroup medium, and the impact of groups for cosmological parameter estimation.  Cosmological simulations provide evolutionarily consistent predictions of these observationally difficult-to-define objects, and have untapped potential to accurately model their gaseous, stellar, and dark matter distributions.  }

\keyword{black holes; galaxy groups; galaxy surveys; intragroup medium/plasma; hydrodynamical and cosmological simulations; active galactic nuclei; X-ray observations; UV observations; cosmological parameters}

\begin{document}


\section{Introduction}\label{sec:1}



Galaxy groups are versatile laboratories to study a range of astrophysics spanning non-gravitational, baryonic processes associated with galaxy formation to large-scale structure statistics constraining cosmology.  Their intermediate scale between galactic halos and clusters offers a unique set of theoretical challenges that are often overlooked relative to adjacent mass bins, but~this scale offers crucial constraints for how a large proportion of galaxies evolved to their present state.  The~perspective of this review, focusing on halos with masses $M_{\rm halo}\approx 10^{13}-10^{14}\ \msolar$, differs from the companion reviews because a cosmological simulation tracks the evolution of all gas, stars and~dark matter, not just the X-ray emitting intragroup medium (IGrM).  Therefore we consider all phases of gas---from the extended, hot IGrM that stretches beyond the virial radius to the cold interstellar medium (ISM) within individual group galaxies---alongside the stars and dark matter in central and satellite group galaxies, as~well as the surrounding cosmological large scale structure (LSS). Simulations have enabled breakthroughs in our understanding of all these components, but~their limited resolution and incomplete physics models still represent a major obstacle on the path to a truly complete understanding of the observed gas, galaxies and~LSS statistics in and around~groups. 

The IGrM is mainly very hot ($T\sim 10^7$ K), but~does not primarily radiate via Bremsstrahlung radiation like the intracluster medium (ICM; see {Figure~1 of the companion review by} \mbox{\cite{lovisari21}}
).  Cooling via line emission offers a greater opportunity to form multi-phase, cool ($T<10^5$ K) gas, and~potentially provide additional fuel to galaxies.  Nevertheless, IGrM observations are currently dominated by hot X-ray probes from especially \Chandra~\citep{sun09,Eckmiller11,Bharad2015} and \XMM~(e.g.,~\mbox{\citet{lovisari15}}).  X-ray-derived profiles of IGrM properties, especially its entropy, provide rigorous tests for simulations that often make diverging predictions (e.g.,~\mbox{\citet{mitchell09, lebrun14}}).  Yet these simulations also produce lower mass groups that can be tested against the Complete Local Volume Groups Sample (CLoGS) survey \citep{osullivan17}, and~sometimes higher mass objects for comparison with observed clusters (e.g., \cite{cavagnolo09, pratt09}).  With~large-scale simulations projects containing multiple volumes and/or zoom-in simulations of massive objects, simulations of the IGrM can be compared and contrasted to lower mass galaxy halos and higher mass~clusters.  

Similar to a cluster, most groups contain a dominant ``brightest group galaxy'' (BGG) near the halo center, usually where the X-ray-traced IGrM peaks in brightness. In~contrast to brightest cluster galaxies (BCGs), BGGs are observed to less likely be quenched early-type galaxies (ETGs; e.g.,~\mbox{\cite{Wetzel_et_al_2012, Davies_et_al_2019}}), and~more likely to have disc-like morphologies (e.g., \cite{Moffett16a}). Understanding the---likely intimate---connection between the BGG and IGrM involves studying how the galactic baryon cycle (the interplay of gas accretion, outflows, and~recycling that is understood as fundamental for the co-evolution of field galaxies and their circumgalactic medium; \mbox{\cite{Somerville_Dave_2015, Tumlinson_et_al_2017}}) transitions to the cluster version of precipitation, jet-driven active galactic nuclei (AGN) feedback, and~chaotic cold accretion (e.g., \cite{voit15, gaspari17}).  

The IGrM also includes cool and warm ($T\sim 10^5-10^6$ K) gas phases. Although~sub-dominant to the hot IGrM component by mass as simulations clearly demonstrate \mbox{(e.g.,~\cite{emerick15,butsky19})}, these phases are thought to represent the link between the IGrM and individual group galaxies: from them, gas can accrete onto the BGG and fuel further star formation (e.g., \cite{McDonald11b,tremblay18}), whereas gas stripped from satellite galaxies (e.g.,~{\cite{Tonnesen_et_al_2011, yun19, Campitiello_et_al_2021}}) or ejected from them through superwind feedback (e.g.,~\mbox{\cite{McGee_et_al_2014}}) is also initially less hot than the virialized IGrM halo. Furthermore, observations of quasar absorption lines in the UV indicate a substantial reservoir of \HI{} and metals in the IGrM at temperatures of $10^{4}${--}$10^{5.5}$ K \citep{burchett18, stocke19}, and~21-cm emission shows extended IGrM structures, at~least for more compact, spiral-rich groups \citep{borthakur10}.  The theoretical modeling of cool/warm gas is crucial to understand for how groups diverge from clusters owing to their lower temperatures promoting more cooling.  

Lower IGrM pressures and galaxy velocities process satellite galaxies differently than in clusters: simulations find both the atomic hydrogen and star forming gas is removed less rapidly after infall \citep{Oman_et_al_2021}, providing an explanation for observed group galaxies being less \HI{} deficient than in clusters \citep{Brown_et_al_2017} and having lower quenched fractions (e.g.,~\mbox{\cite{Wetzel_et_al_2012, Davies_et_al_2019}}). However, the~lower velocity dispersion of group satellites makes dynamical friction more efficient, so that mergers---in particular between satellites and the BGG---are more common \citep{Bahe_et_al_2019}. Finally, although~groups are (typically) dynamically older than clusters, their galaxies are more likely to have been accreted directly from the field, rather than via an intermediate ``pre-processing'' phase in a lower-mass halo (e.g.,~\mbox{\cite{Fujita_2004}}). The~total time that $z = 0$ galaxies have spent as satellites is therefore $\approx$50\% shorter for groups than massive clusters (ca. 4 vs.~6 Gyr; \mbox{\cite{Wetzel_et_al_2013}}, see also \mbox{\citet{Donnari_et_al_2021}}).  


Relative to clusters, the~$\sim$10$\times$ higher number density of groups (e.g.,~\mbox{\cite{Jenkins_et_al_2001}}) combined with their shallower potential wells---which make it easier for AGN feedback to eject baryons---gives groups particular significance for cosmological parameter estimates that rely on the total mass distribution, for example,~lensing measurements, cosmic shear, and~redshift space distortions.  The~distribution of baryons in halos corresponding to groups ($10^{13-14}\ \msolar$) is known to be significantly affected by feedback processes.  Therefore, the~self-consistent modeling of the entire baryonic (gas+stellar) and dark matter distributions of groups, in~way that realistically captures the effects of galaxy formation, is a necessary tool for precision~cosmology.  

In this review, we focus on how groups process their baryons in a cosmological context, and~therefore discuss predominantly cosmological simulations that contain groups. Entire separate reviews could be written about idealized simulations and analytical models at the group scale, but~these are not able to confront the intersection of galaxy formation and cosmology.  We will, however, refer to idealized simulations and other methods simulating important physics when they are relevant to groups, especially in regards to how cosmological simulations can improve.  Ultimately, our understanding of how gas and galaxies evolve together to create the observed distribution of groups would not be as far along without these~tools.  

\textls[-5]{Throughout this review, we use a theorist's definition of a group as a system of galaxies and gas hosted by a halo within a certain mass range. Following common convention, we define these masses as $M_\Delta$, the~sum of all matter species (i.e.,~dark matter and baryons) within a spherical aperture $r_\Delta$ inside which the mean density equals $\Delta$ times the critical density of the universe; we adopt the value $\Delta = 500$ as commonly used in X-ray studies of the IGrM/ICM and thus define a group halo as one with $M_\mathrm{500(c)} = 10^{13}$--$10^{14}\ \msol$ ($r_\mathrm{500(c)} \approx 340$--$600$ kpc)  \footnote{In the remainder of this review, we omit the `c' suffix that identifies the overdensity as measured with respect to the critical, rather than for example,~mean, density of the universe}. 
 This definition places groups between lower-mass galactic halos ($M_{500} \lesssim 10^{13}\,\msol$) and more massive clusters ($M_{500} \gtrsim 10^{14}\,\msol$). For~the benefit of readers more used to masses within other overdensity thresholds, we note that in this halo mass range masses calculated within spheres of average density equal to 200 times the critical, or~200 times the mean, density ($M_{200}$ and $M_{200m}$, respectively), or~to the virial overdensity $\Delta_\mathrm{vir} \approx 18\pi^2 + 82(\Omega(z) - 1) - 39 (\Omega(z) - 1)^2$ \citep{Bryan_Norman_1998} that is based on the analytic solution of spherical top-hat collapse ($M_\mathrm{vir}$), are offset from $M_{500}$ by +0.16 dex ($M_{200}$), +0.31 dex ($M_{200m}$), and~+0.25 dex ($M_\mathrm{vir}$), respectively (and the corresponding radii from $r_{500}$ by factors of 1.5, 2.6, and~2.1) \footnote{All conversions are median differences obtained from the IllustrisTNG300 simulation}. 
  A~typical group with mass of $M_{500} = 10^{13.5}\ \msolar$ has a radius $R_{500}=480$ kpc, a~virial temperature $T_\mathrm{X}\sim1$ keV, and~a velocity dispersion $\sigma \approx 440\ \kms$ at $z=0$.}

This review is organized into sections as follows: We begin with an overview of the cosmological simulations we discuss in Section~\ref{sec:overview}.  The~methods used in simulation of groups are discussed in a series of subsections throughout Section~\ref{sec:modules}.  Section~\ref{sec:results} comprises the results of current simulations creating groups, and~is divided into five main subsections: the baryonic content of groups (Section~\ref{sec:baryons}), the~connection between the central galaxy and the IGrM (Section~\ref{sec:BGG}), the~multiphase IGrM (Section~\ref{sec:multiphase}), satellite galaxies in groups (Section~\ref{sec:satellite}), and~the impact of galaxy group astrophysics on LSS cosmology (Section~\ref{sec:cosmology}).  We discuss future directions in Section~\ref{sec:future} and~make a short final statement in Section~\ref{sec:final}.

\section{Overview of Simulations That Model~Groups} \label{sec:overview}  

Cosmological simulations use a variety of hydrodynamics schemes, span many orders of magnitude in mass and spatial resolution, and~model volumes of vastly different sizes and contents. This rich diversity in modeling approaches reflects the wide range of astrophysical processes and objects that different simulations attempt to model. In~Table~\ref{tab:sims}, we list recent simulations that include galaxy groups and are therefore of particular interest to our~review.

\begin{table}[H]
\caption{Modern cosmological hydrodynamic simulations with groups that run to $z=0$. }
\centering
\tablesize{\scriptsize} 
\begin{tabular}{llllllll}
\toprule
\textbf{Simulation}	& \textbf{Simulation} & \textbf{Hydrodynamic}  &\textbf{Baryon }	& \textbf{Volume}     & \textbf{AGN}      &  \textbf{$f_{\rm gas,500}$ at} & \textbf{$f_{*,500}$ at}    \\
                    & \textbf{Code}       & \textbf{Scheme}        &\textbf{ Resolution}&  \textbf{(Mpc$^3$)} & \textbf{Feedback} &  \textbf{$M_{500}=$} &  \textbf{$M_{500}=$}   \\
                    & \textbf{}       & \textbf{}                  &\textbf{ ($\msolar$)}&                    & \textbf{Scheme}   &   \textbf{$10^{13.5}\msolar$}     &   \textbf{$10^{13.5}\msolar$}            \\

\midrule
cosmo-OWLS$^a$           & \gadget               & Classical SPH           & $1.2\times10^9$ & $2.1\times10^8$ & Thermal & 0.05 &   \\
Illustris$^b$		   & \arepo			    & Moving Mesh   & $1.3\times10^6$ & $1.2\times10^6$ &  Dual   & 0.01 & 0.04 \\
EAGLE$^c$		       & \gadget			    & Modern SPH  & $1.8\times10^6$ & $1.0\times10^6$ &  Thermal & 0.11 & 0.01   \\
\citet{liang16}      & \gadtwo			& Classical SPH           & $9.0\times10^7$ & $2.9\times10^6$ &  None   & 0.09 & 0.06 \\
Horizon-AGN$^d$        & \ramses            & AMR           & $1\times 10^7$  & $2.9\times10^6$ &  Dual & 0.09 & \\
BAHAMAS$^e$            & \gadget               & Classical SPH  & $1.2\times10^9$ & $2.1\times10^8$ &  Thermal & 0.04 & 0.02   \\
C-EAGLE/Hydrangea$^f$  & \gadget			    & Modern SPH  & $1.8\times10^6$ & 30 zooms          &  Thermal & 0.08 &  0.02 \\
FABLE$^g$              & \arepo             & Moving Mesh   & $9.4\times10^6$ & $2.0\times10^5$+6 zooms & Dual & 0.07 & 0.02 \\
The Three Hundred$^h$      & \gadget            & Modern SPH    & $3.5\times10^8$ & 324 zooms  & Dual & 0.10  & 0.02 \\ 
IllustrisTNG100$^i$       & \arepo             & Moving Mesh   & $1.4\times10^6$ & $1.4\times10^6$ & Dual  & 0.08 & 0.02 \\
IllustrisTNG300$^j$       & \arepo             & Moving Mesh   & $1.1\times10^7$ & $2.8\times10^7$ & Dual  & 0.08 & \\
ROMULUS$^k$                & \changa            & Modern SPH & $2.1\times 10^5$ & $1.5\times10^4$+3 zooms & Thermal & 0.11 & 0.04 \\
SIMBA$^l$              & \gizmo             & \tiny{Meshless Finite Mass} & $1.8\times10^7$ & $2.9\times10^6$ & Dual & 0.04 & 0.02 \\
IllustrisTNG50$^m$       & \arepo             & Moving Mesh   & $8.5\times10^4$ & $1.4\times10^5$ & Dual  & 0.09 &  \\
Magneticum-Box2/hr$^n$ &       \gadget & Modern SPH &  $1.4\times 10^8$ & $1.3\times10^8$ & Dual & & \\

\bottomrule
\end{tabular}
\parbox{25cm}{
\scriptsize
$^a$ \citet{lebrun14} $^b$ \citet{vogelsberger14}, $^c$ \citet{schaye15}, $^d$ \citet{dubois16}, $^e$ \citet{mccarthy17}, $^f$ \citet{bahe17} \\ 
\& \citet{barnes17}, $^g$ \citet{henden18}, $^h$ \citet{cui18}, $^i$ \citet{pillepich18a}, $^j$ \citet{nelson18a} $^k$ \citet{tremmel17, tremmel19}\\ 
$^l$ \citet{dave19}, $^m$ \citet{nelson19} $^n$ http://www.magneticum.org 
}
\label{tab:sims}
\end{table}

This list encompasses simulations with three broad classes of hydrodynamics schemes, as~indicated in the third column of Table~\ref{tab:sims}. Adaptive Mesh Refinement (AMR; \mbox{\cite{Teyssier_2015}}) uses a fixed (Eulerian) grid whose cells are locally and dynamically (de-/)refined, typically depending on the local gas density.  Smoothed Particle Hydrodynamics (SPH; \mbox{\cite{Springel_2010SPH, Price_2012}}), on~the other hand, is a Lagrangian scheme in which gas \emph{mass} is discretized into a finite number of particles that move under the influence of gravity and hydrodynamic forces; the latter are calculated by smoothing over neighboring particles. The~third class is a hybrid of these two: in the ``moving mesh'' approach \citep{springel10}, an~unstructured grid is used that moves with the local gas flow, whereas in ``meshless finitie mass'' simulations (\cite{Hopkins_2015}, see also \mbox{\cite{Koshizuka_Oka_1996}}), mass is discretized into particles like in SPH, but~with explicit accounting of mass flows between neighboring particles  \footnote{We note that the ability of particles to move through the simulation volume is not particular to this approach, and~is also an integral feature of the SPH approach}. 
 We note that, as~we discuss in Section~\ref{sec:hydro}, early SPH implementations suffered from systematic problems that have motivated improved ``modern'' formulations of SPH (see e.g.,~\mbox{\cite{Hu_et_al_2014, schaller15, beck16, Borrow_et_al_2020}}); most SPH simulations that we discuss use one of these modern~variants.


Many simulations listed in Table~\ref{tab:sims} evolve periodic, cosmological cubes with side length $\approx$100 Mpc, large enough to contain at least several dozen groups.  In~addition to modeling hydrodynamics and gravity, they all contain ``subgrid'' prescriptions for astrophysical processes that originate on unresolved scales, such as gas cooling, the~formation of stars and black holes, and~feedback associated with it (see Section~\ref{sec:modules}). The~``galaxy formation model'' formed by these prescriptions is nowadays often calibrated to reproduce a particular set of galaxy properties. The~Illustris \citep{vogelsberger14}, EAGLE (Evolution and Assembly of GaLaxies and their Environments; \cite{schaye15}) \footnote{See also \mbox{\citet{crain15}}. for calibration and \mbox{\citet{mcalpine16}} for public release}, 
and~Horizon-AGN \citep{dubois16} simulations are representative examples of this approach. Their ability to (broadly) reproduce galactic stellar mass functions, galaxy colors and star formation rates (SFRs), and~even galaxy morphologies signified a transformational advance over previous generations of simulations. Particularly relevant for groups, almost all the simulations listed in Table~\ref{tab:sims} model the accretion of gas onto SMBHs, and~the resulting AGN feedback. In~some recent simulations, including the IllustrisTNG \mbox{project \citep{nelson18a, pillepich18a}} \footnote{See also \mbox{\citet{Marinacci_et_al_2018, Naiman_et_al_2018, Springel_18}}}  
 \mbox{and SIMBA \citep{dave19}}, these models are explicitly calibrated against gaseous properties of group-scale halos.

\textls[-5]{Simulations using the same or slightly modified codes as in some of the above-mentioned projects target volumes $\gg$$10^6$ Mpc$^3$, with~the aim to model clusters and/or LSS. For~example, the~C-EAGLE/Hydrangea simulations \citep{bahe17, barnes17} extend EAGLE with 30 ``zoom-in'' simulations centered on clusters with $M_{200}=10^{14.0}$--$10^{15.4}\ \msolar$, 24 of which (the Hydrangea suite) with high-resolution regions that extend to 10 $r_{200}$ at $z = 0$ \citep{bahe17}. The~zoom technique allows these simulations to reach the same resolution ($\approx$2$\times 10^6\,\msol$ for baryons) as the 100 Mpc EAGLE ``Reference'' run, but~with slightly adjusted AGN feedback parameters for more realistic gas fractions in groups as we detail in \mbox{Section~\ref{sec:AGNfeedback}}. The~only simulation that reaches a comparable resolution in a full $\gg 10^6\,\mathrm{Mpc}^3$ volume is the TNG300 run of the IllustrisTNG suite: it evolves a ($\approx$300 Mpc)$^3$ volume with a baryon mass resolution of $\approx$1.1$\times 10^7\,\msol$, that is,~$\approx$6 or 8 times lower resolution than \mbox{(C-)EAGLE} or the $\approx$100 Mpc TNG100 simulation, respectively. We note that, irrespective of this resolution difference, all TNG simulations use the same subgrid model and parameters. The~degree of numerical convergence between these different resolution levels is discussed in for example,~\mbox{\citet{pillepich18b}} and \mbox{\citet{Donnari_et_al_2020}}; we refer to \mbox{\citet{schaye15}} for the opposite ``weak convergence'' approach of re-calibrating parameters for different resolution~levels.  }


The cosmo-OWLS \citep{lebrun14} and BAHAMAS (BAryons and HAloes of MAssive Systems)~{\citep{mccarthy17}} simulations run larger periodic volumes at lower resolution.  cosmo-OWLS varied different aspects of the subgrid models, switching different physics models on and off as well as changing the efficiencies of stellar and AGN feedback. No attempt was however made to calibrate the simulations to match observations, even though the default model reproduces different aspects of groups relatively well.  BAHAMAS, on~the other hand, explicitly calibrated the stellar and AGN feedback to reproduce the gas fractions of galaxy groups and the galaxy stellar mass function in order to ensure a realistic treatment of the effects of baryons on the matter power spectrum.  The~Magneticum simulations \citep{dolag16} are a series of volumes that mainly concentrate on LSS and cluster astrophysics.  We list their Box2/hr simulation in Table~\ref{tab:sims}, described in \mbox{\citet{castro21}}, which is capable of resolving a statistical sample of groups in a volume 500 Mpc on a side \footnote{We note the work of  \mbox{\citet{ragagnin19}} who measured the ``fossil-ness'' of Magneticum groups in the even larger 900 Mpc Box2b/hr volume, which apparently ran to $z=0$}.  
Lower resolution, Gpc-scale Magneticum volumes employ a strategy of calibrating subgrid modules to a {\it Planck} cosmology, and~then exclusively varying cosmological parameters to explore the impact on baryonic properties.  The~Three Hundred project \citep{cui18} simulated 324 clusters with $M_\mathrm{vir} (z = 0) \gtrsim 1.2 \times 10^{15}\,\msol$ out to a radius of $15\ \hmpc$; similar to Hydrangea, the~simulations therefore also contain many groups in the periphery of the central~clusters. 
 
We also list the \mbox{\citet{liang16}} simulations and the \mbox{\citet{henden18}} FABLE (Feedback Acting on Baryons in Large-scale Environments) simulations, both of which were run at lower resolution than their contemporary counterparts but with the aim to reproduce properties of groups and clusters. The~former is an example of a simulation suite that does not include AGN feedback. In~contrast, FABLE calibrated their subgrid models to reproduce the galactic stellar mass functions and, specifically, the~gas and stellar contents of $M_{500}\approx 10^{13-15}\ \msolar$ halos.  They simulated one smaller volume, supplemented by 6~zoom simulations extending up to a $M_{500} = 10^{15}\ \msolar$ cluster.  

Finally, we will also discuss two higher-resolution simulations that include groups: the ROMULUS suite \citep{tremmel17,tremmel19,chadayammuri21,jung21} and the IllustrisTNG50 (TNG50; \cite{nelson19,pillepich19}) simulation. ROMULUS includes a small-volume run, ROMULUS25, and~three group/cluster zooms called ROMULUSG1, ROMULUSG2 and~ROMULUSC, all at a baryon mass resolution of $2.1 \times 10^5\,\msol$. TNG50 is a larger volume, $\approx$50 Mpc on a side, which contains 20 halos with $M_{500}>10^{13}\,\msolar$ and reaches an even higher mass resolution of $8.5\times10^4\,\msol$ for baryons.  

Many other simulations have modeled groups from cosmological initial conditions, often with the zoom-in approach. These include the three $10^{13}\ \msolar$ group zooms by \mbox{\citet{feldmann10}}, which focus on the evolution and morphology of the central group galaxy; the 10 EAGLE-CGM zooms of $\sim$$10^{13}\ \msolar$ groups by \mbox{\citet{oppenheimer16}} to understand how the circumgalactic medium (CGM) around passive galaxies differs from star-forming galaxies at $10\times$ lower halo mass, and~the \mbox{\citet{Joshi_et_al_2019}} 2--3$\times 10^{13}\ \msolar$ zoom to study environmental processing of satellite~galaxies.

These simulations span a factor of 10,000 in both mass resolution and volume, and~were run to study objects on a wide range of scales---from $M_\star < 10^8\,\msol$ dwarf galaxies to the $\gg$Mpc scale LSS. However, they all overlap at the mass scale of our fiducial intermediate-mass group with $M_\mathrm{500} = 10^{13.5}\,\msol$. In~Table~\ref{tab:sims}, we therefore list the approximate total gas and stellar mass fractions inside $R_{500}$, defined as
\begin{equation}
f_{{\rm gas}, 500}\equiv\frac{M_{\rm gas}(<R_{500})}{M_{\rm tot}(<R_{500})}
\end{equation}
and
\begin{equation}
f_{\star, 500}\equiv\frac{M_\star(<R_{500})}{M_{\rm tot}(<R_{500}).}
\end{equation}

\textls[-10]{For a more detailed discussion of these fractions, and~their dependence on halo mass, we refer the interested reader to Figures~14 and 15 in the companion review by \mbox{\citet{Eckert2021}}.}


\section{Computational Methods Relevant for Simulations of~Groups} \label{sec:modules}

The cosmological simulations we discuss are N-body+hydro simulations beginning from cosmological initial conditions at $z\ga 100$.  They all contain cooling, star formation, and~stellar feedback, which are necessary to form realistic galaxies.  All but the \mbox{\citet{liang16}} simulation include the SMBH seeding, SMBH accretion, and~AGN feedback, which are necessary to reproduce key properties of groups.  All Magneticum simulations include a passive magnetic field \citep{dolag09} and~IllustrisTNG uses a magnetic hydrodynamic (MHD) solver that self-consistently follows the magnetic field, which is initially seeded in the initial conditions \citep{pakmor13}.  Separate code modules are written for these different aspects and~while our discussion here is not an exhaustive list of code modules in every simulation, we focus on specific methods, prescriptions and~subgrid models that impact the group scale, while providing some additional context.

\subsection{Hydrodynamics} \label{sec:hydro}

Historically, the~choice of hydrodynamic scheme has had significant effects on the appearance of 
simulated clusters, in~particular the entropy profiles near cluster centers.
Grid-based, Eulerian mesh codes tend to produce higher entropy, flatter cores than particle-based schemes as first shown by \mbox{\citet{frenk99}}. 
This behavior is believed to be a result of numerical diffusion and over-mixing in mesh codes and/or the absence of heat diffusion in ``classical'' SPH simulations \citep{wadsley08}.  \mbox{\citet{mitchell09}} explored these phenomena using idealized cluster merger simulations in both FLASH (grid-based) and \gadtwo~(SPH), finding that the former creates flatter high-entropy cores by mixing entropy through vortices and turbulent eddies.  They argued that the suppression of mixing and fluid instabilities (e.g., Kelvin-Helmholtz) in SPH simulations appear to preserve the entropy low of individual gas particles resulting in a power-law entropy distribution reminiscent of a cool core~cluster.  

Recent numerical improvements in modern SPH and AMR codes have greatly improved their ability to recreate a wider range of entropy profiles.  For~example, adding artificial conduction to \gadget~simulations allowed for the creation of cored profiles in SPH simulations by mimicking thermal diffusion, resulting in the production of both cool core (CC) and non-cool core (NCC) entropy profiles across a sample of massive cluster simulations \citep{rasia15}.  \mbox{\citet{hahn17}} ran a set of \ramses~AMR simulations which also produced the dichotomy of CC/NCC clusters.  The~nIFTy simulation code comparison project compared one $10^{15}\ \msolar$ cluster \citep{sembolini16}, finding that modern SPH methods could create entropy cores just as grid-based and moving mesh methods do.
These updated SPH implementations, such as ANARCHY \citep{schaller15} used by EAGLE and the \mbox{\citet{beck16}} scheme used by Magneticum, often include a combination of pressure-entropy formulations of SPH, higher-order SPH kernels, and~new treatments for thermal conduction and artificial viscosity.  Cluster simulation comparison projects using modern SPH schemes (like the nIFTy project and The Three Hundred project \citep{cui18}) now find that hydrodynamic scheme makes less of a difference than the inclusion of subgrid models on the appearance of entropy cores. While the appearance of entropy cores does not necessarily depend on the AGN feedback scheme at the cluster scale, the~lower potentials of galaxy groups may yield a different~answer. 

The majority of the simulations explored here use either \arepo~moving mesh or \gadtwo/\gadget~SPH, either classical or modern as listed in Table~\ref{tab:sims}.  The~\changa ~code has a number of SPH updates putting it in the modern~category.  


\subsection{Gas Cooling and~Heating} \label{sec:cooling}

All simulations we discuss include radiative cooling and photoionization heating by the meta-galactic UV/X-ray background (UVB, for e.g., \cite{haardt12}).  A~module accesses cooling/photo-heating rate tables usually as a function of density, temperature and~redshift.  These lookup tables are usually calculated from CLOUDY \citep{ferland13} models element-by-element, where the redshift dependence accounts for the evolving UVB.  Most of the \gadget~simulations use \mbox{\citet{wiersma09a}} rates calculated for 11 elements, SIMBA uses the GRACKLE library \citep{smith17} that tabulates self-shielding from the UVB for dense gas, and~the \arepo-based simulations track additional photo-heating from local AGN that suppresses cooling in nearby gas \citep{vogelsberger13}.  

For IGrM temperatures and densities that emit in the X-ray, cooling is driven by the balance of recombination and collisional ionization, with~photo-ionization/heating usually playing an insignificant role.  Often ionization equilibrium is assumed, but~some simulations follow non-equilibrium rates of primordial elements (e.g., SIMBA).  \mbox{\citet{cen06}} integrated non-equilibrium tracking of high oxygen ions ($\OV-\OIX$) in uniform mesh simulations, following how these ion species could deviate from equilibrium conditions when the recombination time became significant compared to the cooling time.  \mbox{\citet{oppenheimer16}} ran zoom SPH simulations following all ions from 11 metal species and including collisional \citep{gnat07}, photo-ionization  \citep{gnat12}, Auger ionization and charge exchange~\citep{oppenheimer13a} processes in low-mass groups.  While non-equilibrium processes involving metals are not likely to significantly alter dynamics of cooling at IGrM temperatures, they can have an impact on the observational diagnostics, such as $\OVII$ and $\OVIII$ line emission \citep{gnat07, oppenheimer13a}.  Further exploration of IGrM simulations integrating non-equilibrium ionization and cooling is necessary in light of the coming launch of \XRISM\ that will allow the measurement of line emission in the IGrM and~ICM.

In contrast to simulations with radiative cooling, non-radiative (or adiabatic) simulations do not contain gas cooling (or any galaxy or SMBH formation).  ~\mbox{\citet{frenk99}}, \mbox{\citet{lewis00}}, and~\mbox{\citet{voit05}} ran such simulations, finding that a baseline $K\propto R^{1.1}$ entropy profile is created in the absence of cooling, feedback, and~other processes related to galaxy formation. \footnote{The $\sim$$R^{1.1}$ scaling of the gas entropy profile is set by gravitational infall-related processes, including accretion shocks and subsequent thermalization.  As~shown by \mbox{\citet{lewis00}}, it echoes the pseudo-phase space density profile of the dark matter: $\rho(R)/\sigma^3(R) \propto R^{-1.8}$ \citep{taylor04,Barnes07}.  The~entropy scaling is  $\propto [ \sigma^3/\rho(R)]^{2/3}$}. 

We also note that the ROMULUS simulations do not include metal-line cooling at $T > 10^4\, \mathrm{K}$ \citep{tremmel17}.

\subsection{Star Formation, Stellar Evolution, and~Nucleosynthetic~Production} 

\textls[-10]{The formation of stars, their evolution and~the release of elements are included in subgrid models for all the simulations we discuss.  The~multi-phase ISM is never resolved in separate phases, but~relies on applying subgrid models that encompass multiple phases in a single gas parcel.  A~density threshold, above~which the multi-phase ISM forms, is often the only criterion used for determining if gas can form stars.  Different subgrid ISM models (e.g., \cite{springel03,schaye08,dave16}) use different motivations for their choices of unresolved phases, but~all are calibrated to reproduce the \mbox{\citet{kennicutt98}} SFR surface density as a function of gas surface density relation.  Star particles are spawned from gas parcels probabilistically and~represent a population of individual equal-age stars, since they are at minimum $\sim$$10^5\,\msolar$.  Stellar death is modeled as a time-dependent return of a star particle's mass to surrounding gas as a function of star particle age, based on expectations from stellar evolution.  The~initial mass function (IMF) determines the proportion of stars that die and~can induce a systematic shift in the stellar mass formed per amount of stellar luminosity emitted.  Fortunately, the~simulations used in the comparisons of $M_*$ in Section~\ref{sec:BGG} use either a \mbox{\citet{kroupa01}} IMF (ROMULUS) or a \mbox{\citet{chabrier03}} IMF (every other simulation), which are fairly similar in their~shapes. } 

The release of elements from stars is the source for the enrichment of the IGrM, about which an entire review by \mbox{\cite{Gastaldello21}} is written in this Special Issue ``The Physical Properties of the Groups of Galaxies''.  In~their Section~3.2, they include a discussion of chemical evolution models in cosmological hydrodynamic simulations.  We note here that the simulations we discuss frequently all use chemical evolution models that include the elemental yields taken from stellar evolution models of Type II Supernovae (SNe), Type Ia SNe, and~AGB stars.  We note that yields have a great deal of uncertainty associated with them as explained in \mbox{\citet{wiersma09b}}.  Uncertainties arise in the production of iron from Type Ia SNe since the Type Ia rate is not well constrained, in~the transition stellar mass from Type II SNe to AGB stars, in~the shape of the IMF, and~in the calculations of the elemental yields themselves.  These uncertainties can directly affect the rates of gas cooling (Section~\ref{sec:cooling}), which can be dominated by metal-line emission at many IGrM temperatures (i.e., between $T\approx 10^{4.5}$--$10^{6.5}$ K for a $Z=0.3\ \Zsolar$ \mbox{plasma, \cite{gnat07,oppenheimer13a}).}

\subsection{Metal~Spreading}

The distribution of metals in the IGrM depends on (1) how metals are released from star particles into the surrounding gas; and~(2) how metals diffuse between gas elements.  This is a separate issue from stellar feedback that applies mechanical feedback energy to gas elements, which we discuss in the next~section.  

Metal enrichment to large distances can occur without mechanical feedback if the number of enrichment neighbors extends over a large volume, which may unrealistically enrich diffuse environments.  In~cases where the number of neighbors is large and/or the resolution is low, simulated star particles may artificially enrich diffuse environments like the IGrM without superwinds.  This effect is most obvious in the uniform mesh simulations of \mbox{\citet{cen06}}, where $\OVII$ and $\OVIII$ absorption statistics of diffuse gas remain at similar levels with and without superwinds.  Modern simulations typically release metals over several dozen neighboring gas elements, based on either the smoothing kernel in the case of SPH simulations (e.g., \cite{schaye15}), or~mesh cells in the case of moving mesh simulations (e.g., \cite{pillepich18a}), which appear capable of enriching diffuse environments in simulations without invoking superwind feedback (e.g., \cite{nelson18b}).  \mbox{\citet{tornatore07}} tested the effect of the number of neighboring SPH particles on ICM properties, finding only moderate effects on the ICM metallicity.  On~the other hand, the~\mbox{\citet{dave08}} simulations spread metals over very few (i.e., 3) neighboring SPH particles, resulting in much lower IGrM oxygen metallicity arising from Type II SNe without~mechanical feedback in contrast to their stellar feedback runs that have $\sim$1 dex higher oxygen levels that better agree with observational constraints \citep{Peterson03}.  
However, the~\mbox{\citet{dave08}} IGrM iron enrichment remains similar with and without stellar feedback, indicating enrichment from a different source---Type Ia SNe from older intracluster stars.  Higher IGrM/ICM enrichment levels can also be achieved through increased numerical resolution as reviewed in Section~2.1.2 of \mbox{\citet{biffi2018b}}, owing primarily to the ability to resolve metal-enrichment from smaller halos at high-$z$.  

While metal diffusion occurs naturally in mesh-based codes via advection between gas cells, metal diffusion between gas elements must be explicitly modeled in modern SPH implementations. The~EAGLE simulations use SPH kernel-smoothed metallicities to calculate cooling rates \citep{wiersma09b} to estimate the dynamical effects of metal spreading in a manner that is consistent with the SPH formalism.  The~ROMULUS simulations apply the \mbox{\citet{shen10}} metal diffusion algorithm to mimic turbulent diffusion based on the velocity shear between particles.  Variations of metal mixing implementations are explored using \gizmo~meshless finite mass codes in the FIRE-2 simulations \citep{Hopkins_et_al_2018} and the \mbox{\citet{rennehan21}} simulations, which found that the mixing algorithm significantly affects the resultant CGM and warm-hot IGM metallicity distributions.

In summary, the~algorithm that describes how stars release metals into surrounding gas, the~simulation resolution, and~the metal mixing algorithm (if any is applied), are important considerations when simulating IGrM metallicities as discussed in the companion review by \mbox{\cite{Gastaldello21}}.

\subsection{Stellar~Feedback}

Stellar feedback, directly associated with stellar mass loss and enrichment, is discussed independently and is usually modeled by separate subgrid prescriptions.  This owes to its uncertainty and range of possible outcomes that dramatically change the properties and appearances of galaxies in cosmological simulations (e.g., \cite{oppenheimer10, schaye10, scannapieco12, haas13, torrey14, crain15}).  Stellar feedback has been identified as a key component of the overall solution to the overcooling problem whereby gaseous baryons are too efficiently converted into stars (e.g., \cite{white91}).  Additionally, this feedback provides a pathway to enriching the high-redshift intergalactic medium (IGM; e.g., \cite{oppenheimer09b, keating16, finlator16, rahmati16, bird16, cen11}), which is observed to have metal absorption far from the locations of galaxies (e.g., \cite{cooksey10, becker09, songaila01, dodorico13, matejek12}).  The~enriched high-$z$ IGM certainly contributes to the IGrM metal content of simulated $z\sim0$ groups that show enrichment levels consistent with observations (e.g., \cite{dave08, mccarthy10,planelles14}).  

\textls[-15]{Early cosmological simulations took the expected energy from Type II SNe and imparted velocity kicks to gas articles in star-forming regions \citep{springel03}.  \mbox{\citet{oppenheimer06}}} applied a momentum-driven wind scaling \citep{murray05} to their simulations to mimic the acceleration of dust-driven winds by stellar UV radiation pressure, resulting in greater mass-loading for lower-mass galaxies.  The~large range of stellar feedback prescriptions explored in the OWLS (OverWhelmingly Large Simulations;  \cite{schaye10}) project, including varying the wind mass-loading (as a proportion of the SFR), wind velocity ($\vw$), and~attempting thermally heated wind prescriptions, demonstrated very different outcomes for galaxies and~gas.  

Many modern cosmological simulations treat stellar feedback explicitly as a tunable subgrid model to reproduce key observations of galaxies including the galactic stellar mass function, galaxy sizes, and~their central SMBH masses.  Different physical motivations are used to justify parameter choices, but~the resultant models are often far from the physical mechanisms of feedback.  EAGLE tunes their feedback to heat SPH particles to $10^{7.5}$ K \citep{schaye15}, justifying the choice of temperature to prevent catastrophic cooling and allow feedback to efficiently apply mechanical work \citep{dallavecchia12}.  IllustrisTNG uses a group finder to estimate the dark matter halo velocity dispersion ($\sigma_{\rm DM}$) based upon \mbox{\citet{oppenheimer08}}, and~scale their kinetic wind velocities in proportion to $\sigma_{\rm DM}$ and the mass loading (i.e., the proportion of mass launched relative to SFR) in inverse proportion to $\sigma_{\rm DM}$ to a power that was calibrated   \citep{pillepich18a}.  SIMBA applies the wind velocity and mass loading scalings derived by \mbox{\citet{muratov15}} from FIRE zoom simulations \citep{hopkins14} that followed multiple sources of stellar feedback (radiation pressure, SNe, stellar winds, and~photo-ionization from stars) in their kinetic wind model \citep{dave19}.  

\textls[-15]{The complexity of these modern subgrid stellar feedback prescriptions can be quantified in the amount of energy returned per unit of stellar mass formed.  \mbox{\citet{davies_jj20}} calculated that EAGLE dumps $1.74\times10^{49}$ erg$\,\msolar^{-1}$ via thermal heating, and~IllustrisTNG dumps $1.08\times 10^{49}$ erg$\,\msolar^{-1}$ with 90\% going to a velocity kick and 10\% going to thermal heating.  This is the approximate energy injection rate expected if all SNe energy input into winds, that is,~the expectation if there is $\sim$1 SN with $10^{51}$ ergs per 100 $\msolar$ formed.  SIMBA limits their wind velocity to the total supernova energy available, which applies only to high-$z$ small galaxies. The~ROMULUS simulations apply the  \mbox{\citet{stinson06}} `blastwave' feedback model at an efficiency of $\approx\! 0.75$, with~cooling temporarily disabled to prevent radiative~losses.}  

It should also be noted that metal mass loading is treated differently than total mass loading in some simulations.  Illustris and the TNG simulations apply {\it reduced} metal-loading that is $2.5\times$ lower than mass-loading, because~\mbox{\citet{vogelsberger13}} argued that winds punch through low-density cavities in the ISM, reducing the metallicity of the ejected material.  Alternatively, SIMBA applies higher metal loading that can be up to a factor of twice as high but is more typically 10\%--20\% higher.  

\textls[-10]{While stellar feedback can reduce the efficiency of galaxy formation at lower masses, this mechanism becomes inefficient at preventing star formation in halos greater than $10^{12}\,\msolar$ (e.g., \cite{crain09, oppenheimer10, Bower_et_al_2017}), which is one of the key motivations for applying SMBH~feedback.}

\subsection{Black Hole~Seeding} \label{sec:BHseeding}

\textls[-5]{Subgrid implementations of black hole (BH) seeding (Section~\ref{sec:BHseeding}), accretion \mbox{(Section~\ref{sec:BHgrowth})}}, and~AGN feedback (Section~\ref{sec:AGNfeedback}) in simulations are introduced in Section~5.1 of the companion review by {\citet{Eckert2021}},  
which we expand upon here with discussions of specific implementations applied in the simulations listed in Table~\ref{tab:sims}. Most of these use a version of the \mbox{\citet{Booth_Schaye_2009}} BH seeding module, which applies a friends-of-friends group finder to identify halos, originally performed by \mbox{\citet{dimatteo08}}, then seeding $10^{-3} M_{\rm gas}$ sink particles in halos resolved with 100 dark matter particles (as in OWLS, Cosmo-OWLS, and~BAHAMAS).  

Illustris and EAGLE add $10^5\,\hmsolar$ BH seeds in $10^{10}\,\hmsolar$ halos, while IllustrisTNG uses larger seeds, $8\times 10^5\,\hmsolar$ to avoid the need for boosted Bondi accretion (see next Section).  SIMBA seeds $10^4\,\hmsolar$ BHs when $M_*>10^{9.5}\,\msolar$, skipping the attempt to follow BH accretion in low-mass halos. In~general, dynamical friction is insufficiently resolved, therefore the default of these models is to continually re-position BHs to the local potential minimum (with the exception of Magneticum; see \mbox{\cite{Hirschmann_et_al_2014}}).

The ROMULUS simulations instead seed SMBHs based on local gas properties: they can form whenever and wherever (i) gas density is 15 times the threshold for star formation; (ii) the local metallicity is low ($Z < 3\times10^{-4}$); and~(iii) the temperature is just below the limit for atomic cooling.  Consequently, SMBHs appear in ROMULUS at a higher redshift and in lower mass halos ($10^{8}$--$10^{9}\;\msolar$) than in the other models discussed above. Moreover, SMBHs are not pinned to the local potential minimum; instead, the~effects of unresolved dynamical friction are captured with a subgrid model \citep{tremmel15}.  This degree of freedom alters the SMBHs' growth and feedback trajectories \citep{tremmel17,tremmel19}.


\subsection{Black Hole~Growth} \label{sec:BHgrowth}

Once seeded, black holes primarily grow through the accretion of gas, but~can also grow through SMBH-SMBH mergers.  For~accretion, the~most common implementation uses the Bondi-Hoyle accretion rate limited to the Eddington luminosity.  As~discussed in Section~5.1 of the companion review by \cite{Eckert2021}, initial implementations boosted the Bondi-Hoyle accretion rate by a large constant factor ($\approx$100) to compensate for the lack of dense gas in early low-resolution simulations \citep{springel05}. Instead, \mbox{\citet{Booth_Schaye_2009}} introduced a density-dependent boost to the Bondi-Hoyle accretion rate that only affects black holes surrounded by gas with an (unresolved) dense~phase.  

The \mbox{\citet{Booth_Schaye_2009}} boost formula is used by many simulations, including cosmo-OWLS, BAHAMAS, Horizon-AGN, ROMULUS, while Illustris and FABLE use the older constant \mbox{\citet{springel05}} boost.  Magneticum uses a temperature-dependent \mbox{boost \citep{steinborn15}} that increases for cooler gas to approximate turbulent-driven chaotic cold accretion rates \citep{Gaspari13}.  (C-)EAGLE and IllustrisTNG forgo the need for boosted Bondi at their higher resolutions.  The~former modify their accretion rates by an additional viscous timescale to account for the angular momentum of infalling gas \citep{rosasguevara15}, with~the intention to significantly \emph{reduce} SMBH growth in low-mass halos.  In~this model, a~longer viscous timescale due to high angular momentum of gas around the black hole delays accretion and translates into a higher $M_\star$ threshold where the SMBH accretion rate approaches the $\propto M_{\rm BH}^2$ Bondi limit~(\cite{crain15}, but see \mbox{\cite{Bower_et_al_2017}}).  A~similar motivation is used in ROMULUS to modify the Bondi rate to account for gas rotation in addition to the relative velocity between the gas and the SMBH \citep{tremmel17}.  

\textls[-20]{SIMBA calculates black hole accretion rates from cool gas with the `torque-limited accretion' model \citep{hopkins11,anglesalcazar17}, in~an attempt to diverge from the self-regulated nature of Bondi accretion \citep{Booth_Schaye_2009}. This model results in a much shallower dependence on $M_{\rm BH}$ than the Bondi formula \citep{dave20}, while also recognizing that there are preferred orientations for accretion in addition to feedback. Accretion from hot gas is modeled with the standard Bondi-Hoyle formula, but~the torque-limited mode from cool gas generally dominates SMBH growth in~SIMBA.}


\subsection{AGN~Feedback} \label{sec:AGNfeedback}

\begin{figure}
\includegraphics[width=0.9\textwidth]{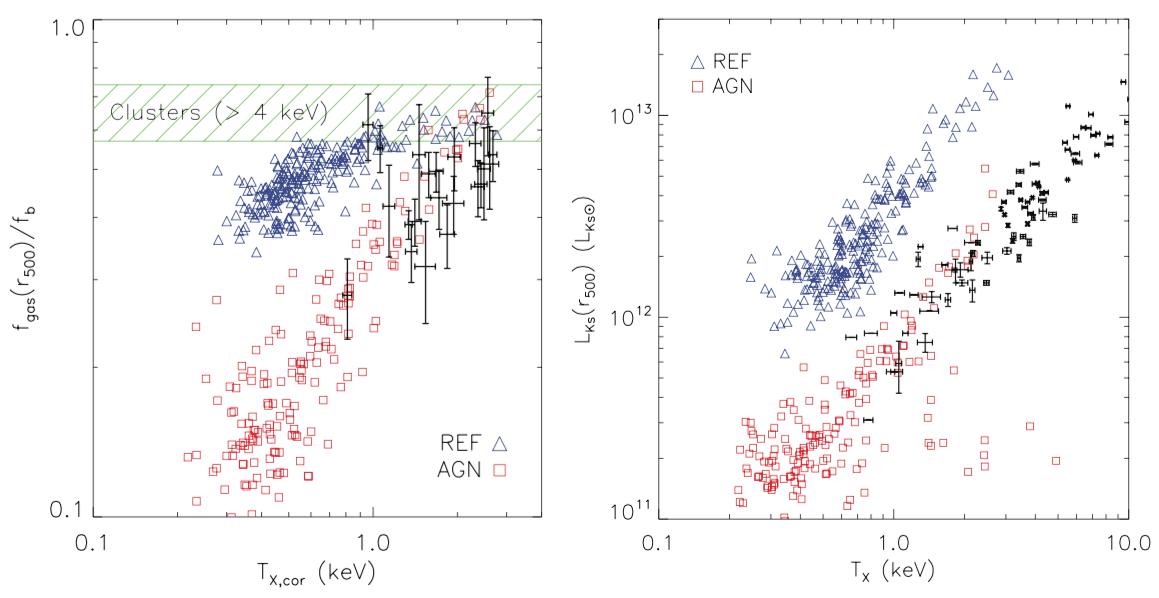}
\caption{{Figure panels adapted with permission from} 
 \mbox{\citet{mccarthy10}} showing OWLS simulations, with~only stellar feedback (REF, blue triangles) and with additional AGN feedback (AGN, red squares).  ({\bf Left})~The AGN feedback provide a better fit to the \mbox{\citet{sun09}} observationally derived $f_{500}$ data (black points) as a function of X-ray temperature; ({\bf Right})~AGN feedback reduces the integrated K-band luminosity within $R_{500}$ to observable values compiled by \mbox{\citet{lin04}}. } 
\label{fig:mccarthy2010}
\end{figure}

Groups provide some of the best evidence that AGN feedback significantly transforms the distribution of baryons, both in terms of the bulk transport of gas outward and the reduction in star formation \citep{mccarthy10, henden18}.  Figure~\ref{fig:mccarthy2010} demonstrates how adding AGN feedback using the \mbox{\citet{Booth_Schaye_2009}} AGN prescription both reduces the gas fraction within $R_{500}$ and the luminosities of galaxies \citep{mccarthy10}.  

Simulations usually apply an AGN feedback model that uses an AGN feedback efficiency term, $\varepsilon$, such that the feedback power is defined as $\dot{E}_{\rm AGN} = \varepsilon \dot{M}_{\rm BH} c^2$.  Physically, $\varepsilon$ is a product of two efficiencies, the~radiative efficiency ($\varepsilon_{r}$), and~the feedback efficiency ($\varepsilon_{f}$).  The~radiative efficiency is the energy radiated away from the accretion onto the black hole, and~is often assumed to be 10\% based on the accretion onto a Schwarzschild BH \citep{Shakura_Sunyaev_1973}.  The~feedback efficiency is the fraction of the radiated energy that is imparted to the gas either thermally and/or~kinetically.  

\begin{table}
\caption{AGN feedback modules used in simulations.}
\centering
\tablesize{\scriptsize} 
\begin{tabular}{lllllll}
\toprule
\textbf{Simulation}	& \textbf{Mode}	& \textbf{Injection} & \textbf{Energy Dump} & \textbf{Efficiency ($\varepsilon$)} & \textbf{Frequency} & \textbf{Loading Factor}\\
\midrule
cosmo-OWLS   & --            & Thermal        & $10^{8.0}$ K & 0.015 & Build-up & 1 particle\\
Illustris    & Quasar        & Thermal        & -- & 0.01 & Continuous & \\
--           & Radio         & Bubble        & -- & 0.07 & $\delta M_{\mathrm BH}=0.15$ \\
EAGLE		 & -- 			& Thermal         & $10^{8.5}$ K & 0.015 & Build-up & 1 particle\\
Horizon-AGN  & Quasar       & Thermal         & $10^7$ K  & 0.015 & Build-up \\
--           & Jet         & Kinetic          & $10^4\ \kms$ & 0.10 & Continuous \\ 
BAHAMAS      & --           & Thermal         & $10^{7.8}$ K & 0.015 & Build-up & 20 particles \\
C-EAGLE/Hydrangea	 & --			& Thermal         & $10^{9.0}$ K& 0.015 & Build-up & 1 particle\\
ROMULUS      & --           & Thermal         & -- & 0.002 & Continuous \\
FABLE        & Quasar       & Thermal         & -- & 0.01   & $\delta t=25$ Myr & \\
--           & Radio        & Bubble & -- &  0.08 & $\delta M_{\mathrm BH}=0.01$ & \\
IllustrisTNG & ``High''       & Thermal         & -- & 0.02  & Continuous & \\
--           & ``Low''       & Kinetic ``Pulse'' & -- & $\leq 0.2$ & Build-Up & {\tiny \citet{weinberger17} Eq. 13} \\
SIMBA     & ``Radiative''  & Kinetic         & $1000\ \kms$ & 0.003$^a$ & Continuous & \\
--           & ``Jet''        & Kinetic ``Jet'' &  $8000\ \kms$ &  0.03$^a$  &  Continuous  & \\
\bottomrule
\end{tabular}
\parbox{20cm}{
\scriptsize
$^a$ Values for a $10^9 \msolar$ SMBH.  
}
\label{tab:AGN}
\end{table}

\textls[-5]{A number of simulations that we list in Table~\ref{tab:AGN} use a single-mode thermal model assuming $\varepsilon_{f}=0.15$ (i.e.,~resulting in $\varepsilon=0.015$), based on the value calibrated by \mbox{\citet{Booth_Schaye_2009}} to reproduce the SMBH mass-halo mass relation derived from observations.  The~EAGLE Reference simulations apply a temperature increase of $\Delta T= 10^{8.5}$ K to a neighboring SPH particle, once enough SMBH feedback energy has been accumulated to heat at least one SPH particle.  The~C-EAGLE/Hydrangea simulations run at the same resolution use a $\Delta T= 10^{9.0}$ K, which leads to less frequent, more bursty feedback that becomes more effective when heating gas above the virial temperatures of massive clusters \citep{mccarthy11}.  The~multiple cosmo-OWLS simulations varied $\Delta T$ while keeping $\varepsilon=0.015$ \citep{lebrun14}.  Their spatial resolution, nearly 1 dex lower than EAGLE, makes their heating temperature not directly comparable to the EAGLE $\Delta T$, since heating a single particle involves injecting a much larger amount of energy. In~both cases, however, larger $\Delta T$ leads to a higher efficiency in transporting baryons beyond the virial radius. In~BAHAMAS, feedback energy is accumulated until multiple gas particles can be heated simultaneously; \mbox{\citet{mccarthy17}} found that heating 20 SPH particles ($\sim$$2\times 10^{10}\ \msolar$) to $\Delta T = 10^{7.8}$ K, which was chosen as a calibration to prevent over-efficient assembly of intermediate-mass galaxies (see {their Figure~3}).} 


ROMULUS, with~its higher resolution, uses a much lower efficiency ($\varepsilon=0.002$), which was selected from a parameter search performed by \mbox{\citet{tremmel17}} using galaxies in $10^{11-12}\,\msolar$ halos and not group/cluster-mass objects. Unlike stellar feedback, AGN feedback in ROMULUS is \emph{not} subject to cooling shutoff.

\mbox{\citet{sijacki07}} introduced the dual AGN model, with~a ``quasar'' mode injecting thermal energy at high accretion rates, which is often defined as a relative fraction of the Eddington accretion rate, $f_{\rm Edd} \equiv \dot{m}_\mathrm{Edd}/\dot{m}_\mathrm{BH}$ where $\dot{m}_\mathrm{Edd}$ is the Eddington accretion limit, and~a ``radio'' mode injecting mechanical energy in the form of bubbles at lower accretion rates.  The~radio mode (at low Eddington ratios) has more energy to impart into the gas since it bypasses the (inefficient) conversion from thermal to kinetic energy. Illustris switches to this mode when $f_{\rm Edd}<0.05$, and~releases built-up bubble events with sizes $\sim$$100$ kpc when the BH grows by 15\% ($\delta M_{\rm BH}=0.15$).  The~aggressiveness of this scheme over-evacuates the IGrM \citep{genel14}.  FABLE applied the same algorithm with similar energy efficiencies, but~using smaller bubble events and a lower $f_{\rm Edd}=0.01$ transition to radio mode that were the result of tuning their AGN feedback to reproduce IGrM gas fractions and galaxy stellar masses.  Finally, more recent Magneticum simulations use the \mbox{\citet{steinborn15}} thermal AGN feedback model that smoothly transitions between radio-mode and quasar-mode based on the fractional Eddington accretion~rate.  

\textls[-5]{Several simulations attempt to simulate jets or jet-like feedback at low Eddington rates.  Horizon-AGN uses the \mbox{\citet{dubois12}} bipolar kinetic jet aligned with the spin of the SMBH.  IllustrisTNG integrates the \mbox{\citet{weinberger17}} model, where randomly-oriented, directional kinetic ``pulses'' predominantly take over once a SMBH grows above a certain mass based on their Equation~(5).  In~practice, this choice yields significant implications as TNG100 SMBHs growing above a threshold mass of $M_{\rm SMBH}\approx 10^{8.1}\,\msolar$ at late times leads to a sharp transition of galaxy properties (reduced SFRs, quenched galaxies, redder colors) and gaseous halo properties (lower $f_{\rm gas}$) \citep{davies_jj19, terrazas20}. Low Eddington ratio feedback in SIMBA uses a bipolar kinetic jet perpendicular to the angular momentum of the local disc that has a constant momentum input rate ($20\times$ the radiative luminosity of the accretion disc divided by the speed of light).  This is meant to complement the preferred orientation of the \mbox{\citet{dave19}} torque-limited accretion model.  This model also decouples for $10^{-4}$ of a Hubble time, which can transport launched jet particles over $\sim\! 10$ kpc.}  

\subsection{Transport Processes and Magnetic~Fields}\label{sec:3.9}

It is fair to say that the effects of viscosity, thermal conduction, and~turbulence (which we collectively refer to as `transport processes') on the IGrM have received significantly less attention in comparison to, for~example, the~overall gravitational and hydrodynamical evolution of groups and the impact of processes associated with galaxy formation (such as radiative cooling and feedback processes).  Indeed, from~the cosmological simulations standpoint, the~vast majority of existing simulations completely neglect the roles of viscosity and conduction (treating the IGrM as an inviscid and non-conducting fluid), while the effects of turbulence are generally only captured on relatively large, well-resolved~scales.  

\textls[-5]{However, there are no a priori compelling physical reasons for neglecting these processes, as~the IGrM and the ICM are both plasmas where, generally speaking, one expects such transport processes to be active \citep{nulsen82}.  To~include their effects in cosmological simulations is non-trivial, though, as~the hydro solvers that normally evaluate the standard (inviscid) hydrodynamic equations must be replaced with more complex solvers capable of evaluating the full Navier Stokes equations in a stable fashion, at~least if one wishes to model the effects of viscosity (e.g., \mbox{\cite{sijacki06}}).  Furthermore, since the transport of heat by viscosity and conduction preferentially occurs along magnetic field lines (and is strongly suppressed perpendicular to the field lines), it really only makes sense to include their effects in the context of MHD simulations that self-consistently follow the evolution of the magnetic fields.  While there is a growing interest in the inclusion of anisotropic thermal conduction, viscosity and magnetic fields in simulations (e.g., \mbox{\cite{springel10,hopkins17}}), we are presently unaware of any large-scale cosmological simulations that include both and have evaluated their impacts on the plasma in groups and clusters \footnote{IllustrisTNG does include anisotropic thermal conduction (and magnetic fields), though~the impact on the evolution on hot gas in groups has not yet, to~our knowledge, been examined in detail.  However, \mbox{\citet{barnes19}} have examined the impact of anisotropic thermal conduction on more massive clusters, concluded that it has the effect of making cool cores more prevalent}. 
  In~terms of modeling turbulence, modern hydrodynamical solvers accurately follow the cascade of energy, momentum, and~mass down to the effective resolution scale of the simulations but generally not below this.  Thus, there are ongoing efforts to model the turbulent cascade to smaller scales using physically-motivated subgrid models (e.g., \mbox{\cite{shen10,rennehan19}}).  Whether turbulence plays a large role or if viscosity is able to strongly damp the cascade depends on the dimensionless Reynolds number but, at~present, this is a very poorly constrained quantity for the IGrM and~ICM.}

What impact might we expect from including anisotropic conduction, viscous heating and/or small-scale turbulence in simulations of the IGrM?  After the launch of {\it Chandra} there was much excitement with the discovery of large bubbles of relativistic plasma that appear, at~least in some cases, to~remain structurally intact even after buoyantly rising to relatively large distances from their inflation points (e.g., \mbox{\cite{fabian00}}).  This is difficult to understand from an unmagnetized, inviscid (and therefore highly turbulent) fluid standpoint, as~the bubbles should have been rapidly destroyed/mixed via Kelvin-Helmholtz and Rayleigh-Taylor instabilities.  High-resolution idealized simulations of viscous and/or magnetized clusters, however, demonstrated that the bubbles were much more structurally stable and long-lived when these processes were included (e.g., \mbox{\cite{bruggen05,dursi08}}).  Thus, how and where bubbles transmit their energy to the IGrM will likely be impacted by the inclusion of transport processes and magnetic fields.  In~addition, damping of sound waves (e.g., produced during the inflation of bubbles) via the viscous friction has been proposed as another way in which the central AGN might couple its energy to the gas over a large volume (e.g., \mbox{\cite{fabian03,ruszkowski04}}).  The~evolution of satellite galaxies (i.e., when and how fast they are stripped) may also be strongly affected by the inclusion of conduction, viscosity, magnetic fields and/or small-scale turbulence (see, e.g.,~Figure~8 of \mbox{\cite{sijacki06}} for a dramatic demonstration of viscous stripping in clusters).  While, theoretically, we expect thermal conduction and viscous dissipation to be considerably more important in massive clusters compared to groups, as~the transport coefficients have strong temperature dependencies, much depends on the geometry of the magnetic field lines and, at~present, the~impact of these transport processes on the evolution of the IGrM and the satellites in groups remains an important unknown.  The~advent of new codes such as \arepo~and \gizmo~that are capable of accurately incorporating their effects is a promising step in the right direction and we expect to see important progress in answering these questions in the coming~years.





\subsection{Cosmic~Rays}\label{sec:3.10}
Although cosmic ray (CR) physics has not yet been included in cosmological simulations of groups or clusters, recent advancements in CR hydrodynamics and results from both idealized simulations of massive galaxies and cosmological simulations of lower mass galaxies demonstrate that cosmic rays may be an important source of pressure and energy in galaxy groups. In~this subsection we briefly describe existing approaches to modeling CR physics and their expected effects in galaxy~groups. 

In hydrodynamics galaxy simulations, cosmic rays are typically modeled as a relativistic fluid of GeV protons, separate from the thermal gas \citep{jiang18, chan19, thomas19}. This CR fluid advects with the gas and can provide non-thermal pressure support, inject momentum, or~heat the gas. Additionally, cosmic rays can move relative to the gas through diffusion or streaming, both of which are approximations of the bulk flow of CR energy density along magnetic field lines (see \mbox{\cite{zweibel13, zweibel17}} for a comprehensive review).  Unfortunately, there is no empirical or theoretical consensus on the CR pressure in galaxy groups or on the correct model for CR hydrodynamics. For~this reason, quantitative predictions of galaxy, outflow, and~halo properties from simulations can vary by orders of magnitude depending on the model parameters \citep{wiener17, butsky18, hopkins21_transport}. However, there are several \textit{qualitative} ways in which cosmic rays alter galaxy and halo properties that are consistently demonstrated in~simulations. 

\begin{figure}
\includegraphics[width=0.65\textwidth]{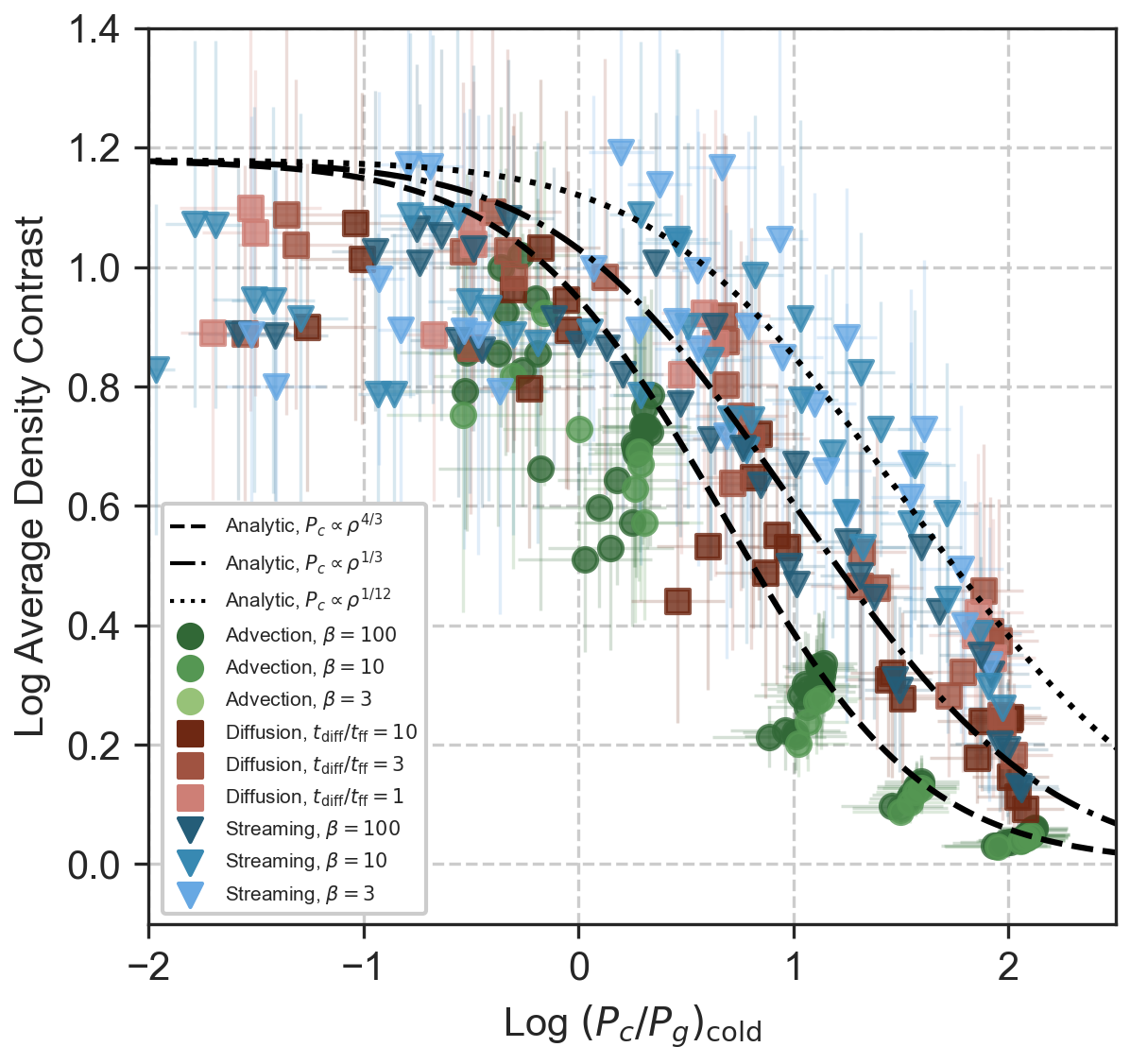}
\caption{The average density contrast between cold and hot gas phases as a function of cosmic ray pressure support in the cold gas \citep{butsky20}. Each point represents a time-averaged measurement from an idealized simulation of thermal instability with varying initial conditions and cosmic ray physics. The~black lines show different analytical predictions for various degrees of coupling between cosmic rays and gas. With~increasing cosmic ray pressure, the~density of cool gas decreases. However, the~detailed quantitative predictions are sensitive to the invoked cosmic ray transport~model. }
\label{fig:cr_density}
\end{figure}

\begin{enumerate}[leftmargin=3em,labelsep=2.8mm]
\item[(1)] \textls[-25]{Cosmic ray transport (either streaming or diffusion) can drive galactic outflows \mbox{\citep{ipavich75, uhlig12, pakmor16, ruszkowski17, hopkins21_outflows}}}. Cosmic rays are injected into the ISM during stellar feedback events (typically $\sim$10\% of the total supernova energy). CR transport redistributes CR pressure out of the galaxy, creating a non-thermal pressure gradient that exerts a force opposing gravity. If~the force exerted by the CR pressure gradient is sufficiently strong,  it will trigger galactic outflows. Relative to thermally driven winds, CR driven winds are cooler, smoother, and~more mass-loaded \citep{girichidis18}. Additionally, since cosmic rays do not suffer radiative losses, CR-driven winds may continue accelerating gas at large distances from the galactic disc. However, since the gravitational force is stronger in more massive galaxies, CR-driven winds may become inefficient in galaxy groups~\citep{jacob18}.  

\item[(2)] Streaming cosmic rays impart energy to heat the surrounding gas. In~massive galaxies, this CR heating rate can efficiently balance radiative cooling, preventing a cooling catastrophe (e.g., \cite{guo08,ensslin11,wiener13,su20}). CR heating may also be a key aspect in the self-regulated AGN feedback cycle. As~cosmic rays lose energy, gas cools more efficiently, fueling AGN feedback which re-injects CR energy into the IGrM \citep{jacob17}.    

\item[(3)] CR pressure qualitatively alters the structure of multiphase gas in galactic halos (e.g., \cite{salem16, butsky18, ji20, buck20}). Non-thermal pressure support enables cool gas to exist at lower densities than expected from purely thermal equilibrium. Figure~\ref{fig:cr_density} demonstrates how the density contrast between cold and hot gas in a two-phase medium diminishes with increasing CR pressure support. In~the extreme case of a CR pressure-dominated galaxy halo, cool and hot gas can exist at the same densities. However, CR pressure is unlikely to be the dominant source of pressure in the halos of massive galaxies. Therefore, the~likely effect of cosmic rays in the IGrM is a modest decrease of cool cloud and cool filament densities \citep{sharma10, ruszkowski18}. 
\end{enumerate}

\section{Results of Simulations at the Group~Scale} \label{sec:results}

\subsection{The Baryonic Content of Group~Halos} \label{sec:baryons} 

The intermediate scale of group halos between galactic halos and clusters provides a unique lever arm on the nature of superwind feedback.  Unlike most galactic halos, the~temperature, density, and~metallicity of their gaseous baryons can be probed via soft X-ray emission without stacking, while their shallower gravitational potential wells relative to clusters allow energetic feedback to remove a significant fraction of baryons.    Twenty years ago, IGrM observations from missions including {\it ROSAT} provided only singular data points per group, which were then compiled into the $L_{X}-T_{X}$ relation.  The~steep $L_{X} \propto T_{X}^{4.9}$ relation observed for groups \citep{helsdon00} relative to clusters \citep{white97} indicated a significant deviation from the $L_{X} \propto T_{X}^2$ relationship expected for virialized gaseous halos retaining all their baryons in hydrostatic equilibrium (e.g., \cite{balogh99}, see companion review by \mbox{ \cite{lovisari21}}). 

With the launch of more powerful X-ray missions, such as {\it Chandra} and {\it XMM-Newton}, hot gas profiles around groups could be resolved. This revealed that the IGrM had far fewer gaseous baryons than the cosmic ratio, $\fb\equiv\Omega_{\rm b}/\Omega_{\rm_M}\simeq 0.16$, inside $R_{2500}$ and a still low proportion inside $R_{500}$ \cite{sun09,lovisari15}.  Altogether these results indicated that many of the baryons are removed from the hot phase.  Earlier theoretical work argued that more efficient line cooling at cooler group temperatures, as~opposed to clusters where Bremsstrahlung cooling dominates ($\Lambda_{\rm cool}\propto T^{0.5}$), could more efficiently build the stellar component of \mbox{galaxies \citep{bryan00,dave02}}.  While \mbox{\citet{gonzalez07}} measured that the integrated group stellar masses from the BGG, the~intragroup light (IGrL), and~satellite galaxies could account for the missing gaseous baryon resulting in a groups retaining all baryons inside $R_{500}$, this was later challenged.  \mbox{\citet{balogh08}} argued that groups could not be far more efficient than clusters in converting their gas to stars due to the hierarchical requirement that clusters are assembled from progenitor groups.  More recent compilations of group stellar masses summed from BGGs, satellites and~IGrL \citep{kravtsov18} find lower stellar mass fractions, strongly suggesting that groups are missing~baryons.

Cosmological volume simulations with periodic volumes 100{--}150 comoving Mpc on a side contain populations of groups that can be statistically compared to observations.  Initial simulations by \mbox{\citet{dave08}}, including only stellar superwind feedback, were able to enrich the IGrM to observed levels \citep{helsdon00} and additionally add entropy to group halos as observed \citep{ponman03}; however the continued late-time star formation in these groups is a telltale sign that these simulations fail to solve the cooling flow problem \citep{fabian84}.  \mbox{\citet{liang16}} used \gadtwo~SPH simulations similar to \mbox{\citet{dave08}} to successfully fit a range of X-ray observations, including IGrM masses inside $R_{500}$.  Nevertheless, the~total stellar masses exceeded observations by at least a factor of two, and~the total baryon content (gas+stars) of groups exceeded 80\% of the cosmic fraction.  These simulations demonstrated that it is possible to reproduce a wide range of IGrM properties, while assembling the wrong galaxies, which in these cases had too much late-time star~formation.   

\mbox{\citet{mccarthy10}} made the case for AGN feedback by using simulations from the OWLS \citep{schaye10} suite of simulations by comparing to \Chandra~observations from \mbox{\citet{sun09}} that resolved X-ray emission profiles out to $R_{500}$ in group-scale objects.  Comparing OWLS simulations without and with \mbox{\citet{Booth_Schaye_2009}} AGN feedback, \mbox{\citet{mccarthy10}} demonstrated that the latter could much better reproduce $f_{\rm gas}(R_{500})$, the~$L_X-T_X$ relation, as~well as the stellar K-band luminosity of the BGG and all stars within $R_{500}$ for the approximately 200 group-sized objects in $100\ \hmpc$ boxes with gas element resolution of $1.2\times 10^8\ \msolar$  (see Figure~\ref{fig:mccarthy2010}).  We refer the reader to the {\citet{Eckert2021}} companion review (their Section~5.2) for a discussion of a broader range of simulations and their baryon fractions within $R_{500}$. 

\subsubsection{Gaseous and Stellar Masses in Recent~Simulations} \label{sec:barcontent}

\begin{figure}
\includegraphics[width=0.33\textwidth]{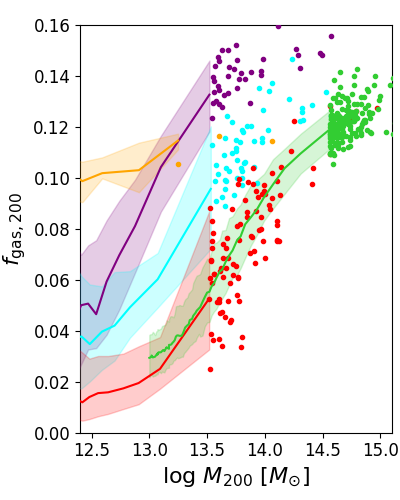}
\includegraphics[width=0.33\textwidth]{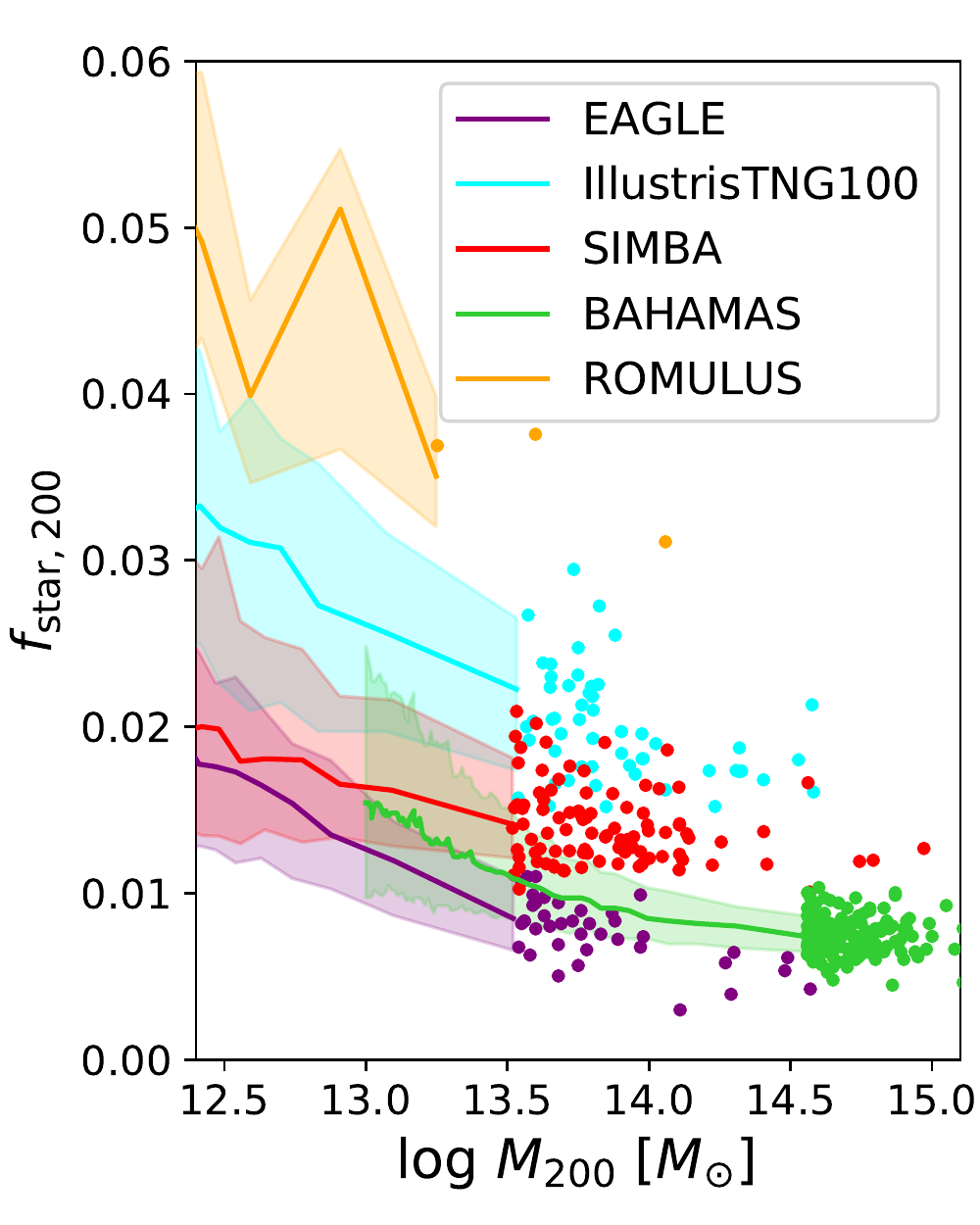}
\includegraphics[width=0.33\textwidth]{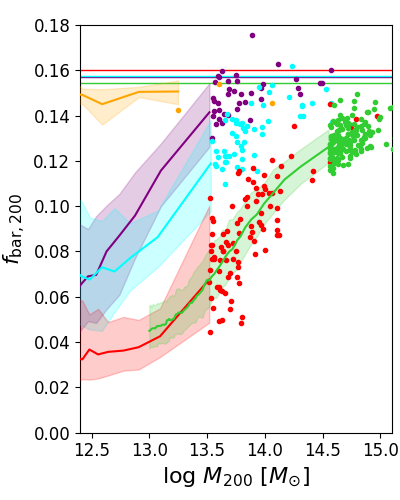}
\caption{Gas, stellar and~combined baryon fractions inside $R_{200}$ from a variety of widely-used, contemporary cosmological hydrodynamic simulations.  Medians and 1-$\sigma$ spreads are shown with lines and shading up to a mass where individual objects are plotted as points.  The~cosmic baryon fraction $f_b = \Omega_{\rm b}/\Omega_{\rm M}$ is plotted in the right-hand panel for the given simulation cosmology.  IGrM measurements rarely extend to $R_{200}$, therefore this plot does not include observations.  It is notable that stellar contents do vary by a factor of more than $3\times$ between different simulations in the group regime, and~more than double between EAGLE and IllustrisTNG.  ROMULUS retains almost all of its baryons within $R_{200}$ while BAHAMAS and SIMBA eject more than half their baryons at $M_{200} \la 10^{13.5}\ \msolar$. For~reference, $M_{500}$ is typically 0.16 dex lower than $M_{200}$ for this halo mass~range.}
\label{fig:mhalo_fgas}
\end{figure}

One can consider as ``contemporary'' intermediate resolution simulations those that resolve gas resolution elements at $\sim\! 10^6\ \msolar$ mass resolution in $\sim\! 10^6$ Mpc$^3$ volumes. The~first simulations to satisfy these criteria were Illustris \citep{vogelsberger14} and EAGLE \citep{schaye15}.  Both simulations follow SMBH growth and AGN feedback, and~are tuned to match the $z=0$ stellar mass function, some other galaxy characteristics such as galaxy sizes, and~SMBH demographics.  However, they both were not tuned to reproduce the properties of X-ray emitting halos, and~fail to reproduce gaseous properties of groups and poor clusters.  \mbox{\citet{genel14}} showed that Illustris severely underpredicts the IGrM mass within $R_{500}$, by~as much as a factor of $10\times$ compared to observations by \mbox{\citet{giodini09}}.  EAGLE produces group halos with much higher gas fractions, exceeding 80\% of $\fb$ at $M_{200}>10^{13.5}\ \msolar$ \citep{schaller15b}.  We plot gas, stellar, and~baryon fractions inside $R_{200}$ as a function of $M_{200}$ in Figure~\ref{fig:mhalo_fgas} for EAGLE and other contemporary simulations \footnote{The companion review by \cite{Eckert2021}, {Figures~14 and 15}, 
plots gas fractions, $f_{\rm gas,500}$, as~a function of $M_{500}$, which are more evacuated for a given halo than $f_{\rm gas,200}$ as plotted here.  The~gas, stellar, and~baryon fractions within $R_{200}$ are less well constrained due to group X-ray measurements not extending out to $R_{200}$ yet, and~therefore represent predictions for future observations}.  
\mbox{\citet{schaye15}} found the IGrM masses derived from virtual X-ray observations to be too high by a factor of two and the $L_{X}-T_{X}$ relationship to be too luminous for a given $T_{X}$.  This reveals that simulations tuned to fit galaxies can deviate significantly for gaseous properties on group scales, which is why accurately simulating the IGrM can provide orthogonal constraints on the processes governing galaxy formation and~evolution.

BAHAMAS explicitly tuned their AGN feedback prescriptions to reproduce properties of groups/clusters and massive galaxies as explained in Section~\ref{sec:AGNfeedback}.  The~FABLE simulations (\cite{henden18}, not shown in Figure~\ref{fig:mhalo_fgas}) also explicitly tuned their feedback to reproduce massive halos, at~$\sim\!\!100\times$ higher mass resolution than BAHAMAS but in a $1000\times$ smaller volume that is augmented by a series of zooms extending up to cluster masses.  These simulations have $f_{\rm gas, 500}$ for a $M_{500}=10^{13.5}\ \msolar$ between 0.04 and 0.07 in Table~1, which reflects the uncertainty in the \mbox{\citet{sun09}} and \mbox{\citet{lovisari15}} observations ({see Figure~5 of the companion review} by {\citet{Eckert2021}}).  

The C-EAGLE/Hydrangea zooms \citep{bahe17, barnes17} use the EAGLE model with a higher AGN heating temperature ($\Delta T=10^{9.0}\,$K). As~shown by \mbox{\citet{schaye15}}, this change lowers (i.e., improves) $f_{\rm gas, 500}$ in $M_{500} \lesssim 10^{13.5}\ \msolar$ groups, but~more massive objects remain too baryon rich, especially in the regime of rich groups/poor clusters \citep{barnes17}. Similarly, the~IllustrisTNG AGN feedback was calibrated to $f_{\rm gas,\, 500}$ of groups at TNG100 resolution \citep{weinberger17}, though~the simulations still predict $f_{\rm bar,\,200}\ga 0.12$ for halos with $M_{200} > 10^{13.5}\ \msolar$, substantially higher than the explicitly calibrated BAHAMAS simulation (see the right-hand panel of Figure~\ref{fig:mhalo_fgas}). SIMBA \citep{dave19} predicts $f_{\rm gas,200} \approx 0.05$ at $M_\mathrm{500} = 10^{13.5}\,\msol$ and agrees well with BAHAMAS in this metric within the overlapping halo mass range. Interestingly, the~companion review by \citet{Eckert2021}, however, shows that $f_{\rm gas,500}$ in SIMBA increases more rapidly with halo mass in SIMBA compared to BAHAMAS ({their Figure~15}), 
so that more baryons are contained near the center of poor clusters with $M_\mathrm{500} = 10^{14-14.5}\ \msolar$. We  will return to the importance of the radial range for assessing the baryon content of simulated groups~below.

Higher resolution simulations ($m_\mathrm{bar} \sim 10^5\,\msol$) currently contain at best a handful of group halos, but~still yield constraining results. The~ROMULUS suite \citep{tremmel17, tremmel19} predicts $f_{\rm gas,200}=0.10-0.12$ between $M_{200}=10^{12.5-14.0}\ \msolar$ and a higher $f_{*,200}$ than all the aforementioned simulations (orange lines in Figure~\ref{fig:mhalo_fgas}). These halos therefore retain nearly all their baryons, in~strong conflict with the observational evidence outlined above. It is unlikely that this shift is a direct consequence of the higher resolution, since the higher-resolution simulation of the IllustrisTNG family, TNG50 \citep{nelson19}, predicts $f_{\rm gas, 500}$ rising from 0.06 to 0.12 over the mass range $M_{500} = 10^{13.0-14.0}\ \msolar$ (not shown), in~excellent convergence with TNG100 (we find a similar result for the lower-resolution version, TNG300).


\begin{figure}
\includegraphics[width=0.33\textwidth]{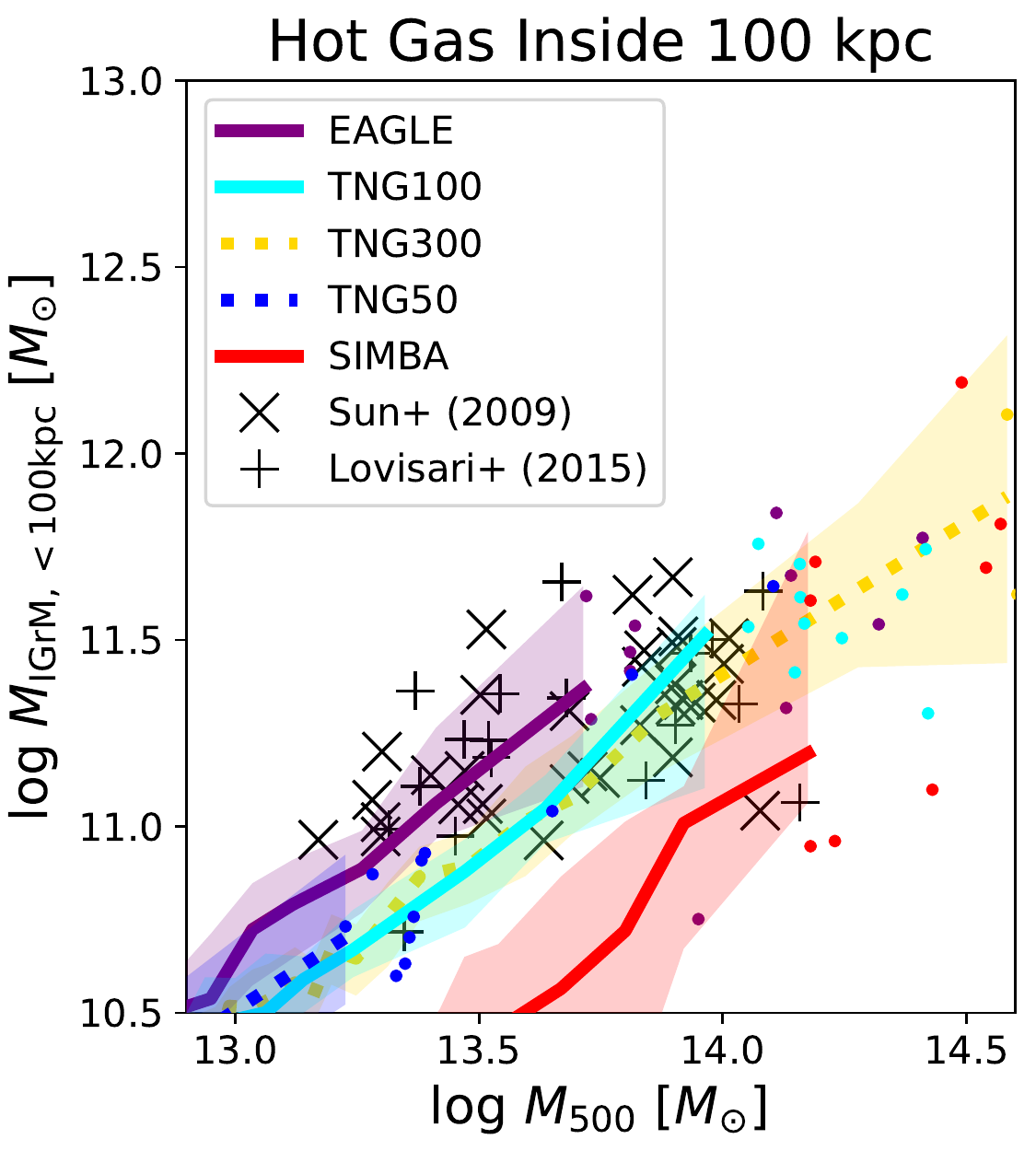}
\includegraphics[width=0.33\textwidth]{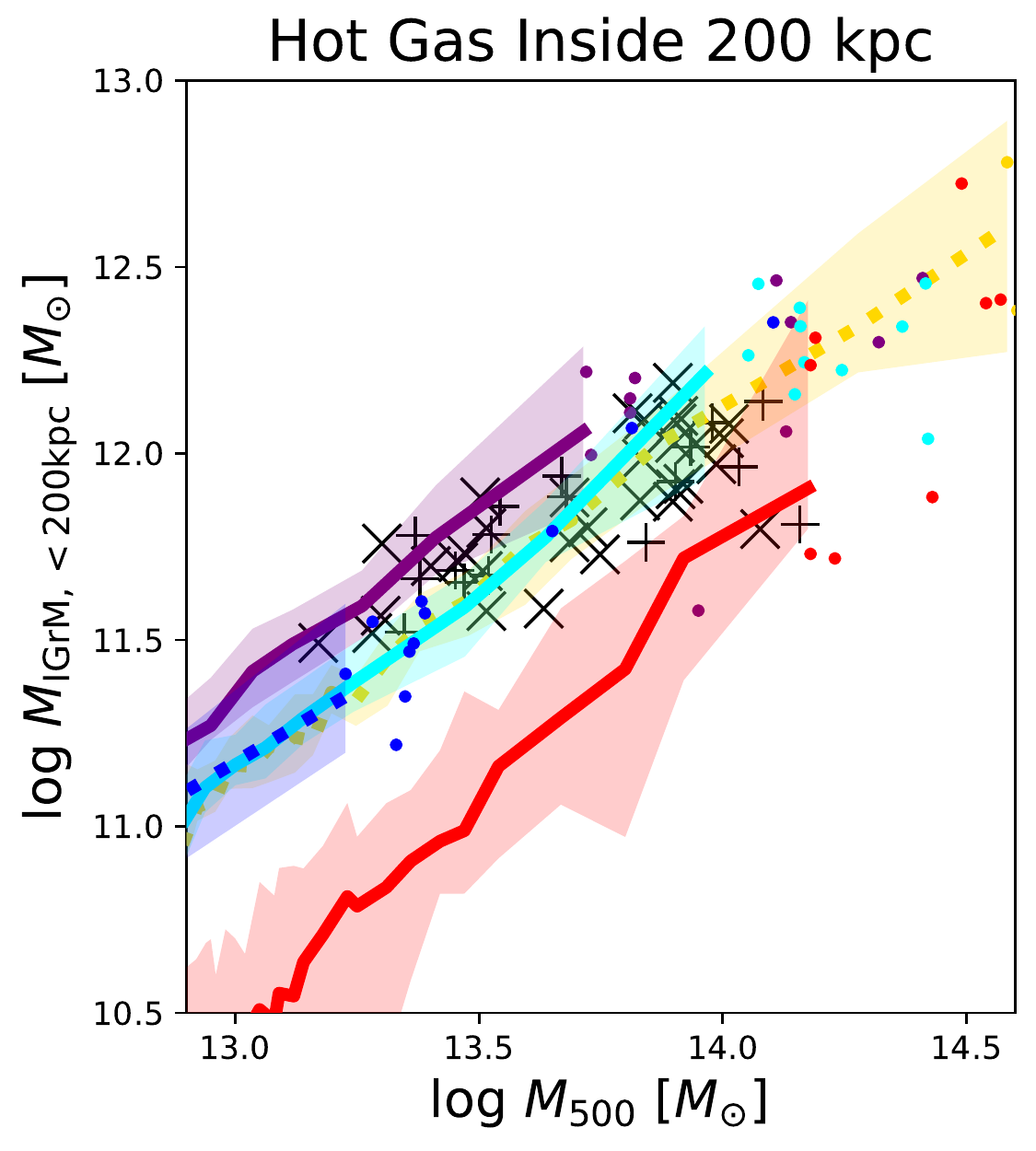}
\includegraphics[width=0.33\textwidth]{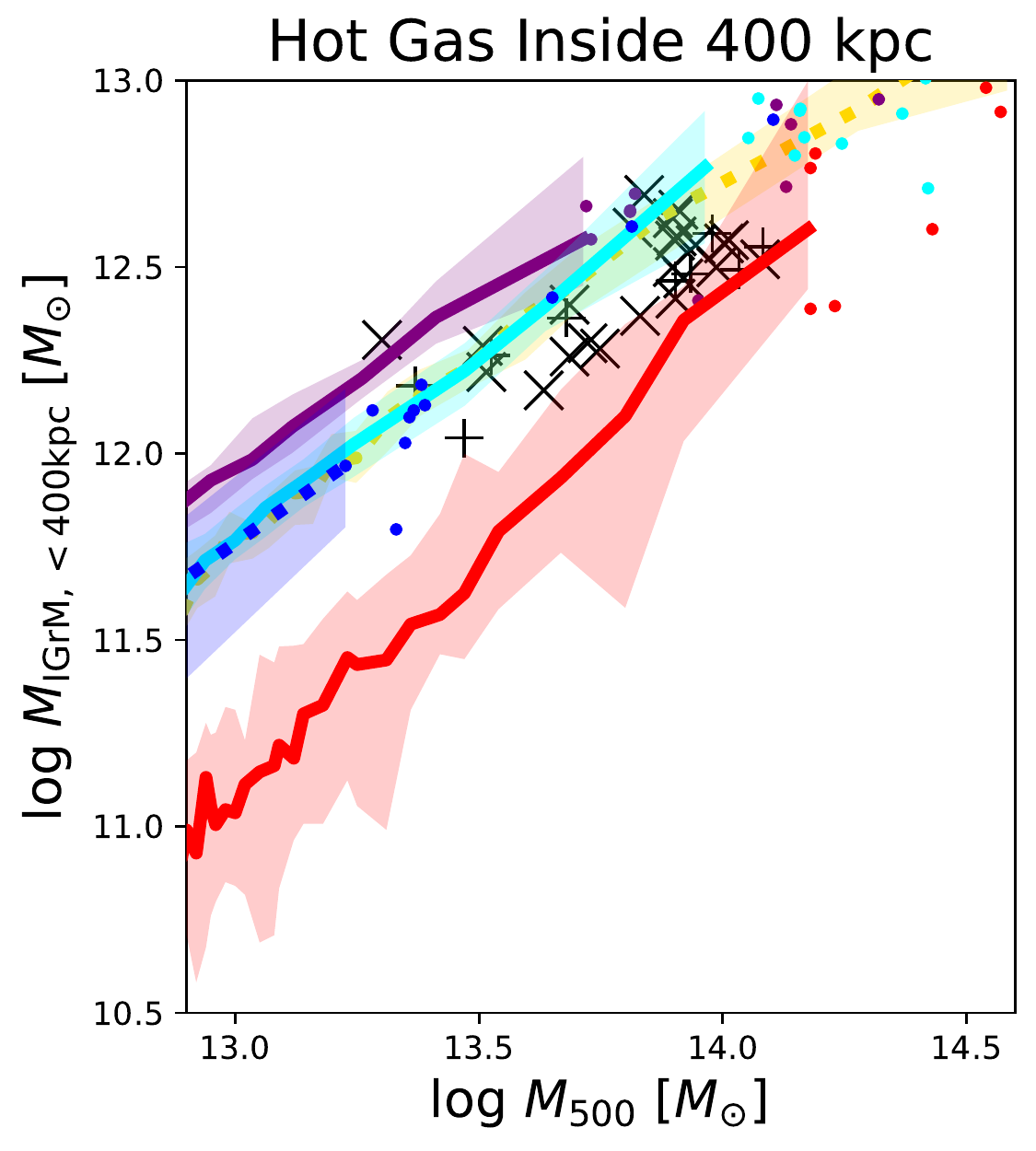}

\caption{IGrM masses (gas at $T\geq 10^6$ K) within three fixed 3D apertures plotted for EAGLE (purple), SIMBA (red), and~all three IllustrisTNG volumes (dark blue, cyan, and~yellow). Medians and 1$\sigma$ scatter are indicated by lines (dotted for TNG50 and TNG300, solid for others) and shaded regions, respectively. Different panels correspond to limiting radii of 100 kpc (\textbf{left}), 200 kpc (\textbf{middle}), and~400 kpc (\textbf{right}). Observed IGrM masses of individual groups within these radii, obtained from the (3D) profiles of \mbox{\citet{sun09}} and \mbox{\citet{lovisari15}}, are shown as black symbols. While differences between simulations are approximately consistent across the three limiting radii, the~agreement with observations differs noticeably between the group center (left-hand panel) and outskirts (right-hand~panel).}
\label{fig:fixed_mass_cuts}
\end{figure}

The determination of IGrM properties within $R_\mathrm{500}$ is physically motivated, but~measurements within fixed apertures are observationally more straightforward. In~Figure~\ref{fig:fixed_mass_cuts}, we therefore show the integrated IGrM masses as a function of halo mass as predicted by EAGLE, SIMBA, and~the three IllustrisTNG boxes out to fixed radii of 100, 200, and~400 kpc in the 3 subpanels. For~comparison, we obtain the equivalent masses from the observations of \mbox{\citet{sun09}} and \mbox{\citet{lovisari15}} by integrating their measured electron density profiles out to the same radii. While the relative differences between the simulations shown in Figure~\ref{fig:fixed_mass_cuts} are consistent across the three radial cuts---remarkably close agreement between the three TNG runs, with~EAGLE and SIMBA offset by $\approx$ +0.1 and $-$0.5 dex, respectively, at~$M_\mathrm{500} = 10^{13.5}\,\msol$---the comparison to the observations reveals additional details in each panel. Close to the group center ($r \leq 100$ kpc), EAGLE shows promising agreement with the observations, owing to its comparatively less aggressive AGN feedback. SIMBA, on~the other hand, despite having calibrated its AGN model to match the IGrM fraction inside $R_\mathrm{500}$, is evacuating the central region too efficiently. Within~the larger 400~kpc aperture (right-hand panel), the~IGrM masses of EAGLE are higher than observed, while SIMBA is at least marginally consistent with the observations at the high-mass end ($M_\mathrm{500} \gtrsim 10^{14}\,\msol$). The~IGrM masses of IllustrisTNG are consistent with the observations across radii, albeit with a tendency of being too low in the center ($r < 100$ kpc) of low-mass groups ($M_\mathrm{500} \lesssim 10^{13.5}\,\msolar$).

\subsubsection{Gaseous Profiles in Recent~Simulations} \label{sec:gasprofiles}

\begin{figure}
\includegraphics[width=0.49\textwidth]{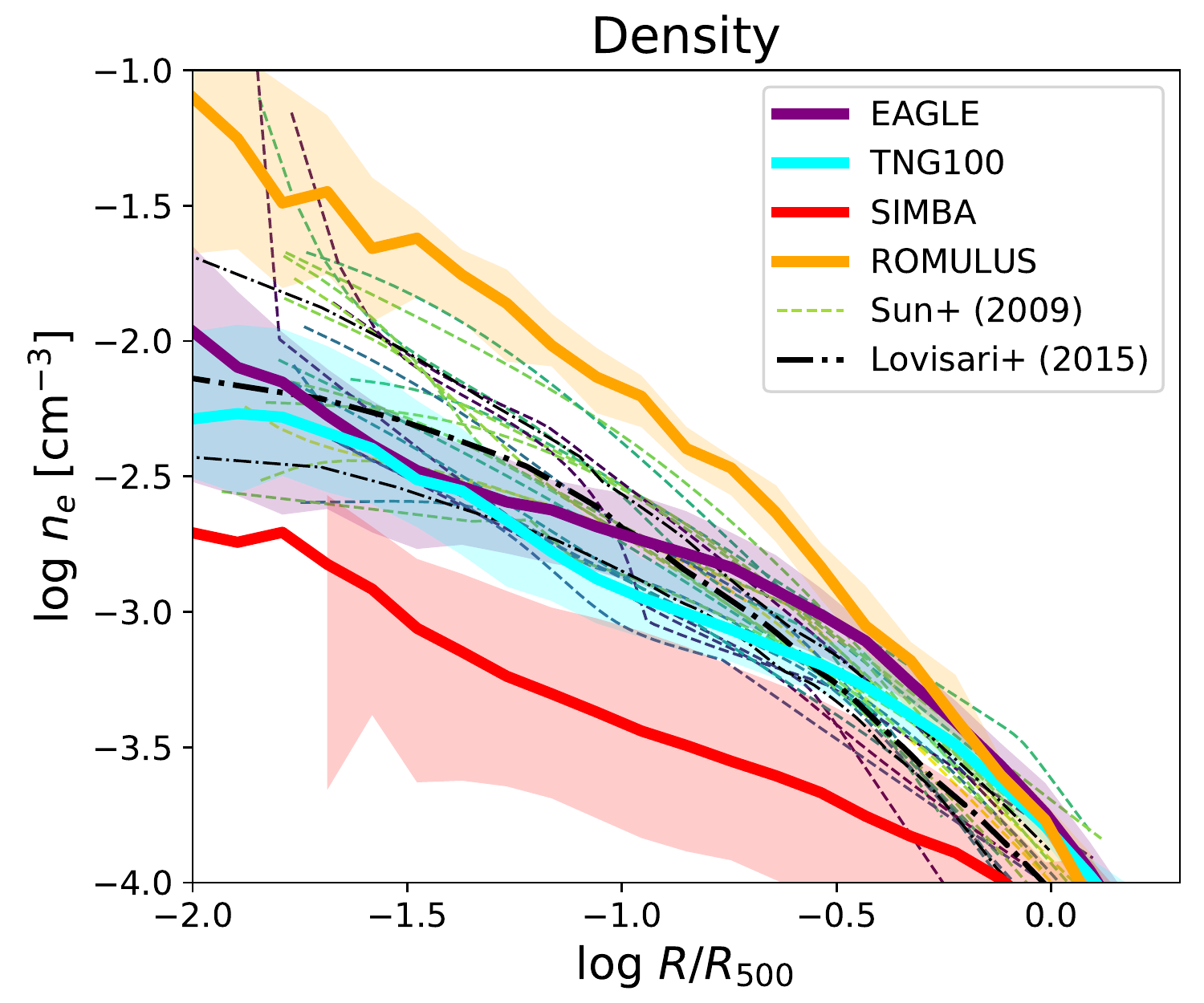}
\includegraphics[width=0.49\textwidth]{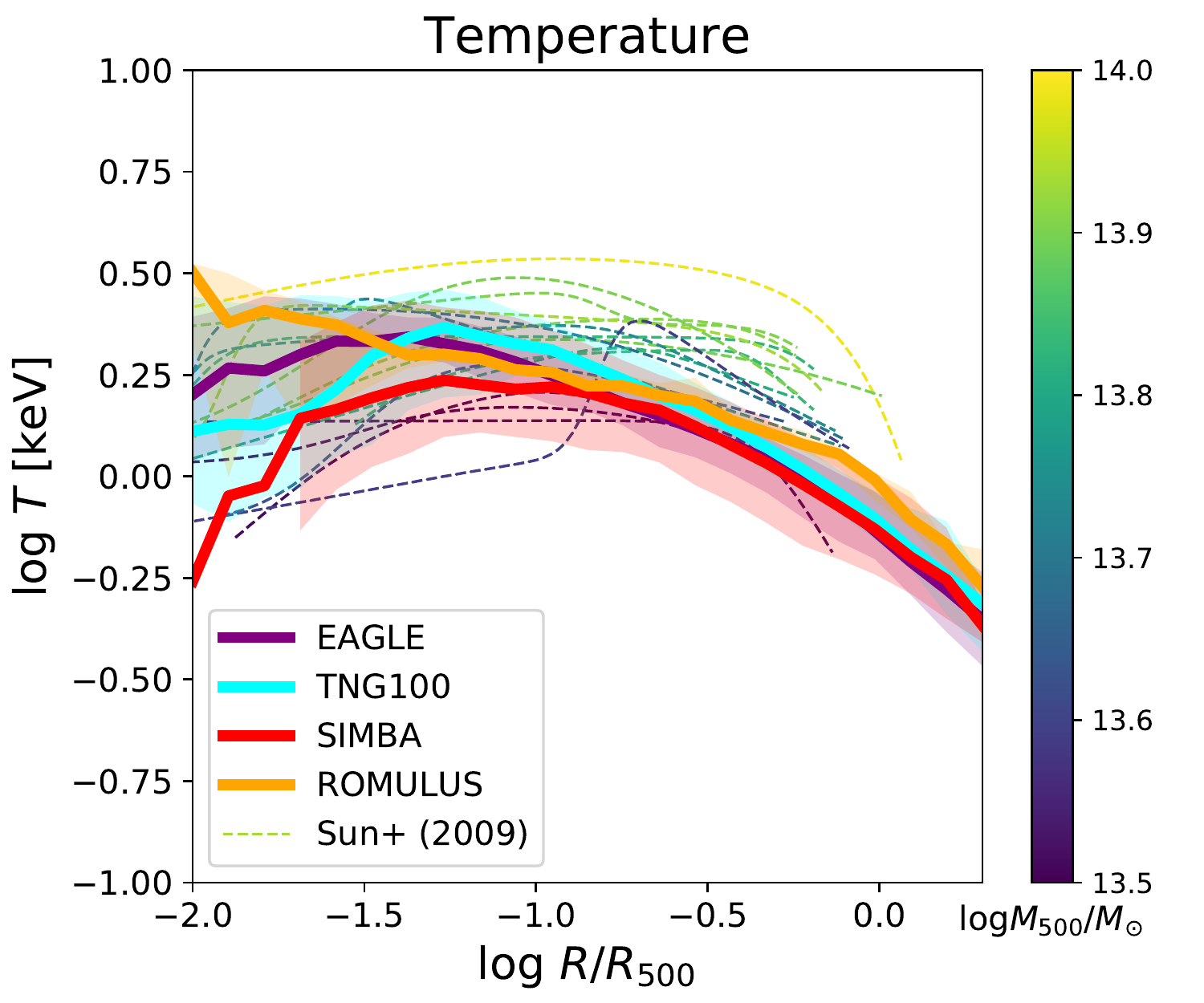}\\
\includegraphics[width=0.49\textwidth]{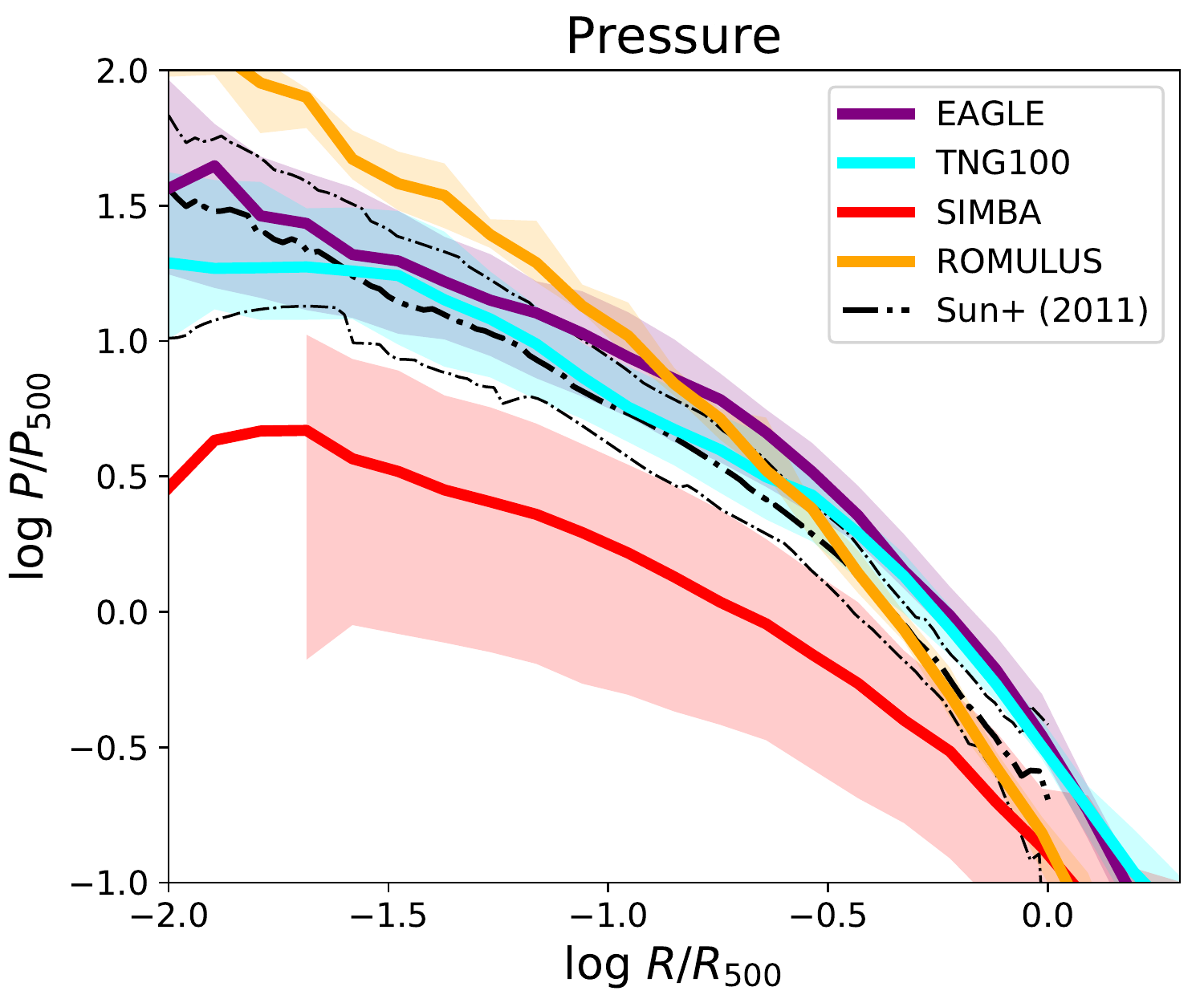}
\includegraphics[width=0.49\textwidth]{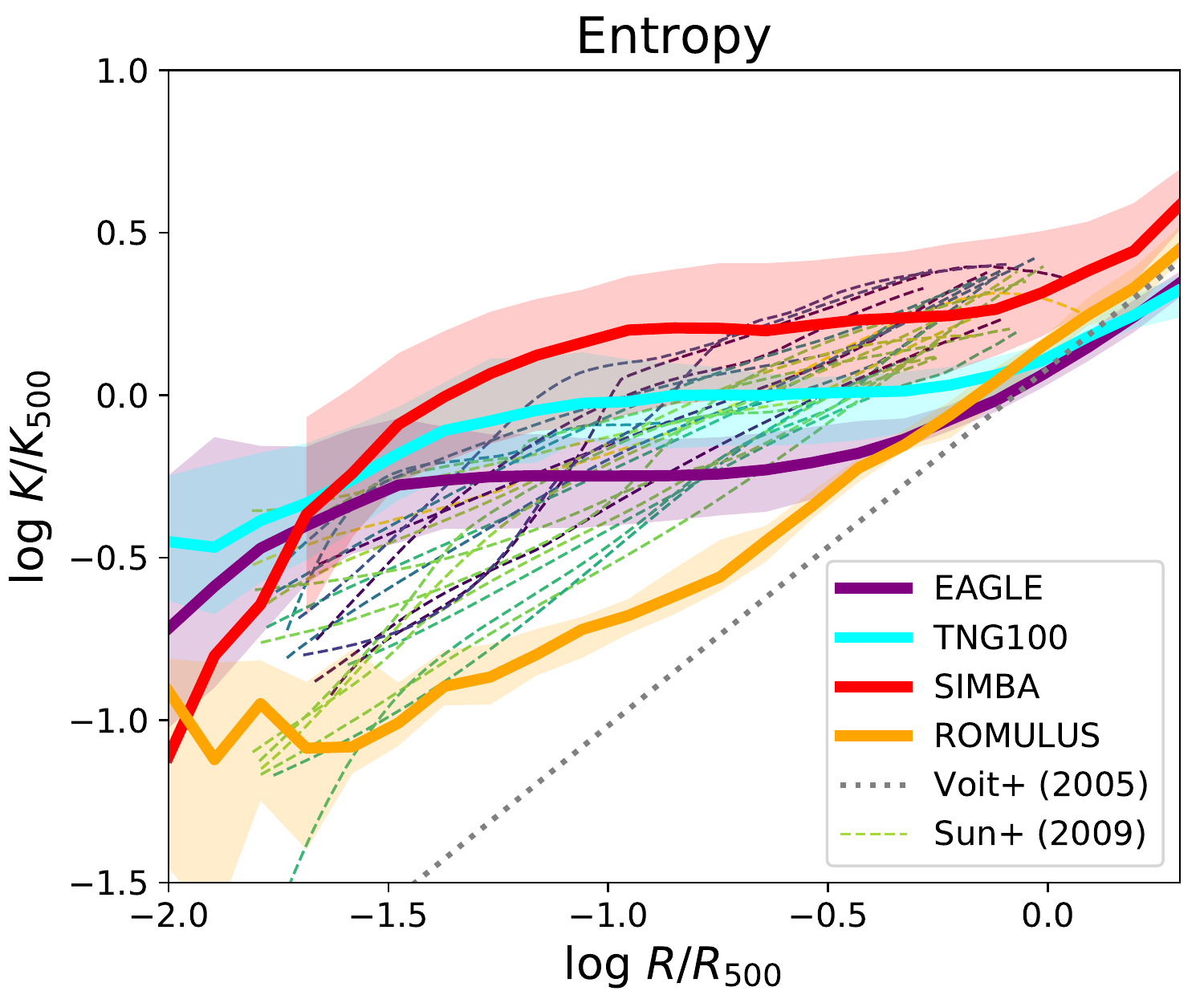}
\caption{Mass-weighted 3D radial profiles of IGrM density (\textbf{top left}), temperature (\textbf{top right}), pressure (\textbf{bottom left}) and entropy (\textbf{bottom right}) in groups with $M_{500}=10^{13.5}-10^{14.0}\ \msolar$ from the EAGLE ($N = 18$ groups, purple), TNG100 ($N=35$, cyan), SIMBA ($N=80$, red), and~ROMULUS ($N=1$, orange) simulations. For~the first three, we show profiles at $z = 0$, while for ROMULUS we have stacked the profiles obtained from the single group in this mass range at five snapshots with $z \leq 0.36$. Running medians within each simulation are plotted as solid lines, with~shaded bands representing the $1\sigma$ scatter. For~comparison, profiles of individual groups in the same mass range derived from X-ray observations \citep{sun09} are shown as thin dashed lines in three panels, colored by (temperature-derived) halo mass; for pressure, we instead show the SZ-based profiles of the same groups by \mbox{\citet{sun11}} as black dash-dotted lines (thick and thin ones for the median and $1\sigma$ scatter, respectively). In~the top-left panel, the~stacked density profile from X-ray observations of \mbox{\citet{lovisari15}} is shown in the same fashion, while the grey dotted line in the bottom right panel represents the ``base line'' $K\propto r^{-1.1}$ entropy profile seen in non-radiative simulations (e.g.,~\mbox{\citet{lewis00,voit05}}). With~the exception of temperature at large radii, there is little agreement between simulations. The~entropy profiles in particular are also clearly different from what is observed in all four cases, even for simulations such as EAGLE and TNG100 that approximately reproduce the observed pressure profile. In~general, stronger AGN feedback prescriptions raise entropies and reduce densities and pressures.}  
\label{fig:radial_profiles}
\end{figure}

\textls[-10]{We have seen above that even simulations that were calibrated to match gas and stellar fractions within the virial radius can make discrepant predictions about the IGrM content within other radii. A~more challenging test of the simulations is therefore provided by their predicted IGrM profiles, which are plotted for $z \approx0$ groups of $M_{500}=10^{13.5}-10^{14.0}\ \msolar$ from EAGLE, TNG100, SIMBA, and~ROMULUS \footnote{For EAGLE, TNG100, and~SIMBA, all profiles shown are at $z = 0$; for ROMULUS we combine five snapshots at $z \leq 0.36$ because the simulation contains only a single halo in this mass range} 
in Figure~\ref{fig:radial_profiles}. In~the top left panel, we show stacked profiles of electron density $n_\mathrm{e}$, calculated as $n_\mathrm{e} = \rho_\mathrm{IGrM} / (\mu_\mathrm{e}\,m_\mathrm{H})$ with $\rho_\mathrm{IGrM}$ the density of the hot IGrM gas (here defined as $T > 10^6$ K), $\mu_\mathrm{e} = 1.14$ the mean molecular weight per free electron, and~$m_\mathrm{H}$ the proton mass. Stacked IGrM temperature profiles are shown in the top-right panel. In~both cases, we compare to observed profiles of individual groups in the same mass range from thin dashed lines color-coded by mass (\citet{sun09}); for density we also plot the stacked observed profile of black dash-dotted lines \mbox{(\citet{lovisari15}).}}  

All simulations (except ROMULUS, which may be affected by its small sample size) agree closely with each other for the temperature profile beyond $\approx$0.1 $R_\mathrm{500}$, and~are broadly consistent with the observations of \mbox{\citet{sun09}}.  This similarity is interesting in the context of AGN feedback prescriptions that heat the gas differently resulting in similar temperature profiles throughout most of the IGrM, indicating that virialization primarily sets the IGrM temperature.  Significantly more variety is seen in the density profiles (top-left); all simulations predict a comparable and realistic density around $R_\mathrm{500}$, the~increase towards smaller radii is substantially stronger for ROMULUS, and~weaker for SIMBA, than~observed. TNG100 and EAGLE, on~the other hand, agree quite closely with each other and the observations, albeit with a slight ($\lesssim$0.2 dex) excess of gas towards the outskirts---especially for EAGLE---and a more substantial deficit ($\approx$0.2--0.5 dex) near the center that is stronger for~TNG100.  

Of particular interest are two physically motivated combinations of density and temperature: the IGrM pressure $P \equiv n_\mathrm{e}\,T$ and its entropy $K \equiv T/(n_\mathrm{e}^{2/3})$. These not only highlight additional discrepancies between simulations and observational data, but~reveal imprints of the subgrid prescriptions, most notably AGN feedback schemes. We show their radial profiles, normalized to their analytically expected values within $R_\mathrm{500}$: for pressure, $P_\mathrm{500} \equiv k_{\rm B} T_{500} n_{\rm e,500}$ and for entropy, $K_\mathrm{500} \equiv k_{\rm B} T_{500} (n_{\rm e,500})^{-2/3}$, where $T_\mathrm{500}$ is taken as the virial temperature $k_{\rm B} T_{500} \equiv G M_{500}\mu m_{\rm H}/R_{500}$ with mean molecular weight $\mu = 0.59$ and the electron density $n_\mathrm{e,500}$ as the ideal value of $n_\mathrm{e,500} \equiv 500 \fb \rho_{\rm crit}/(\mu_{\rm e} m_{\rm H}$) in the bottom row of Figure~\ref{fig:radial_profiles}.

At large radii, all simulations follow a power-law entropy profile with an index close to 1.1, as~expected from accretion and associated shocks \citep{babul02} and as found in non-radiative simulations (e.g.,~ \cite{lewis00,voit05}). Within~$\approx$0.3 $R_\mathrm{500}$, the~median profiles of three simulations (SIMBA, TNG and EAGLE) flatten to an extended entropy core, while ROMULUS entropy profiles keep decreasing and only flatten in the very center. The~magnitude of these simulated entropy cores appears to scale broadly with the aggressiveness of the AGN feedback. A~factor of 100 difference in the energy coupling efficiency $\varepsilon$ of AGN feedback progressively raises the entropy levels from ROMULUS ($\varepsilon=0.002$) to TNG ($\varepsilon=0.2$ for their pulse mode). SIMBA, however, exemplifies that the situation is more complex; it has the highest entropy core levels ($\gtrsim$$K_\mathrm{500}$ beyond 0.03 $R_\mathrm{500}$) despite an AGN coupling efficiency comparable to EAGLE (see Table~\ref{tab:AGN}). It is plausible that the high entropy core of SIMBA is, at~least in part, due to its decoupled kinetic AGN feedback scheme \citep{dave19}, which can efficiently inject entropy at large radii. At~the same time, it still allows transport of low-entropy gas towards the center, leading to the strong drop in central entropy (which, to~a lesser extent, is also seen for EAGLE).


Although the core entropy levels in TNG100 and EAGLE are lower than for SIMBA, their AGN feedback schemes---randomly-oriented, pulsed feedback \citep{weinberger17} and highly energetic thermal injection \citep{schaye15}, respectively---are still creating extended cores. \textls[-5]{The~only simulation without a clear high-entropy core is ROMULUS, which injects AGN energy thermally but with a much lower temperature increase. As~we have seen above, this low entropy injection correlates with higher group baryon fractions than in the other~simulations.}

A comparison of these predictions to the \mbox{\citet{sun09}} profiles clearly reveals that {\it none of the scaled entropy profiles from any of the simulations resemble the observations.} The latter do not have large extended plateaus; they typically scale with radius as $R^{0.7}$ for $R<R_{500}$ (see also Figure~10 of \mbox{\citet{sun09}} and Figure~4 of \mbox{\citet{osullivan17}} for the CLoGS group sample). This is despite reasonable agreement in terms of the pressure profiles (especially for EAGLE and TNG100), which indicates that the simulated IGrM remains approximately in pressure equilibrium throughout the (significant) AGN energy and entropy injection \citep{barnes17}. The~FABLE simulations (not shown \cite{henden18}) show more promising agreement of IGrM profiles (including entropy), although~\mbox{\citet{henden18}} discuss that even their explicitly calibrated (bubble) feedback might be too energetic at late times, since the $z = 0$ FABLE groups fall within the scatter, but~mostly below the median, of~the \mbox{\citet{sun09}} density profiles (see {their Figure~11}). It will be interesting to see whether future simulations can overcome this shortcoming with more sophisticated AGN feedback models and higher resolution, or~whether observational selection biases (i.e., preferential inclusion of cool-core systems with dense, bright centers in X-ray selected samples as discussed by, e.g.,~\mbox{\citet{henden18}}) are responsible for at least part of the discrepancy (see Section~\ref{sec:future}).

\begin{figure}
\includegraphics[width=0.54\textwidth]{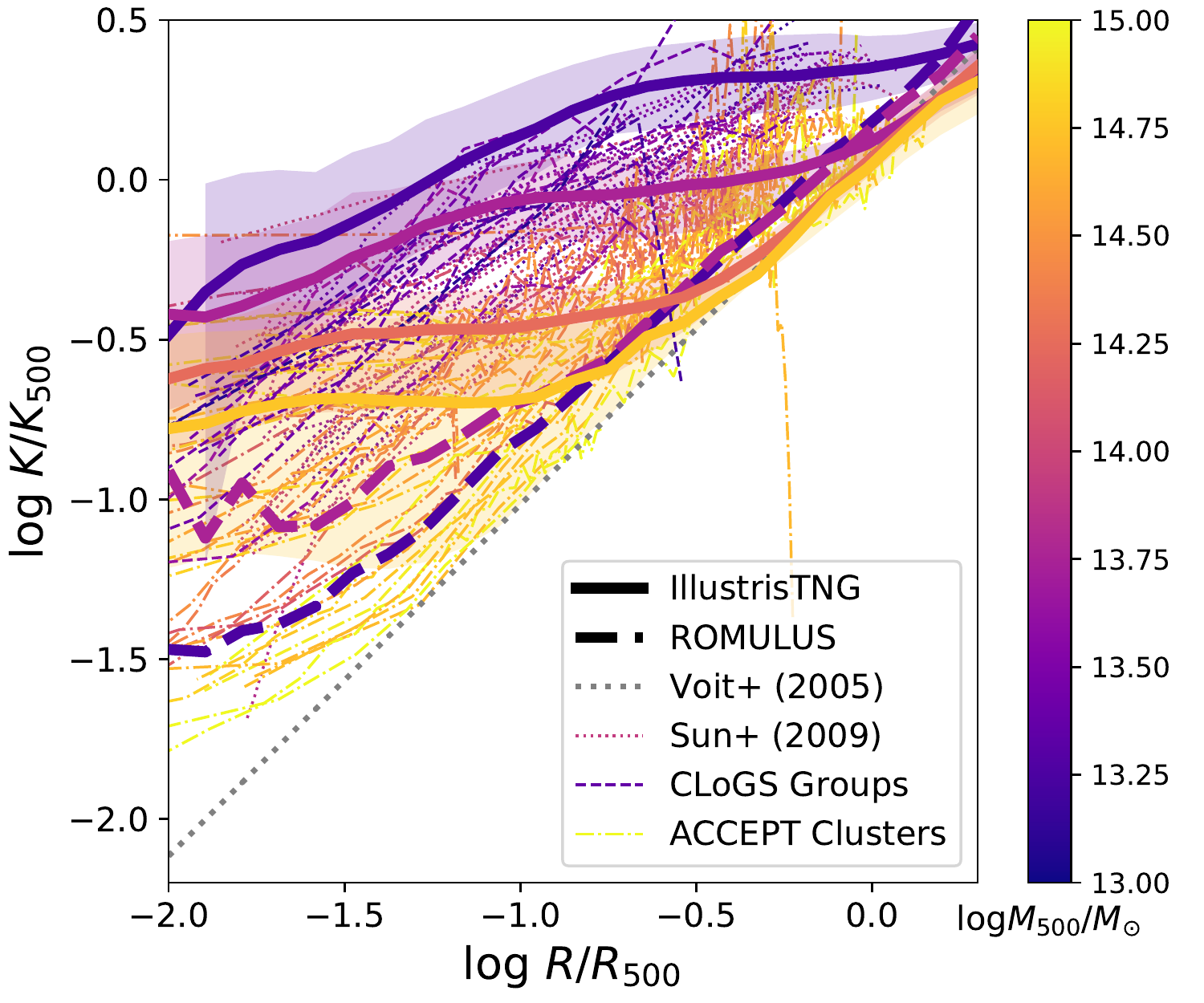}
\includegraphics[width=0.41\textwidth]{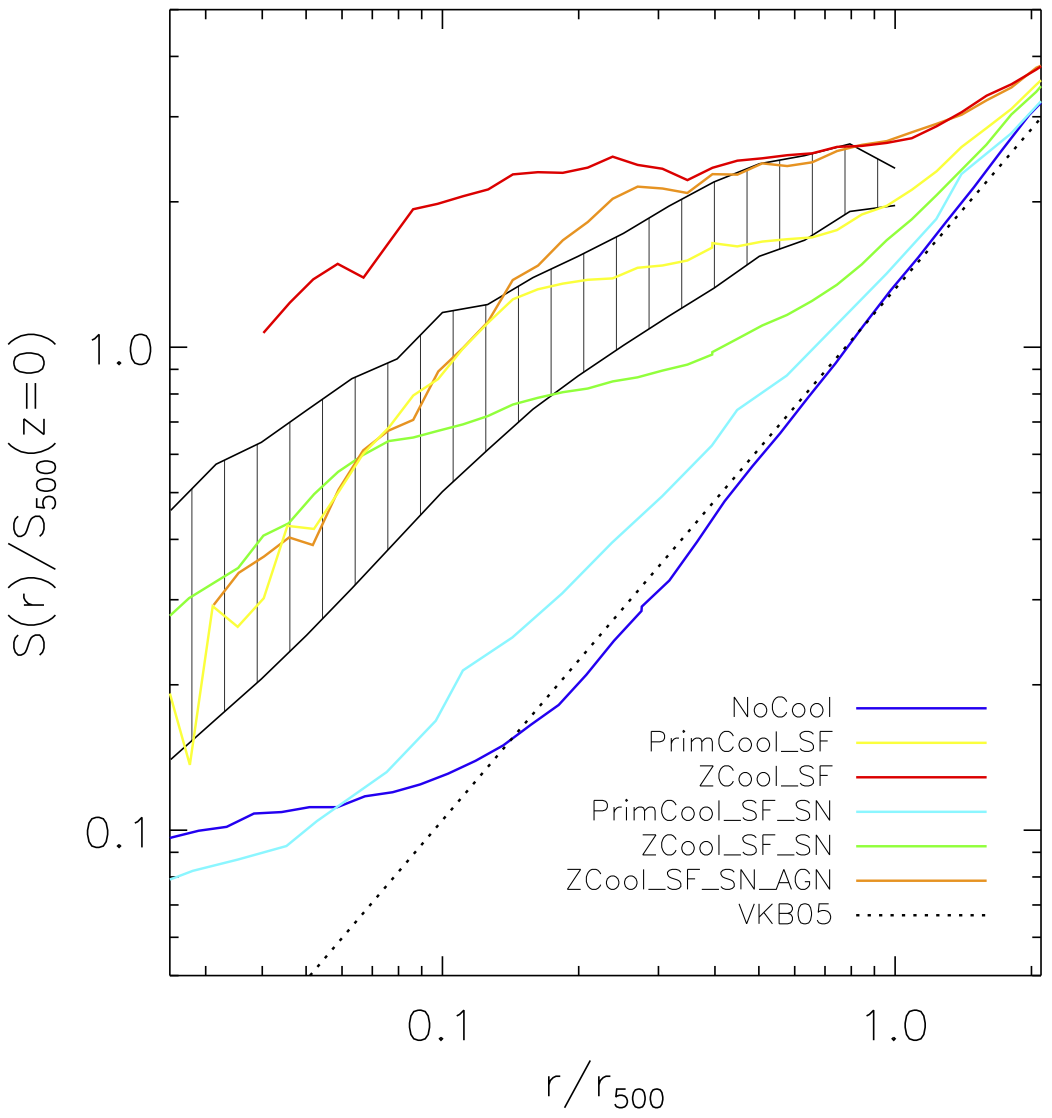}
\caption{(\textbf{Left}) Normalized mass-weighted 3D entropy profiles at $z \sim 0$ for groups and clusters with $M_\mathrm{500} = 10^{13}$--$10^{15}\,\msolar$ in TNG100 and TNG300 (combined, thick solid lines) and ROMULUS (thick dashed lines). We show median-stacked profiles of halos in 0.5 dex bins, with~264 (7), 110 (5), 91, and~15 halos per bin for TNG (ROMULUS) in order of increasing mass, color-coded by mass from low-mass groups (purple) to massive clusters (yellow). For~TNG, we also indicate their $1\sigma$ scatter with shaded bands. These predictions are compared to individual halos from three observed samples, all colored analogously by halo mass: low-mass CLoGS groups (dashed lines \cite{osullivan17}), massive groups   (dotted lines \cite{sun09}), and~a random 20\% subset of the ACCEPT clusters (dash-dotted lines \cite{cavagnolo09}). The~gray dotted line indicates the $K \propto r^{1.1}$ scaling found in non-radiative simulations \citep{voit05}. In~agreement with observations, TNG (but not ROMULUS) predicts higher normalized entropy at lower mass, but~with higher-entropy cores than observed. In~contrast, ROMULUS groups have steeper and lower entropy profiles that are more typical of observed massive clusters. (\textbf{Right}) Normalized mass-weighted 3D entropy profiles of $M_\mathrm{500} = 10^{13.25}$--$10^{14.25}\,\msol$ groups predicted by OWLS simulations with different subgrid \mbox{models \citep{schaye10}}, reproduced with permission from \citet {mccarthy11} (note that their symbol for astrophysical entropy is $S$, rather than $K$). Simulations include a run without any cooling or feedback (blue); without feedback but with star formation and cooling without (yellow) or with (red) the contribution from metal lines; with the addition of SNe feedback (cyan and green, respectively); and with AGN feedback added in addition to metal-line cooling, star formation, and~SNe feedback (brown). The~black-hatched band indicates the $1\sigma$ scatter of the {\citet{sun09}} observations, the~black dotted line a $r^{1.1}$ power law. While AGN feedback has a clear effect on the IGrM, the~entropy profiles are also sensitively affected by cooling, star formation, and~SNe~feedback.}
\label{fig:entropy_profiles}
\end{figure}

For deeper insight into the predicted IGrM entropy profiles, it is instructive to consider them in context with their more massive cluster counterparts. This comparison is shown in the left panel of Figure~\ref{fig:entropy_profiles}, where we plot scaled $z \approx 0$ entropy profiles in analogy to the bottom-left panel of Figure~\ref{fig:radial_profiles} but over a wide range of halo mass, $M_\mathrm{500} = 10^{13.0-15.0}$. For~clarity, only IllustrisTNG (TNG100 and TNG300 combined, with~TNG300 dominating because of its larger volume; solid lines) and ROMULUS (dashed lines) are shown, with~halos median-stacked within 0.5 dex bins in $M_\mathrm{500}$. These are compared to three observational samples that together span a similar mass range: the \mbox{\citet{sun09}} and (lower-mass) CLoGS~\citep{osullivan17} groups as well as clusters from the ACCEPT survey \citep{cavagnolo09}.

There are two features of this comparison that are particularly worth highlighting. First, the~median entropy profiles of IllustrisTNG feature prominent entropy cores that are significantly higher than observed across the selected mass range, from~poor groups (purple) to massive clusters (yellow). The~same is true for EAGLE and SIMBA clusters (not shown). On~cluster scales, these high-entropy cores corresponding to non-cool-core (NCC) systems have previously been highlighted by C-EAGLE~(\citet[]{barnes17}), IllustrisTNG~(\citet[]{barnes18}), and~SIMBA~(\citet[]{robson20}); in the case of IllustrisTNG, \mbox{\citet{barnes18}} found a rapid decline in the fraction of cool-core (CC) systems at $z < 1$ and identified the cause to be the AGN feedback implementation, which appears too efficient at removing baryons from the inner $0.01 R_{500}$.

Second, there is a clear trend towards lower (normalized) entropy with increasing halo mass at fixed radius, in~both the observations and the IllustrisTNG simulations  \footnote{At first sight, this offset may be surprising given the explicit normalization of our profiles by the integrated $K_\mathrm{500}$. The~reason for this apparent contradiction lies in the definition of $K_\mathrm{500}$, which is calculated under the assumption that the baryon fraction within $R_\mathrm{500}$ is equal to the cosmic average $f_b$. As~shown in Figure~\ref{fig:mhalo_fgas}, the~actual gas fraction, and~hence electron density, is lower by a factor of up to $\approx$3 at $M_\mathrm{500} = 10^{13}\,\msol$ even within $R_\mathrm{200}$, so that $<\!\!K\!\!> > K_\mathrm{500}$}. 
We speculate that these features may be a signature of quantized feedback dumping a fixed amount of entropy per AGN feedback event; further investigation of this topic is clearly~warranted. 
Interestingly, such a trend is not seen in ROMULUS, although~the small number of groups and absence of massive clusters prevents strong conclusions here. If~anything, the~lower-mass groups (blue dashed line) have entropy profiles that are even closer to observed CC clusters, plausibly also because of a late-time merger in the more massive ROMULUSC group \citep{tremmel19}. As~discussed further below, it is plausible that this CC-like behaviour of ROMULUS groups is due to the absence of metal-line cooling---which may limit the removal of low-entropy gas from its IGrM---as well as the highly collimated nature of their AGN-driven~outflows.

There are additional reasons why the ubiquitous conversion from CC to NCC halos in most simulations is problematic. \mbox{\citet{mccarthy08b}} showed that the cost, in~energy, is much greater than the typical jet power of an AGN outburst \citep{Gitti11}, with~most of the energy being expended to lift the gas rather than raise its entropy. This suggests that the AGN outbursts in the simulations are likely injecting much more energy than observed radio AGNs in galaxy clusters. Observations also show little evidence of AGN feedback affecting CC-to-NCC transformations. The~entropy profiles of NCC as well as  CC CLoGS groups (with and without observed jet activity) are all, 
to first order, similar in shape and normalization \citep{sun09, osullivan17}. Moreover, none of the clusters showing evidence of having experienced an {\it extreme} AGN feedback event in the recent past---e.g.,~MS 0735.6+7421 \citep{Vantyg14} and Hydra A~\citep{David01,Gitti11}---have core entropies as high as the median value in the simulations. We therefore speculate that the action of the  AGN feedback models in SIMBA, TNG, and~(C-)EAGLE may be too aggressive in the low-$z$ Universe (see also the discussion in \mbox{\citet{barnes17}}). Although~we are comparing individual observed profiles to medians from simulations, our examination of simulated individual profiles (not shown) indicates that the median profiles are fairly representative, so that this is unlikely a source of significant~bias.

Given the mismatched entropy profiles in contemporary simulations, it is helpful to refer to \mbox{\citet{mccarthy11}}, who demonstrated using multiple OWLS simulations that adding cooling to non-radiative simulations can alter the entropy slopes as much as feedback (Figure \ref{fig:entropy_profiles}, right panel).  While non-radiative simulations fall on the baseline $K\propto R^{-1.1}$ relationship,  adding only cooling without feedback ``cools out'' low-entropy gas, raising the entropy of the remaining IGrM.  Metal-line cooling, which is comparatively more important for groups than clusters, raises entropy even more.  Adding SNe feedback without metal-line cooling actually lowers the entropy nearly to the $R^{1.1}$ line, which may explain the ROMULUS entropy slopes that have weak AGN feedback that is more similar to SNe feedback and no metal-line cooling.  While \mbox{\citet{mccarthy11}} confirm that adding AGN feedback indeed increase the entropy (cf. orange vs.~green lines), the~difference is smaller than changing the cooling physics. Furthermore, they showed that the AGN-induced entropy increase is, for~the largest part, also an indirect effect: in their simulation, it is \emph{not} (primarily) caused by heating gas that remains in the IGrM, but~by the selective ejection of low-entropy gas from the less tightly bound $z \approx 2$--4 progenitor halos.

\begin{figure}
\includegraphics[width=0.95\textwidth]{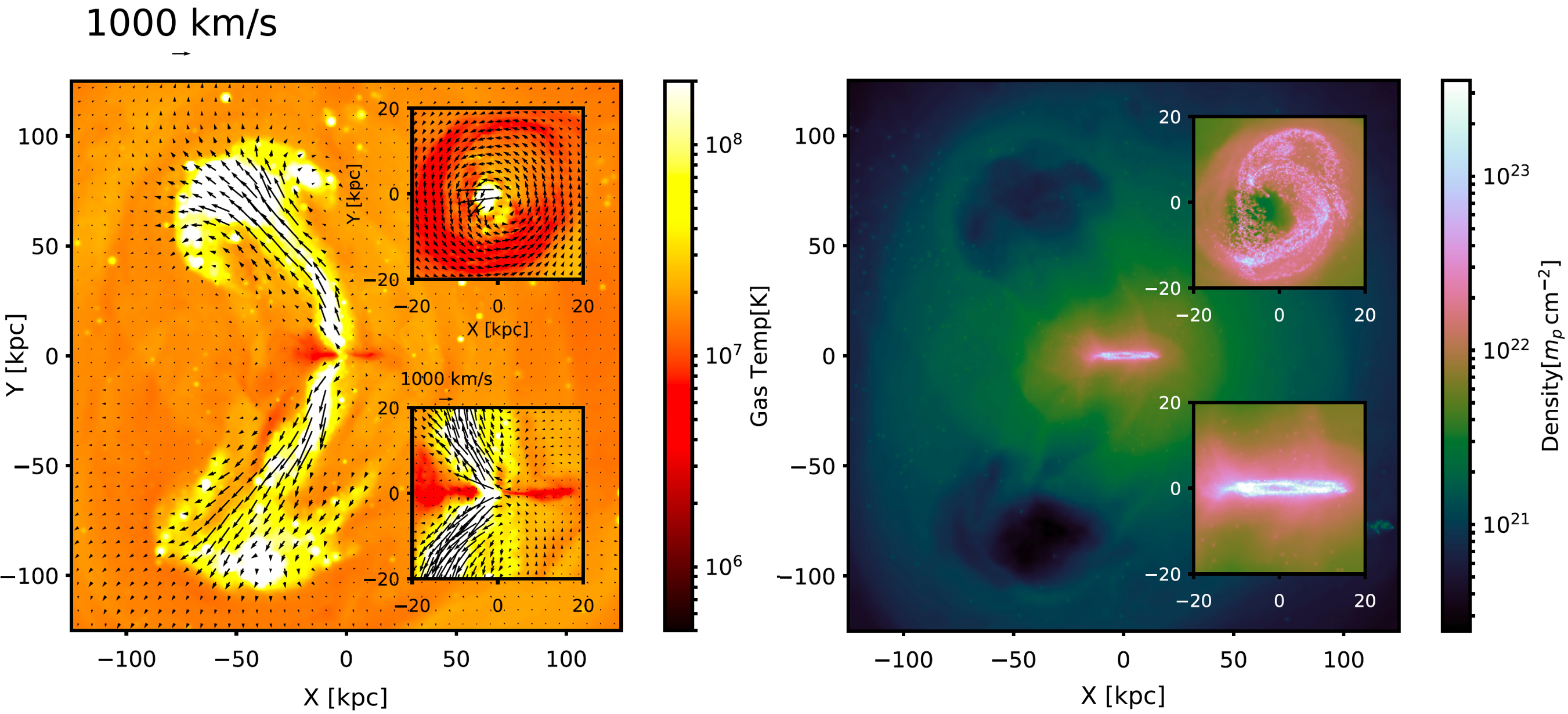}
\caption{Wind-blown bubbles driven by AGN feedback from the central galaxy in the ROMULUSC simulation at $z=0.53$, adapted with permission from \mbox{\citet{tremmel19}}.  This $M_{500}=10^{13.6}\ \msolar$ group produces a cool core that is elusive in other simulations, although~the central gas densities are high compared to observations.  This AGN feedback is a thermal dump model without \emph{explicit} collimation, and~is the most obvious example of bubble-driven feedback at the group scale found in the cosmological simulations we~explore.  }
\label{fig:romulusc_central}
\end{figure}

We end our discussion of IGrM properties by pointing out that even radial profiles convey only a simplified view of the processes shaping the gaseous halos of groups. This is exemplified particularly clearly by high-resolution simulations such as ROMULUSC, for~which we show 2D maps of the gas temperature and density near the center of its most massive group at $z=0.53$ in Figure~\ref{fig:romulusc_central} (adapted from Figure~10 of \mbox{\citet{tremmel19}}). A~bi-conical jet is clearly seen in the temperature map, which emerges naturally by collimation of their (intrinsically isotropic) AGN feedback by a central gas disc. This jet evacuates bubbles in the vicinity of the group center---not unlike X-ray cavities observed in real clusters (e.g.,~\mbox{\citet{Hlavacek-Larrondo_et_al_2015}})---but does not strongly affect gas away from its axis, thus preserving a steep $R^{1.1}$ entropy slope down to $0.04 R_{500}$ in a spherically averaged sense (Figure~\ref{fig:entropy_profiles}).

\subsection{Brightest Groups~Galaxies} \label{sec:BGG}

For more than two decades, attempts to model the formation and evolution of galaxy groups and to identify the role of different physical processes that shape these systems have focused primarily on the properties of the hot IGrM/ICM ( e.g., \cite{balogh99,babul02,mccarthy08b,cavaliere98,Voit01,Voit02}, and references therein). Even with the advent of cosmological simulations, this tendency has largely continued 
\citep{dave02,borgani04,dave08,mccarthy10,lebrun14,schaye15,liang16,henden18,robson20}. However, observational studies find that a number of BGG properties (e.g., stellar mass, size, morphology, the~nature of the surface brightness and stellar velocity dispersion profiles, whether the BGG is a fast or slow rotator,  etc.) are also correlated with the properties of their host system \citep{lin04,brough06,brough08,vonderlinden07,Weinmann06,Gonzaliasl16,Cougo20,Loubser18}.  Such correlations are not entirely unexpected given that BGGs are typically found at the bottom of their host halo's gravitational potential well. As~such, the~BGGs are thought to have experienced numerous mergers and close tidal encounters with other group galaxies over cosmic time.  Simulations suggest that such interactions result not only in the growth of the BGGs' total stellar mass but also, induce structural and kinematic transformations \citep{murante07,Dubinski98,DeLucia07,Cooper15,Nipoti17,R-F18}.  Mergers can also potentially transport in cool gas and fuel {in-situ} star formation in disc-like structures.  Additionally, as~explicitly demonstrated by \mbox{\citet{lewis00}} using the first hydrodynamic cosmological simulation of a rich galaxy group/poor cluster that allowed for radiative cooling, any gas cooling out of the IGrM/ICM  flows towards the group center and ends up in the central galaxy.  Consequently, the~study of the BGGs, their properties, and~the existence of any correlations between the latter and the properties of the host groups offer a complementary window onto the physical processes underlying the evolution of galaxy groups.  Just as importantly, in~the context of the present review, detailed comparisons of the observed BGG properties and the characteristics of those in numerical simulations offers an equally powerful opportunity for testing the efficacy of latest generation of cosmological simulations.  In~recent years, a~handful of papers have assessed the characteristics of BGGs forming in hydrodynamic simulations (e.g., \cite{schaye15,clauwens18,lebrun14,dave19,tremmel17,tremmel19,barnes17}) although this was done as part of a broad survey of the general properties of simulated galaxies.  There are however notable recent exceptions,  like \mbox{\citet{davidson20,Henden20,Jackson20}}, \mbox{\citet{Katsianis21,pillepich18b}}, \mbox{\citet{Remus17,Tacchella19}} and \mbox{\citet{jung21}}, that have treated BGGs as a distinct class, investigating both the structural and the kinematic properties of the simulated BGGs and comparing these to~observations.

\subsubsection{Central Galaxy Stellar~Masses}

\begin{figure}
\includegraphics[width=0.92\textwidth]{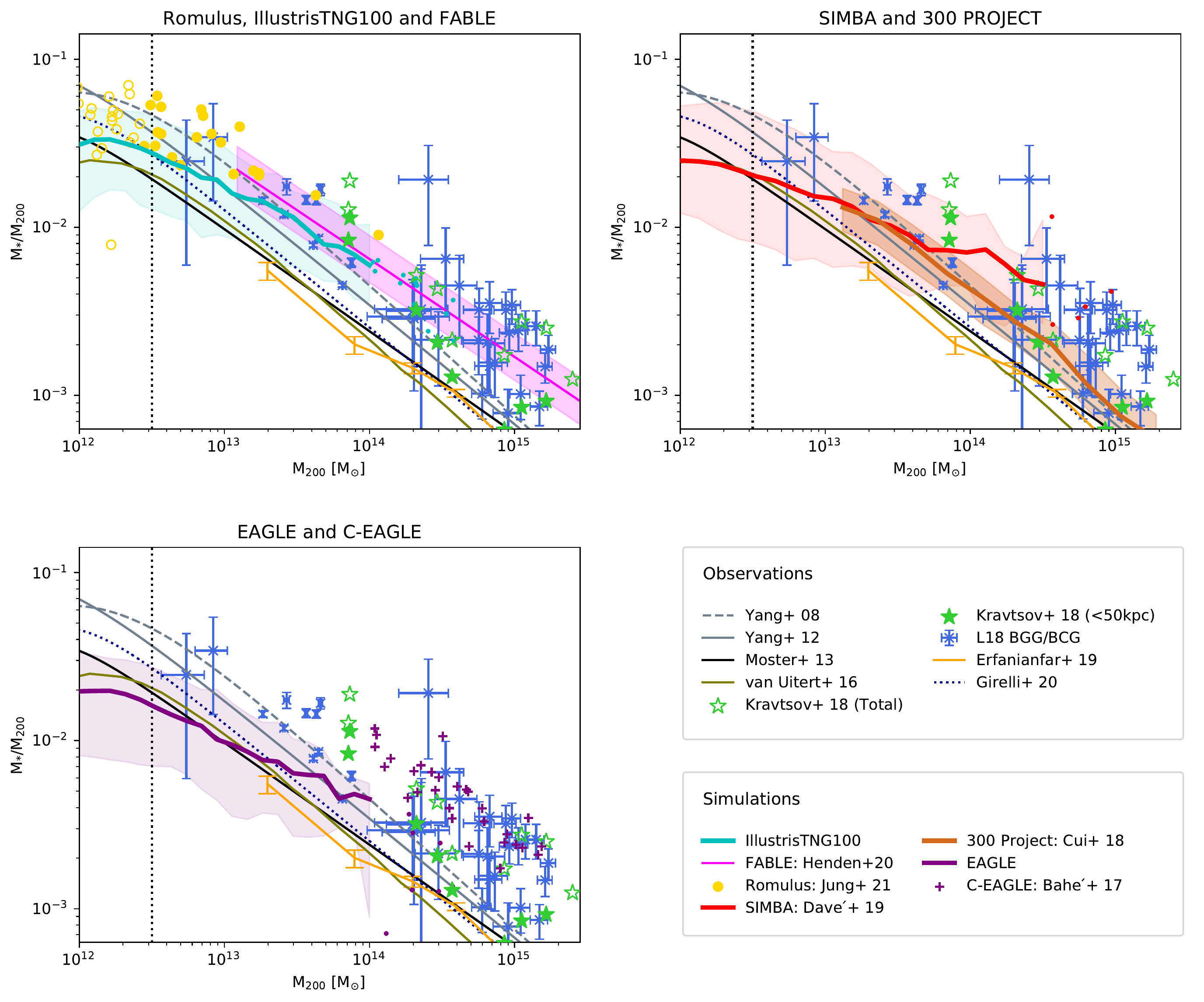}
\caption{{{Figure adapted from} \mbox{\citet{jung21}}:} \textls[20]{The $z=0$ stellar mass$-$halo mass relationship ($M_\star/M_{200}$) versus $M_{200}$ for central galaxies in observed and simulated galaxy groups and clusters.  The~top left panel shows results for TNG100 (turquoise line/band and dots; \cite{pillepich18b}), FABLE (magenta line/band; \cite{Henden20}), and~ROMULUS suite of simulations (open and filled yellow points; \cite{jung21,tremmel19,chadayammuri21}). The~thick solid lines are the median SMHM; the shaded region spans 95\% of the systems. The~top right panel shows results for SIMBA (red line/band and dots; \cite{dave19}) and The Three Hundred project (brown line/band; \cite{cui18}). The~bottom left panel shows results for the EAGLE (purple line/band and dots; \cite{schaye15}) and C-EAGLE (purple crosses; \cite{bahe17}).  We identify systems with $\log(M_{200}/\msolar) \geq 12.5$ (to the right of the dotted vertical line) as groups/clusters.  The~ROMULUS, FABLE, SIMBA, C-EAGLE and The Three Hundred project stellar masses are $M_{\star,50,2D}$;  EAGLE and TNG stellar masses are $M_{\star,30,3D}$ corrected to $M_{\star,50,3D}$.  For~comparison, we also plot the same nine observationally determined SMHMs in all three panels.  These are described in the text. There is considerable spread between the nine relationships but jointly, they show how $M_\star/M_{200}$ scales with $M_{200}$. We interpret the spread as a measure of the uncertainty. }}
\label{fig:fstarcentral}
\end{figure}

The first BGG (and BCG) property we consider is their stellar mass$-$halo mass (SMHM) relationship.  This is shown in Figure~\ref{fig:fstarcentral}, where we plot $M_\star/M_{200}$ against $M_{200}$.  $M_\star$ is the mass of the stars associated with a BGG/BCG.  Stellar observations show that once the light from individual resolved galaxies other than the central BGG/BCG is excluded, the~resulting surface brightness can be separated into two components: that due to light from a localized concentration of stars usually identified with the BGG/BCG proper and that due to light from a diffuse, often extended, component: the intragroup/intracluster light (IGrL/ICL) (e.g.,~see \cite{gonzalez13}).  The~$M_\star$ we quote here is the sum of the mass in the two components.  The~SMHM offers a window onto not just the efficiency with which the gas in the central galaxies is turned into stars but also the tug-of-war between heating and cooling in the group cores as well as the role of mergers in the build-up of the central plus IGrL stellar~mass.

For the purposes of clarity, we plot the simulation results across the three panels in Figure~\ref{fig:fstarcentral} but for reference and comparison, we show the same set of nine representative observationally determined SMHMs \footnote{readers interested in seeing a more complete set of available SMHMs results are referred to Figure~10 of \mbox{\citet{Coupon15}} or Figure~9 of \mbox{\citet{Girelli20}}} 
 in all three panel.  All of the observed results are based on the projected stellar light distribution and correspond to the central+IGrL stellar mass within a cylinder aligned along the line of sight with some aperture of radius $R$ (i.e.,~$M_{\star,R,2D}$).  This mass estimate explicitly {\it excludes} the contribution from resolved satellite galaxies.  \textls[-5]{The~dashed, dotted and solid curves are results from \citet{Yang08,Yang12,moster13,vanuitert16,Erfanianfar19,Girelli20}}.  There are two sets of \mbox{\citet{kravtsov18}} datapoints: the filled stars are based on $M_{\star,50,2D}$ and the open stars are based on ``total'' aperture masses, derived by extrapolating and integrating the ICL profiles to large radii in the sky ($M_{\star,{\rm tot},2D}$).  The~blue crosses with error bars are total aperture masses from \citet{Loubser18} and \mbox{\citet{Kolokythas21}}: points with $\log(M_{200}/\msolar) < 14.0$ are for BGGs from the high richness subset of CLoGS groups \citep{osullivan17} while the BCG results are from the Multi Epoch Nearby Cluster Survey  and the Canadian Cluster Comparison Project (MENeaCS and CCCP, respectively \cite{Sand12,Bildfell08,Mahdavi13,Hoekstra15,Loubser16,Herbonnet20}). There is a considerable spread in the published observationally derived SMHM results. This spread is due to a variety of factors, including (i)  the groups and clusters are sourced from deep targeted observations as well as large surveys (e.g.,~SDSS, COSMOS and GAMA) of varying depth; (ii) the use of apertures of different sizes; (iii) the use of different strategies to estimate the background around large massive galaxies in crowded environments \citep{kravtsov18}, as~well as to model and extrapolate their observed light profiles of interest; (iv) the use of different estimates s of $M_\star / L$ ratio to convert measured light distribution into stellar mass; and (v) the use of different approaches for estimating the halo mass, ranging from the use of X-ray observations under the assumption of hydrostatic equilibrium and abundance matching, to~weak gravitational lensing estimates.  The~observational results shown in Figure~\ref{fig:fstarcentral}, taken together, show how $M_\star/M_{200}$ scales with $M_{200}$.  We interpret the spread in the SMHM determinations as a measure of the uncertainty when comparing to the simulation~results.

\textls[-20]{As for the simulations, we first consider the results from the ROMULUS simulations (i.e.,~ ROMULUS25 \citep{tremmel17}, ROMULUSC \citep{tremmel19,chadayammuri21,jung21}, ROMULUSG1 and ROMULUSG2 \citep{jung21}). These are plotted in the first (top left) panel.  Following \mbox{\citet{liang16}}, \mbox{\citet{robson20}},} and~\mbox{\citet{jung21}}, we identify systems with $\log(M_{200}/\msolar) \geq 12.5$ as groups/clusters; these are shown as filled yellow circles. The~lower mass systems are plotted as open yellow circles.  The~$M_\star$ for ROMULUS systems is the projected stellar mass within an $R=50\;$pkpc aperture (i.e., $M_{\star,50,2D}$).
The $M_\star/M_{200}$ for the ROMULUS groups is in very good agreement with the observations; however, the~overall trend suggests a drop with increasing $M_{200}$ that is not as steep as the observed~trend.

In the same panel, we also show the results for TNG100 (cyan line/band; \cite{pillepich18b}) and FABLE (magenta line/band; \cite{Henden20}) simulations.  The~available TNG stellar masses are $M_{\star,30,3D}$, that is,~stellar mass within a \emph{sphere} of radius $R=30\;$pkpc. We have corrected these masses to stellar mass within a $50\;{\rm pkpc}$ sphere using mass profiles from \mbox{\citet{pillepich18b}}.  We suggest that it is preferable to compare this corrected stellar mass to the observed aperture masses.   The~FABLE results are $M_{\star,50,2D}$, like ROMULUS. The~thick solid line is the median SMHM relationship and the shaded region encompasses 95\% of the systems.   Both TNG and FABLE results are in very good agreement with the observations on the group and low-mass cluster scales (i.e.,~$\log(M_{200}/\msolar) < 14.0$ for TNG and $\log(M_{200}/\msolar) < 14.3$ for FABLE).  However, neither the TNG nor the FABLE median curve decreases as steeply as the observed relationship on the cluster scales. We also note that in the case of TNG, the~median curve may become shallower still if one were to use a cylindrical volume. 

In the second (top right) panel, we show the results from SIMBA (red line/band; \cite{dave19}) and The Three Hundred project (brown line/band; \cite{cui18}).  Both results are $M_{\star,50,2D}$.  Considering the SIMBA results first, we find that on the group scale, that is,~$\log(M_{200})=[12.5, 13.8]$, the~distribution of   $M_\star/M_{200}$ points for individual SIMBA BGGs (not explicitly shown) is consistent with the observations. The~median curve, however does not have as steep a slope as the SMBHs based on observations. Consequently, the~SIMBA central galaxies on the cluster scale (BCGs) have larger stellar mass than their observed counterparts and the discrepancy grows with increasing halo mass.  
Overall, SIMBA is in modest agreement with the observations when compared with observational results across the entire span, from~low mass groups to massive clusters.  The~Three Hundred SMHM does not extend to low mass groups but over the mass range covered ($\log(M_{200})=[13, 15]$), the~relationship is in excellent agreement with the observations.

The third (bottom left) panel shows the results from C-EAGLE (purple crosses \citet{bahe17}) and the EAGLE reference run (purple line/band; \cite{schaye15}). The~C-EAGLE stellar masses are $M_{\star,50,2D}$.  \textls[-15]{The~results for the lowest mass C-EAGLE systems are consistent with the observations but in massive clusters, as~\mbox{\citet{bahe17}} notes, the~C-EAGLE stellar masses exceed the $M_{\star,50,2D}$ (filled stars) from \mbox{\citet{kravtsov18}} by up to 0.6 dex.   As~for EAGLE, the~reported $M_\star$ is $M_{\star,30,3D}$; we
used a mean stellar mass-dependent correction factor, derived from plots in \mbox{\citet{schaye15}} and \mbox{\citet{mccarthy17}}, to~map $M_{\star,30,3D}$ to $M_{\star,50,3D}$. The~correction factor is negligible for galaxies with $\log(M_{\star,30,3D}/\msolar) < 10.7$ \citep{schaye15}}; on the other hand, the~stellar mass in the most massive  BGG/BCGs  approximately doubles.  Over~the mass range $\log(M_{200})=[13, 14]$,
the distribution of individual $M_\star/M_{200}$ EAGLE points (not explicitly shown) is consistent with the observed SMHM results; however, like the SIMBA results, the~EAGLE median curve is shallower than the observed trend.  Consequently, for~$\log(M_{200}/\msolar) < 13$, more than half of EAGLE BGGs fall below \mbox{\citet{moster13}} line, which is the lowest of the observationally-derived SMHM curves. Overall, the~EAGLE SMHM is in modest  agreement with the observations across the entire span, from~low mass groups to massive clusters.  Finally, we point out that the EAGLE median curve in Figure~\ref{fig:fstarcentral} is flatter than that in \mbox{\citet{schaye15}} because of the mass-dependent correction, and~we expect that the corresponding results based on $R=50\;{\rm pkpc}$ aperture mass will be shallower still since our correction maps to a sphere, not a~cylinder.

\subsubsection{Central Galaxy Star Formation~Rates}

\begin{figure}
\includegraphics[width=0.9\textwidth]{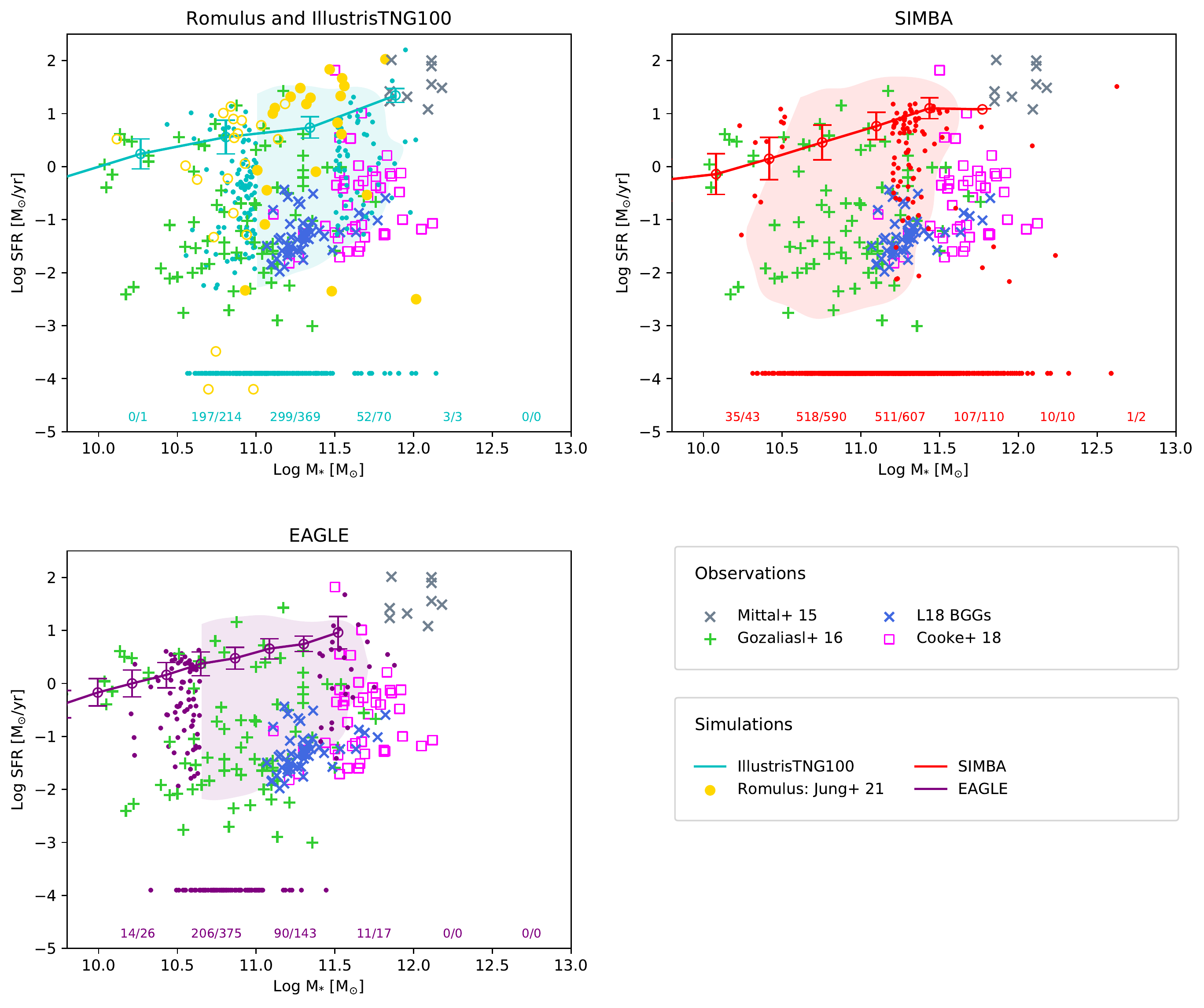}
\caption{
{\textls[-20]{{Figure adapted from} 
 \mbox{\citet{jung21}}:} Comparison of the $z=0$ star formation rate (SFR)-$M_\star$ relationship for observed and simulated BGGs and BCGs.  Arranged as Figure~\ref{fig:fstarcentral}, each panel plots the same four sets of observational results (see text for details).}
The top left panel shows the results for ROMULUS (yellow filled and open circles; filled circles represent systems we identify as groups/clusters) and TNG100 simulations (cyan); the top right panel shows the results for SIMBA (red); and the bottom left panel shows the results for EAGLE (purple).  The~simulation SFRs are extracted from a sphere of radius 30 pkpc encompassing the central BGG/BCG while the stellar masses as the same as in Figure~\ref{fig:fstarcentral}. 
The line connecting the open circles with error bars shows the star forming main sequence (SFMS) for TNG100, SIMBA, and~EAGLE from \mbox{\citet{dave20}}. The~shaded regions, and~the handful of individual points to the left and right, show where the simulation BGGs/BCGs with {\it measurable SFR} lie on this plot. BGGs/BCGs with unresolved/too low SFRs are assigned $\log(\mathrm{SFR}/\msolaryr)=-4$ and plotted accordingly.  We identify all simulation BGGs/BCGs with SFRs $\ga 0.75$ dex {\it below} the observed $z<0.5$ SFMS of \mbox{\citet{Whitaker12}} as ``quenched''.  For~TNG100, SIMBA, and~EAGLE, we specify the fraction of quenched galaxies in 0.5 dex $M_*$ bins at the bottom of the panels.  We do not differentiate between quenched systems with low but measurable SFRs and those with very low/unresolved SFRs.  Measuring very low SFRs using observationally accessible diagnostics is extremely challenging; strictly speaking, the~corresponding published SFRs ought to be treated as upper limits.
}
\label{fig:sfr}
\end{figure}

Figure~\ref{fig:sfr} shows the second trend that we consider here: the star formation rate$-$stellar mass relationship (hereafter, SFR$-M_\star$).  The~SFR is based on star formation within a $R=30\;{\rm pkpc}$ sphere centered on the BCG (very little star formation is observed beyond $30\;{\rm pkpc}$)  while the $M_*$ is the same as that in Figure~\ref{fig:fstarcentral}.  We first consider the four sets of observational data points that appear in all three panels: (i) results based on combined data from XMM–LSS, COSMOS, and~AEGIS surveys for BGGs and BCGs hosted by X-ray bright galaxy groups and clusters (green pluses; \cite{Gonzaliasl16}); (ii) results for BCGs in \mbox{\citet{mittal15}} sample of CC clusters (grey crosses); (iii) results for the high richness subset of the CLoGS sample (blue crosses; \cite{osullivan17,Loubser18,Kolokythas21}), which consists of groups containing at least 4 optically bright ($\log(L_{\rm B}/L_{\odot}) \geq 10.2$) galaxies, of~which the central galaxy is an ETG; and (iv) results for BCGs from the COSMOS survey (magenta squares; \cite{cooke18}).  

\textls[-5]{The \mbox{\citet{Gonzaliasl16}}, \mbox{\citet{Loubser18}}, \mbox{\citet{cooke18}} and \mbox{\citet{mittal15}} data collectively illustrate the approximate dichotomy in the population of BGGs and BCGs with respect to their star formation properties.  On~the cluster scale, this phenomenon has been widely discussed (e.g., \cite{Bildfell08,McDonald11}, and references there in) and linked to the CC/NCC dichomotomy in the X-ray properties, including the X-ray luminosity at fixed halo mass, the~core entropy value and the core hot gas cooling time as highlighted by \mbox{\citet{McCarthy04,mccarthy08b,cavagnolo09}} and others.  Essentially, not all BCGs are ``red and dead''; there exists a population of star-forming BCGs that reside in strong cool core clusters \citep{Bildfell08}.  {\it These BCGs' SFRs are such that they are distributed within approximately $\pm 0.75$ dex of the extension of the observed $z<0.5$
star forming main sequence} (SFMS) of \mbox{\citet{Whitaker12}}.   The~star-forming BCGs comprise $\sim$25\%--30\%  of all BCGs;  this is also the fraction of strong CC clusters \citep{Mahdavi13,Andrade-Santos2017}.  \mbox{\citet{Gonzaliasl16}} find that this dichotomy is also present on the group scale and the fraction of star-forming BCGs is comparable to that of star-forming BCGs; that is,~20\%--25\%.  The~majority of the observed central galaxies in Figure~\ref{fig:sfr} have SFRs less than $0.75$ dex {\it below} the SFMS.  We refer to these as ``quenched'' BGGs/BCGs.    }

\textls[-10]{As for the simulations, we restrict ourselves to groups and clusters whose BGG/BCG satisfy $\log(M_\star/ \msolar)\la 12.0$; there are too few systems with higher stellar masses.  We classify the simulation BGGs/BCGs in the same way as the observed systems: We designate all simulation BGGs/BCGs with SFRs less than $0.75$ dex {\it below} the observed $z<0.5$ SFMS~\citep{Whitaker12}} as ``quenched'' systems, and~the rest of the galaxies as star-forming.  The~numbers at the bottom of the panels specify the fraction of quenched galaxies in 0.5 dex $M_*$ bins for TNG100, SIMBA, and~EAGLE simulations.  Many of the quenched galaxies have low/unresolved SFRs.  In~order to display these galaxies in the panels, we assign them an SFR of $\log(\mathrm{SFR}/\msolaryr)=-4$. To~the extent that it is of interest, we find that the number of galaxies with very low/unresolved SFRs relative to the total number varies considerably from simulation to simulation. However,  we do not differentiate between quenched galaxies with low/unresolved SFRs and those with low but measurable SFRs, and~there is there is no observational basis for doing so. Measuring low SFRs using observationally accessible diagnostics is extremely challenging \citep{Bonjean19,KennicutEvans12ARA&A,Pipino09,Fogarty17,Mittal17,Loubser11}, and~strictly speaking, very low published SFRs ought to be treated as upper~limits.  

Considering the ROMULUS results (yellow points) in the top left panel first, we find that this model gives rise to both star-forming and quenched BGGs; however, the~corresponding fractions are inverse of the observed fractions: $\sim$60\% of the ROMULUS BGGs are star-forming and $\sim$40\% are quenched. The~top left panel also shows the entire TNG100 population of BGG/BCGs. Only $17\%$ of these are star-forming. This is lower than the observed fraction. The~SIMBA results in the top right panel are similar to TNG100 and also on the low side: only about $14\%$ of all the SIMBA BGGs/BCGs are star-forming. The~fraction of EAGLE BGGs (lower left panel) that are star-forming is $\sim$40\%, which is higher than the observed fraction though not as high as the ROMULUS fraction.  We will return to these results in Section~\ref{sec:BGG-IGrM}.

\subsubsection{Central Galaxy~Morphologies}

\textls[-5]{The final trend we consider is the stellar mass-morphology relationship for the central galaxies. Qualitatively, the~observed morphologies of central galaxies span the full continuum, from~disk dominated to pure spheroids. \mbox{\citet{Weinmann06}} found that approximately $50\%$ of the BGGs in low-mass SDSS groups are late-type galaxies (see also~\cite{Li19}); this fraction drops to $\sim$10\% in rich groups/poor clusters.  Conversely, the~fraction of early-type central galaxies rises from $\sim$30\% in low-mass groups to $70\%$ in high-mass systems.  The~morphological mix of the ROMULUS BGGs, which are from mainly low-mass groups, is consistent with \mbox{\citet{Weinmann06}} results in that nearly a half are disky and star forming \citep{jung21}.  Similarly, a~visual inspection of the images of BCGs from the 20 most massive TNG100 systems with halos masses between $9\times 10^{13}\;\msol$ and $4\times 10^{14}\;\msol$ ( \cite{pillepich18b}) also suggests that their morphological types  are also compatible with \mbox{\citet{Weinmann06}} results for low-mass clusters: A significant fraction are ellipsoidal or S0-like and appear to be~red.  }

More quantitatively, a~galaxy's morphology is commonly characterized by the fraction of its total stellar mass that is in the spheroidal component; that is,~its spheroidal-to-total ratio ($S/T$). This ratio has been computed using photometric observations, by~decomposing the 2D projected light profile into disc and spheroidal components, but~it can also be derived using stellar kinematics to identify the spheroidal component.  The~two approaches do not also give similar results.  Simulation studies have shown that photometrically derived $S/T$s tend to be significantly lower than the kinematic $S/T$s \citep{Scannapieco10}, with~\mbox{\citet{Bottrell17}} finding that the former leads to a higher likelihood of a galaxy being classified as ``disky even when stellar kinematics show no ordered rotation''.  In~effect, the~\mbox{\citet{Scannapieco10}} and \mbox{\citet{Bottrell17}} studies show that it is not sufficient to only consider the BGG/BCG properties derived from photometric data.  There is much to be learnt from examining the galaxies' kinematic~properties.  

\begin{figure}

\includegraphics[width=0.63\textwidth]{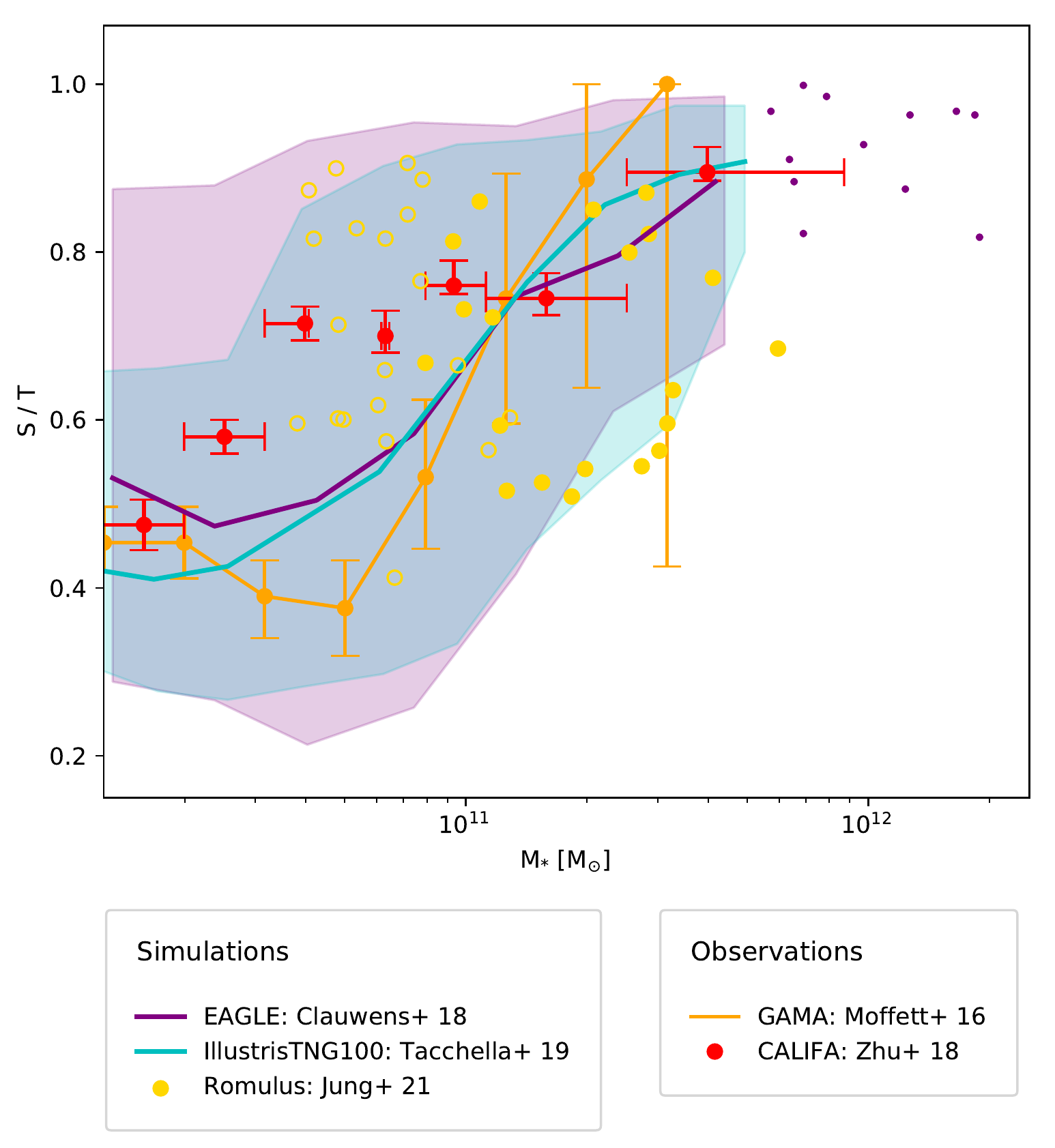}

\caption{\textls[-10]{{{Figure adapted from} \mbox{\citet{jung21}}:} The $z=0$ spheroidal-to-total ($S/T$) for central galaxies in simulated galaxy groups and clusters from three simulations (ROMULUS, TNG100, and~EAGLE), juxtaposed against results from two observational studies.  The~yellow filled and open circles show the ROMULUS kinematically derived results. (As in previous two figures, filled circles correspond to systems we identify as groups and clusters.) The cyan and purple curves show the median kinematically derived $S/T$ results for TNG100 \citep{Tacchella19} and EAGLE \citep{clauwens18}, respectively.  The~purple shaded region indicates the 10--90\% range while the cyan shaded region spans  the 16\%--84\% range.  The~ROMULUS and TNG100 $S/T$ are computed using the same criterion while the EAGLE results are based on a slightly different criterion.  The~two criteria, however, give very similar results.  Of~the two observational results shown, the~CALIFA data are kinematically derived \citep{zhu18} while the GAMA results are based on photometry \citep{Moffett16a}.  The~two approaches are not equivalent.  We discuss this further in the text.  The~simulation results should, in~the first instance, be compared to the CALIFA~results.}}
\label{fig:S2T} 
\end{figure}

\textls[-10]{In Figure~\ref{fig:S2T}, we show the kinematic $S/T$ ratios for the central galaxies from ROMULUS \citep{jung21}, TNG100 \citep{Tacchella19} and EAGLE \citep{clauwens18}, as~a function of galaxy stellar mass.  We use comparable apertures as \mbox{\citet{Tacchella19}} to compute ROMULUS $S/T$ ratios: For BGGs with $\log(M_*/\msol) \leq 10.9$, $S/T$s are computed using star particles within spheres of radius $R$ = 15--20 pkpc; for more massive BGGs, we use spheres of radius $R$ = 25--30 pkpc.  We also use the same criterion to define the spheroidal component as \mbox{\citet{Tacchella19}}: the mass of the spheroid is the sum of the mass of stellar particles with $\epsilon_{\rm J}<0.7$ and the 15\% of the stellar particles with $\epsilon_{\rm J}>0.7$, 
where $\epsilon_{\rm J}=J_{\rm z}/J_{\rm circ}(E)$,  $J_{\rm z}$ is a z-component of the specific angular momentum of a stellar particle, z is the net spin axis of a galaxy, and~$J_{\rm circ}(E)$ is a specific angular momentum of the stellar particle on a circular orbit with the same orbital~energy.}

The EAGLE criterion for determining $S/T$ ratios differs slightly but we have confirmed that the two criteria give comparable results.  Figure~\ref{fig:S2T} also shows the $S/T$ for central galaxies in the GAMA groups sample derived using photometric data (orange points;~\cite{Moffett16a}), as~well as the kinematic $S/T$ for BGG/BCGs from the CALIFA survey (red points;~\cite{zhu18}).  We use the same prescription as \mbox{\citet{Tacchella19}} to compute the CALIFA $S/T$s. The~simulation results should be comparable to the CALIFA~results.   

Overall, there is a broad agreement between the three simulation results and for $M_* \ga 10^{11}\;\msol$, all three simulation results are consistent with the CALIFA results.  In~these galaxies, the~spheroidal component dominates: $S/T$ rises from approximate 0.7 to 0.9 with stellar mass.  Towards lower $M_*$, however, the~median TNG100 and EAGLE curves diverge from the CALIFA results.  At~present, it is unclear how much weight this divergence ought to be given.  Firstly, there is a considerable spread in the simulation results and the CALIFA results are well within the shaded region; secondly, the~CALIFA sample is known to be increasingly incomplete towards at lower masses.  It is possible that the trend in the CALIFA results reflects this~incompleteness.


Finally, we note that $S/T$ is not the only kinematic measure that is correlated with either the central galaxies' stellar mass or the host systems' properties.  Other measures less commonly discussed in the galaxy formation literature include (i) the shape of the velocity dispersion profiles and whether they rise or fall with radius; (ii) the ``anisotropy parameter'' $V_{\rm rot}/\sigma_0$, which characterizes the global dynamical importance of rotation and random motions of stars in a galaxy; and~(iii) whether the galaxy is a fast or slow rotator.   \mbox{\citet{jung21}} investigates some of these measures for ROMULUS galaxies and compare them to observations and we refer interested readers to that paper for further details and~discussion.


\subsubsection{The Link between the BGG and~IGrM}\label{sec:BGG-IGrM}


The properties reported in Figures~\ref{fig:fstarcentral}--\ref{fig:S2T} are the observable byproducts of a myriad of physical processes that the BGGs and BCGs are subject to over the course of cosmic time.  These properties are also sensitive to the details of the subgrid prescriptions used to model those processes that cannot be directly resolved in the simulations.  All three of the properties discussed in this subsection are related to each other and more importantly, they are all correlated with the the current or the recent state of the IGrM/ICM in the group and cluster cores.  Here, we briefly elaborate on this relationship, focusing on the SFR-$M_\star$~results.

As highlighted throughout this review, among~the most important present-day challenges in simulating galaxy groups and clusters is (i) preventing strong radiative cooling flows from forming in the IGrM/ICM and~(ii) ensuring that the resultant population of groups and clusters span the spectrum from CC to NCC systems.  Nearly two decades ago, \mbox{\citet{Valageas99,Nath02,babul02}} advocated for AGN feedback as the key mechanism for addressing these challenges.  
The current generation of simulation models of galaxy formation/evolution (see Table~\ref{tab:AGN}) all allow for SMBH seeding and growth as well as AGN feedback although the manner in which these are implemented vary from one simulation to~another.  

To examine how well the current models fare, we first consider the ROMULUS simulations.   As~noted previously, the~fact that nearly $60\%$ of the ROMULUS BGGs are star-forming (Figure \ref{fig:sfr}) means that while AGN feedback in ROMULUS may temper cooling, it does not entirely halt it and consequently, a~majority of the systems sustain some degree of cooling flow in their cores.  A~closer examination of the ROMULUS entropy profiles, such as those shown in Figure~\ref{fig:entropy_profiles}, offer some insights.  For~the most part, ROMULUS groups behave like CC clusters, not CC \emph{groups}.  Their entropy profiles  are $\sim$2--3$\times$ lower than the \mbox{\citet{sun09}} profiles at large radii, and~the profiles themselves are steeper ($R^{1.1}$ versus $R^{0.7}$).  We suspect this is the result of a combination of AGN feedback being too weak and metal-line cooling being absent.  ROMULUS simulations do not produce stable NCC systems although once in a while, they do produce weak NCC systems \citep{chadayammuri21} and during this phase, the~BGG is quenched \citep{jung21}.  This phase typically arises when the BGG suffers a sizable merger that elevates the core entropy; however, the~phase is temporary, lasting $\lesssim 2$ Gyrs.

Turning to the TNG and SIMBA simulations, we find that 78\% and 91\% of the BCGs, respectively, with~$\log(M_*/\msol) \geq 11.3$ are quenched.  Such high fractions are not surprising given that the vast majority of the $z=0$ TNG and SIMBA clusters are NCC systems with high entropy cores (see, for~example, the~orange and yellow curves in Figure~\ref{fig:entropy_profiles}). Neither TNG nor SIMBA produce clusters with typical CC cluster entropy profiles, and~in both cases, the~main reason appears to be that feedback is both too efficient, resulting in larger and higher entropy central cores than observed (c.f.~ \cite{barnes18,robson20}).

The main takeaway, therefore, is that while all of the simulation models appear to find reasonable agreement with a subset of the available observations of galaxy groups and clusters, none of them correctly describe galaxy groups and clusters in a comprehensive manner.  It is very likely that the problems with the entropy profiles and quenched/star-forming BGGs/BCGs are due to either a problem with the physical models that inform the subgrid modeling of the various aspects of SMBH formation and evolution, and/or due to the implementation these models in the simulations.  All simulation models have ad~hoc features embedded within their  subgrid prescriptions and when the models run into difficulties, quite reasonably a common response is to attempt to patch one or more of the subgrid schemes in the hope that that fixes the problem without causing any breakage elsewhere.  In~the case of AGN feedback, we assert that minimal patches have generally not led to a realistic population of simulated galaxy groups and clusters.  This should be seen as a clear invitation to revisit the  treatment of SMBHs and AGNs in the simulations.  We offer some thought on the subject in Section~\ref{sec:researchtopics}.

\subsection{The Multiphase~IGrM} \label{sec:multiphase}

\begin{figure}[H]
\includegraphics[width=0.95\textwidth]{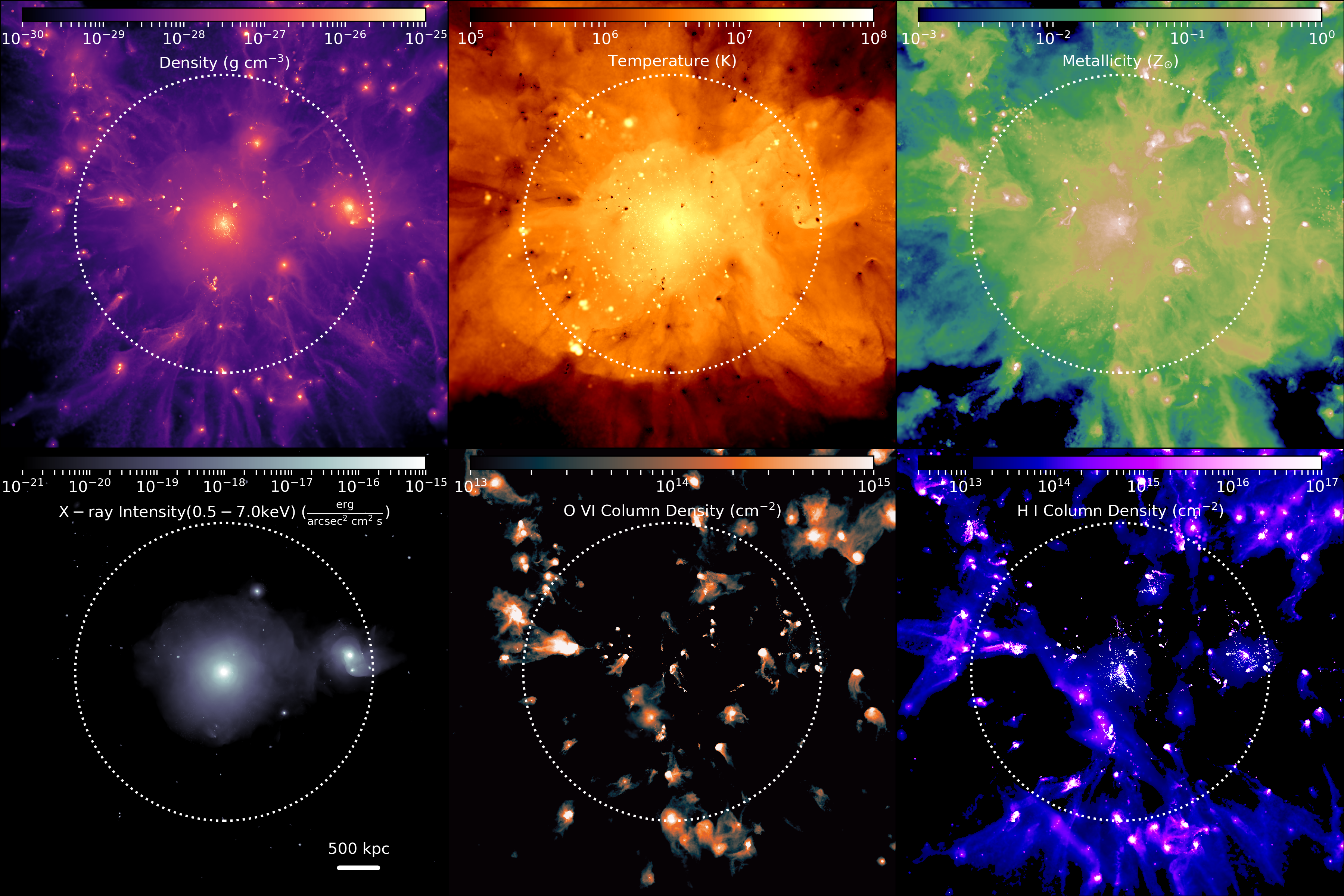}
\caption{The ROMULUSC simulation at z = 0.31 \citep{butsky19}. The~top row shows the projected density, temperature, and~metallicity. The~bottom row shows the predicted X-ray emission as well as the O {\sc VI} and H {\sc I} column densities. Each image spans 5 Mpc across and the white dashed circle has a radius of $3R_{500}$. X-ray emission probes the inner hot, dense region of the IGrM. UV absorption of ions like O {\sc VI} and H {\sc I} provide a highly complementary view of the IGrM, tracing the filamentary structure of cool and warm gas as it is stripped from its host~galaxies. }
\label{fig:ROMULUSC_Maps}
\end{figure}

Observations of the IGrM are dominated by the X-ray emission from hot, dense gas surrounding the BGG. Since X-ray emission scales with the square of gas density, diffuse warm and cool phases of the IGrM are prohibitively difficult to observe directly in emission.  While several groups have pioneered efforts to characterize the warm and cool IGrM/ICM with UV absorption specifically targeting group environments \citep{pointon17, nielsen18, stocke19}, the \mbox{Virgo~\citep{yoon12, manuwal19}} and Coma \citep{yoon17} clusters, other clusters \citep{burchett18, connor19},  cluster outskirts \citep{muzahid17,jaydev19}, as~well as low-mass group halos hosting ETGs \citep{zahedy19}, such studies are limited by the available number of background sources with existing instrument sensitivities.  For~this reason, simulations of groups offer a unique insight into the multiphase structure of the IGrM and the physical origins of its different components.  Because~warm and cool gas is expected to form structures that are significantly smaller than the volume-filling hot phase, simulations that aim to accurately model the multiphase IGrM require extremely high resolution. Recent advancements in the resolution of group-scale simulations (like RomulusC and TNG50) have allowed for unprecedented insights into the nature of the multiphase IGrM. Figure~\ref{fig:ROMULUSC_Maps} demonstrates the multiphase gas structure in the ROMULUSC simulation at \mbox{z = 0.31}. The~top row shows $5 \times 5$ Mpc ($9.5 \times 9.5\ R_{500}$) projections of gas density, temperature, and~metallicity. The~bottom row shows synthetic X-ray emission and UV absorption maps, roughly calibrated to the sensitivity of existing instruments.  This figure clearly demonstrates that X-ray emission and UV absorption studies probe highly complementary regions of the IGrM. Additionally, it highlights the rich multiphase structure of the IGrM that exists out to the edges of the group~outskirts.

\begin{figure}[H]

\includegraphics[width=0.49\textwidth]{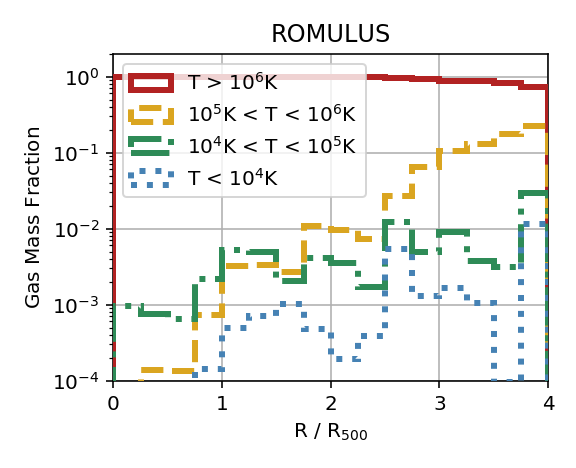}
\includegraphics[width=0.49\textwidth]{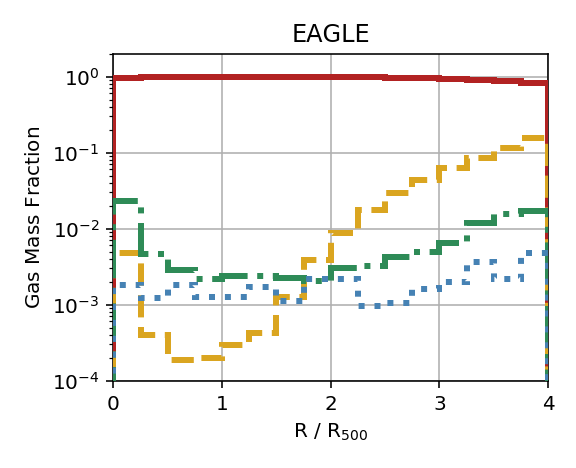}
\includegraphics[width=0.49\textwidth]{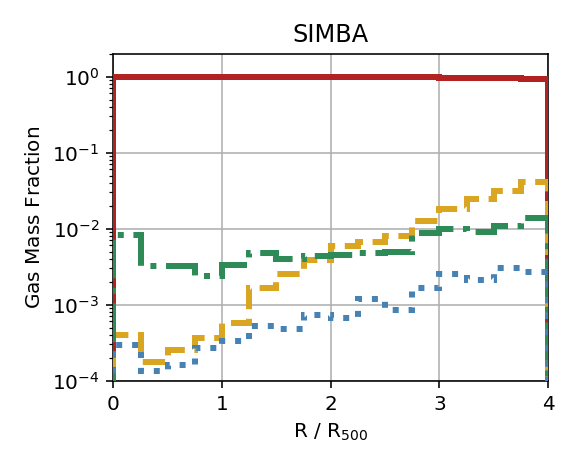}
\includegraphics[width=0.49\textwidth]{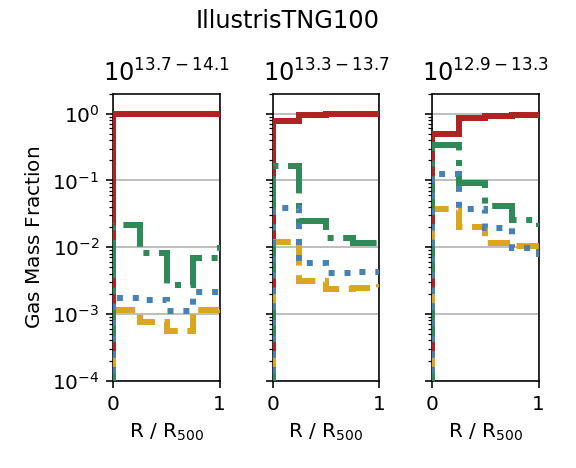}

\caption{The total mass fraction of hot ($T \geq 10^6K$), warm ($10^5K \le T < 10^6K$), cool ($10^4K \le T < 10^5K$), and~cold ($T \le 10^4 K$) gas as a function of distance from the group center at $z=0$ for halos at $M_{500}\approx 10^{13.9}\,\msolar$ for ROMULUS (ROMULUSC in this case), EAGLE ($N=7$ groups), SIMBA ($N=39$), and~the inner regions of TNG100 for this bin (left subpanel, $N=13$) and two lower mass halo bins (middle, $M_{500}=10^{13.3-13.7}\,\msolar$, $N=41$; \& right, $M_{500}=10^{12.9-13.3}\,\msolar$, $N=105$).  Although~hot gas dominates the mass fraction within $4 R_{500}$ of the BGG in all cases, cool gas never constitute less than $10^{-3}$ of the total gas mass within $R_{500}$.}
\label{fig:temperature_profiles}
\end{figure}

We quantify the fractional breakdown of multiphase gas throughout the extended IGrM for $z=0$ snapshots of ROMULUSC, EAGLE, SIMBA, and~TNG100, in~Figure~\ref{fig:temperature_profiles}.  Gas is divided into ``hot'' ($T \geq 10^6K$), ``warm'' ($10^5K \le T < 10^6K$), ``cool'' ($10^4K \le T < 10^5K$), and~``cold'' ($T \le 10^4 K$).  The~IGrM is clearly dominated by the hot phase all the way out to and beyond $ 4R_{500}$, and~the cooler phases are not dramatically different between the 4 simulations.  SIMBA has the least amount of multi-phase gas, and~by total mass even less inside $R_{500}$, since the IGrM is most evacuated in this simulation.  ROMULUSC has the most warm-hot gas at large radii, but~far less than at $z=0.31$ where \mbox{\citet{butsky19}} showed more multi-phase gas that was disrupted by a 1:8 merger \citep{chadayammuri21}.  The~properties of this hot inner region are intimately tied to the feedback from the central BGG. In~particular, the~temperature and entropy of the IGrM near the BGG depends on the recent AGN activity. After~periods of relatively low AGN activity, the~entropy profile of the inner 0.04 $R_{500}$ is low and susceptible to thermal instability, which fuels star formation and AGN activity. After~vigorous AGN activity, the~inner 0.2 $R_{500}$ develops hotter temperatures and a steeper entropy profile \citep{chadayammuri21}.  The~metallicity of the hot phase is around $0.3\ Z_{\odot}$ and remains relatively constant throughout the inner region \citep{butsky19}. The~prevalence and uniformity of metals in the hot phase implies that the hot phase was primarily enriched over time by diffuse gas in the group outskirts and mergers with group satellites (e.g., \cite{biffi2018a, biffi2018b}). 

The statistical samples of EAGLE, SIMBA, and~TNG100 show more similar patterns of cool gas inside $R_{500}$ including a fractional increase inside $0.25 R_{500}$, which we detail below.  Further out, warm gas begins to constitute a substantial fraction of the IGrM mass and beyond $\sim2.5 R_{500}$, but~unlike the smooth, large volume-filling hot phase, the~cool gas distribution is patchy and traces gas that is stripped from satellite galaxies as they move through the IGrM.  In~the most extreme cases, the~satellite galaxies and their stripped tails are referred to as ``jellyfish galaxies'' as seen in ROMULUSC $\OVI$ and $\HI$ maps in Figure~\ref{fig:ROMULUSC_Maps}.  In~ROMULUSC, gas stripped from satellite galaxies tends to be cooler and more metal-enriched than the ambient IGrM gas in the group outskirts, leading to a wider distribution of metallicities traced by cool and warm gas. ROMULUSC satellite galaxy CGMs (i.e., gas within $\la 150$ kpc of satellites) show decreasing covering fractions of O {\sc VI}, C {\sc VI} and H {\sc I} at lower IGrM radii with a significant decline inside $3 R_{500}$ \citep{butsky19}. Eventually, the~stripped halo gas mixes with the rest of the IGrM. Constraining this mixing rate will be important for understanding how galaxies lose their gas and how the group environment is important for galaxy~evolution. 

In addition to being stripped from satellite galaxies, cool gas can also form through thermal instabilities, which is prevalent at a $\sim$$10^{-2}$ fraction in the inner $0.25 R_{500}$ of most simulations of $\sim$$10^{14}\,\msolar$ groups.  When the local cooling time ($t_{\rm cool}$) of gas is roughly less than $10\times$ the gravitational freefall time ($t_{\rm ff}$) \citep{mccourt12}, small perturbations can seed a runaway cooling effect through which cool gas condenses out of the background medium and precipitates onto the central BGG. This process of local thermal instability in a globally stable atmosphere is self regulating and maintains a global cooling to freefall time ratio ($t_{\rm cool} / t_{\rm ff}$) $\simeq$ 10--20 throughout the IGrM (e.g., \cite{esmerian21, Sharma12}).  Any gas with $t_{\rm cool} / t_{\rm ff} \le 10$ will form cool filaments, lowering the density (and cooling time) of the remaining hot IGrM.  This cool gas is accreted onto the central BGG and can trigger stellar or AGN feedback, driving the baryonic feedback cycle and maintaining global thermal stability in the IGrM.   This process of thermal instability promoting the formation of clumps of cool gas was traced by \mbox{\citet{nelson20}} using TNG50, which we check show very similar fractional values as TNG100 for the two most massive group halos in that $50^3$ Mpc$^3$ box.  

Finally, we show the mass trend of fractional IGrM phases for changing group halo mass bins in the subpanels on the bottom right of Figure~\ref{fig:temperature_profiles} for TNG100.  The~uptick in cool gas inside $0.25 R_{500}$ goes from 2\%, to nearly 20\%, to over 30\% as one progresses down the mass scale from $M_{500}=10^{13.9}$ to $10^{13.1}\,\msolar$.  Thus it should not be surprising that ions like $\HI$, $\SiIII$, and~$\CIV$ in halo gas around relatively isolated ETGs that may occupy $\sim$$10^{13}\,\msolar$ halos \citep{tumlinson13, zahedy19}.  However, surveys targeting more massive groups show very little cool ($\HI$) or warm ($\OVI$) gas inside $R_{\rm vir}$ \citep{stocke19}.  

\subsection{Satellite Galaxies in~Groups} \label{sec:satellite}

\begin{figure}[H]

\includegraphics[width=0.7\columnwidth]{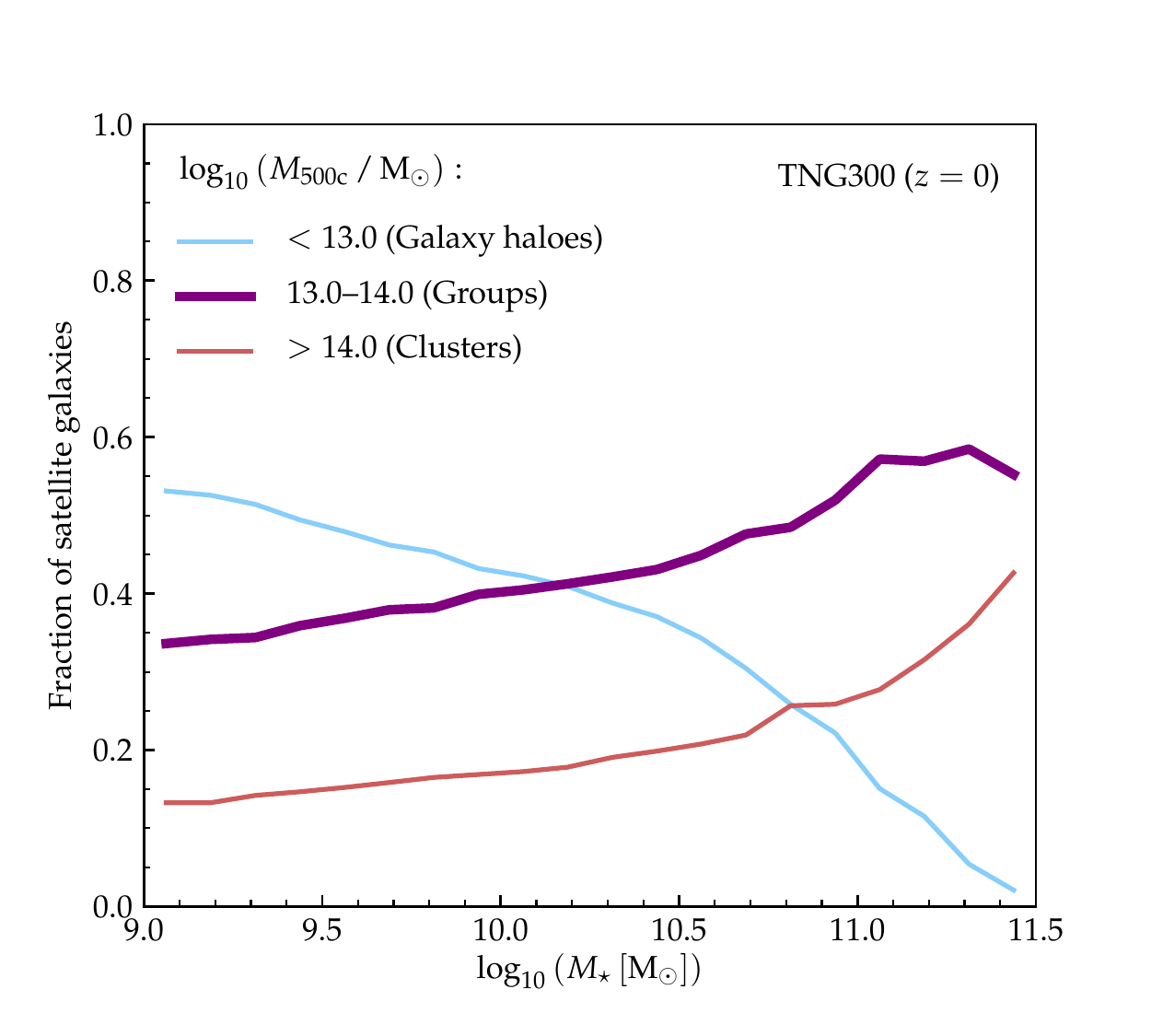}

\caption{Fraction of satellite galaxies in the IllustrisTNG300 simulation that are hosted by galaxy-scale halos ($M_\mathrm{500c} < 10^{13}\,\msol$, light blue), groups ($M_\mathrm{500c} = 10^{13}$--$10^{14}\,\msol$, thick purple), and~clusters ($M_\mathrm{500c} > 10^{14}\,\msol$, light red), respectively, as~a function of their stellar mass. Groups host around a third of low-mass satellites, and~more than half of those with $M_\star \approx 10^{11}\,\msol$; their contribution exceeds that of clusters at all stellar masses shown~here.}
\label{fig:sat_fractions}
\end{figure}

Group satellites have been studied extensively in observational surveys, not least because groups are more than an order of magnitude more common than  massive clusters~\citep{Jenkins_et_al_2001}; even medium-size surveys such as GAMA therefore include more than a thousand of them~\citep{Viola_et_al_2015}. Although~each individual group contains fewer satellite galaxies than a rich cluster, collectively they still host $\gtrsim$2 times as many satellites (see Figure~\ref{fig:sat_fractions}). These observations generally place group satellites between clusters and the field with, for example,~quenched fractions of $\approx$60\% \citep{Wetzel_et_al_2012} and \HI{} mass fraction ratios ($M_{\HI}/M_\star$) of $\approx$0.15~\citep{Brown_et_al_2017} at $M_\star = 2\times 10^{10}\,\msol$, (extended) X-ray detection fractions of bright ETGs as high as $\approx$90\%~\citep{Jeltema_et_al_2008}, and~elliptical galaxy fractions of $\approx$50\% \citep{Bamford_et_al_2009}.

In simulations, stripping of the (extended) warm-hot gas halos of group satellites is robustly predicted (e.g.,~\mbox{\citet{Bahe_et_al_2012, Zinger_et_al_2018, butsky19}}), even for galaxies that are still (well) beyond the virial radius of the group. Ram pressure is strong enough to explain this gas loss, in~particular for galaxies that (temporarily) move through a denser part of the IGrM or with a higher velocity than typical at a given radius \citep{bahe15}. As~discussed above, this gas loss is simultaneously predicted to be an important route of IGrM enrichment \citep{butsky19}.

\textls[-25]{To our knowledge, no simulation that stretches up to group scales is currently able to self-consistently model the evolution of atomic and molecular hydrogen (see e.g., {\citet{Hopkins_et_al_2018, Applebaum_et_al_2021}} for examples of such simulations on smaller scales). Some authors have, however, modelled \HI\ in post-processing with theoretically and/or empirically motivated relations (e.g., \mbox{\citet{Blitz_Rosolowski_2006,Rahmati_et_al_2013,Gnedin_Draine_2014}})} to derive \HI{} and H$_2$ masses of group satellites. For~EAGLE, \mbox{\citet{Marasco_et_al_2016}} demonstrated agreement with ALFALFA observations and found that \HI{} loss on group scales was driven by a complex mixture of tidal stripping, ram pressure, and~satellite--satellite encounters. \mbox{\citet{Stevens_et_al_2019}} created detailed \HI{} mock observations of the TNG100 simulation, showing that this step was critical in achieving a match to the observed \HI{} mass fractions of \mbox{\citet{Brown_et_al_2017}}. They also found, however, that the \HI{} loss in TNG100 parallels the decline in SFR, contrary to the observations that show lower \HI{} mass fractions even at fixed sSFR \citep{Brown_et_al_2017}.

\begin{figure}[H]
\includegraphics[width=0.7\columnwidth]{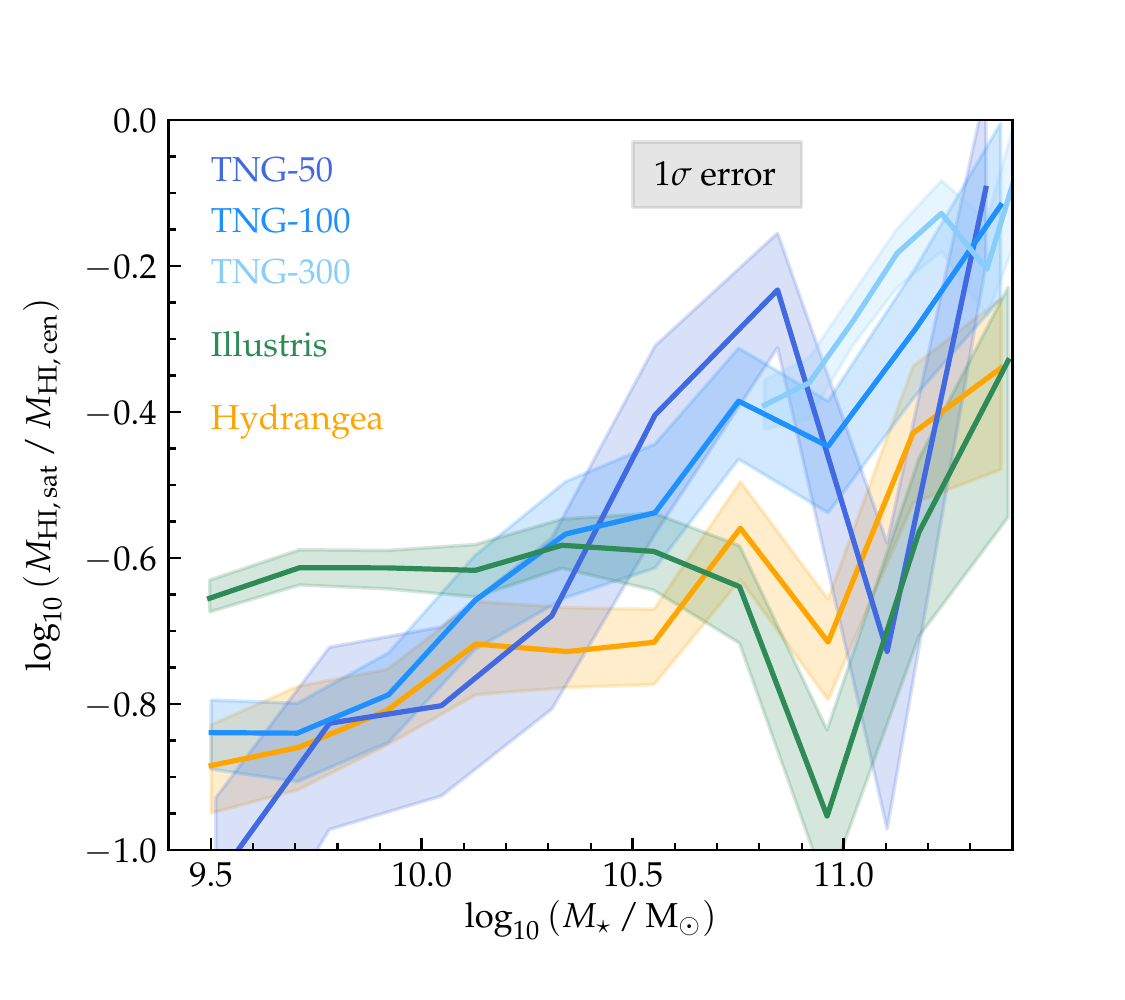}

\caption{Atomic hydrogen (\HI) deficiency of group satellites compared to centrals of the same stellar mass, as~predicted by the Illustris (green), IllustrisTNG (shades of blue), and~Hydrangea (orange) simulations (computed following \mbox{\citet{Diemer_et_al_2018}} and \mbox{\citet{Bahe_et_al_2016}}, respectively). Solid lines show the difference between mean \HI{} masses in each bin, shaded bands the corresponding $1\sigma$ uncertainties obtained from bootstrapping. All simulations predict \HI-deficient group satellites, with~the difference generally largest (almost 1 dex) for the least massive galaxies. Note that for IllustrisTNG300 (light blue), \HI{} masses are only computed for all galaxies with $M_\star > 5\times 10^{10}\,\msol$.}
\label{fig:HI_deficiency}
\end{figure}

In Figure~\ref{fig:HI_deficiency}, we compare the \HI{} deficiency of group galaxies $\Delta\HI{}$ = $\log_{10}(M_\mathrm{\HI{}, sat} / M_\mathrm{\HI{}, cen})$ (where $M_\mathrm{\HI{}, sat}$ and $M_\mathrm{\HI{}, cen}$ are the total \HI{} masses of group satellite and central galaxies in a narrow range of $M_\star$) as predicted by the Illustris, IllustrisTNG, and~Hydrangea simulations. For~the first two, we take the \HI{} masses computed by \mbox{\citet{Diemer_et_al_2018}} on a cell-by-cell basis with the \mbox{\citet{Gnedin_Draine_2014}} \HI{}/H$_2$ partition; for Hydrangea the \HI{} have been computed in analogy to \mbox{\citet{Bahe_et_al_2016}} with the empirical \mbox{\citet{Blitz_Rosolowski_2006}} \HI{}/H$_2$ partition. Despite the variety of simulations, resolutions, and~\HI{} models, the~predictions are remarkably uniform: all simulations (except for Illustris) predict an \HI{} deficiency of $\Delta \HI{} \approx 0.9$ at the low-mass end ($M_\star \approx 10^{9.5}\,\msol$), and~a less extreme difference at high masses ($\Delta \HI{} \approx 0.2$).

\mbox{\citet{Stevens_et_al_2021}} investigated the molecular (H$_2$) masses of satellite galaxies in TNG100 through a similar mock imaging approach as \mbox{\citet{Stevens_et_al_2019}}. Despite the lack of a directly modelled cold ISM phase in these (and other) simulations, they obtained an H$_2$ mass fraction for group satellites that is $\approx$0.6 dex lower than in the field, consistent with data from the xCOLD GASS survey \citep{Saintonge_et_al_2017}.

\begin{figure}[H]
\includegraphics[width=0.98\textwidth]{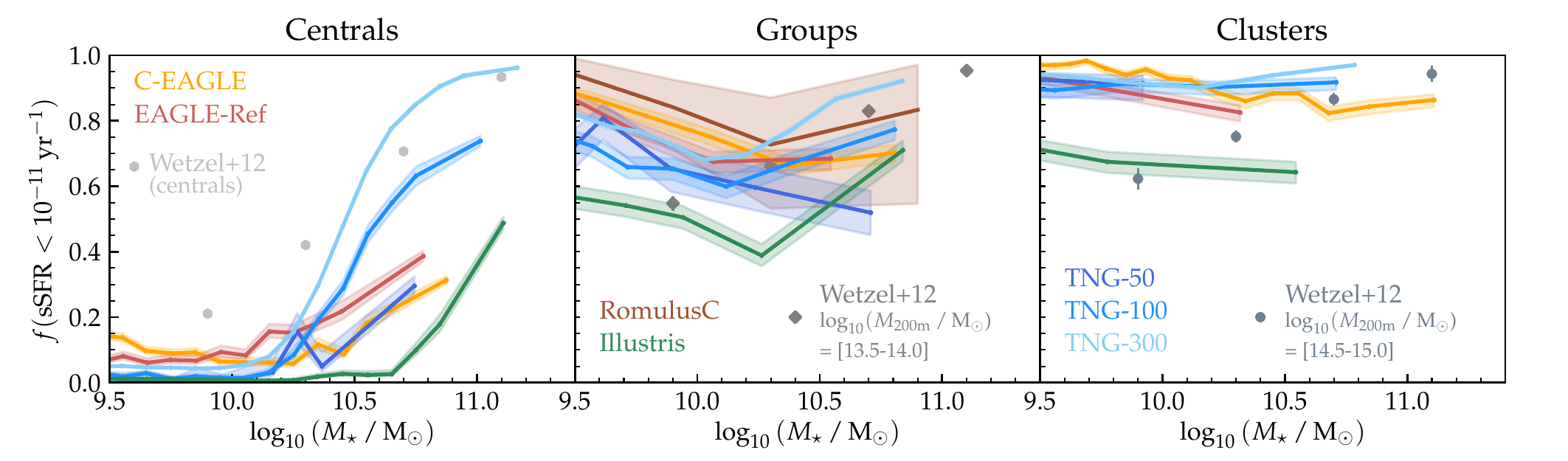}
\caption{\textls[-5]{Quenched fractions $f_q$ of central galaxies (\textbf{left}) and of satellites in groups ($M_\mathrm{500} = 10^{13}-10^{14}\,\msol$, (\textbf{middle})) and clusters ($M_\mathrm{500} > 10^{14}\,\msol$, (\textbf{right})) as predicted by different simulations. For~(C-)EAGLE, Illustris, and~IllustrisTNG, these are computed directly from the simulation outputs; the quenched fractions for RomulusC are taken from \mbox{\citet{tremmel19}}. Especially on the low stellar mass end ($M_\star \lesssim 2 \times 10^{10}\,\msol$), different simulations agree closely for centrals ($f_q \approx 0$) and clusters ($f_q \approx 1$), whereas predictions on group scales show a much larger diversity ($f_q$ between 0.6 and 1). Observational data from \mbox{\citet{Wetzel_et_al_2012}} are shown for approximate guidance, but~neither the satellite selection nor the host mass ranges are matched to their~analysis.}}
\label{fig:galaxies_fq}
\end{figure}

The impact of this gas loss on star formation is arguably the most well-studied aspect of simulation works on group galaxies. Unanimously, simulations predict that the quenched (or red) fractions of group galaxies are significantly higher than for equal-mass field galaxies (e.g.,~\mbox{\citet{bahe17, tremmel19, Donnari_et_al_2021}}). In~simulation suites that include groups, as well as more massive clusters (Hydrangea/C-EAGLE, IllustrisTNG), clear trends with halo mass are seen, at~least for $M_\star \lesssim 10^{11}\,\msol$ (see Figure~\ref{fig:galaxies_fq}): quenched fractions in groups are below those for clusters by up to a factor of $\approx$2 \citep{bahe17, Donnari_et_al_2021}. When only considering satellites that were directly accreted onto their $z = 0$ host and not quenched already, however, \mbox{\citet{Donnari_et_al_2021}} report that this trend reverses for $M_\star \gtrsim 3\times 10^{10}\,\msol$: at the massive end, the~quenched fraction of this subset of satellites is highest for groups (up to 80\%) and~lowest in massive clusters (40\%). \mbox{\citet{Donnari_et_al_2021}} interpret this ``host rank inversion'' as a result of AGN feedback: while this still operates efficiently for massive satellites in groups, it is suppressed in more massive clusters and therefore does not quench massive galaxies as~efficiently.

A second indicator of changes to the baryon cycle in group galaxies, their ISM metallicity, was investigated by \mbox{\citet{Genel_2016}}, \mbox{\citet{Bahe_et_al_2017a}}, and~\mbox{\citet{Gupta_et_al_2018}} with the Illustris, EAGLE, and~TNG100 simulations, respectively. All three studies found elevated metallicities of satellites compared to the field, a~difference that is more pronounced for lower $M_\star$ and  higher halo mass, in~qualitative agreement with observations \citep{Pasquali_et_al_2012}. For~EAGLE, \mbox{\citet{Bahe_et_al_2017a}} showed that this enhancement also agrees quantitatively with the observations, and~is also predicted for stellar metallicities. Together, these studies identified three mechanisms that contribute to the elevated metallicities. Firstly, ram pressure stripping removes predominantly gas at larger radii, where the metallicity is lower \citep{Genel_2016, Bahe_et_al_2017a}. Secondly, suppressed inflows of pristine gas within satellites \citep{vanDeVoort_et_al_2017} prevent the dilution of the ISM \citep{Bahe_et_al_2017a}. Finally, \mbox{\citet{Gupta_et_al_2018}} showed that the gas that is still replenishing the satellite ISM has higher metallicity than for isolated galaxies. Both TNG and EAGLE predict that this enhancement is not restricted to satellites within the virial radius of their group, but~already affects galaxies during their infall \citep{Bahe_et_al_2017a, Gupta_et_al_2018}.

While ram pressure is acting on group galaxies, it distorts their gas into long ``jellyfish'' tails. In~TNG100, these have been studied by \mbox{\citet{yun19}} through visual inspection. Depending on group mass, these authors found that $\approx$25--45\% of gas-bearing satellites show evidence of such tails in their (total) gas density maps, only moderately lower than the equivalent fraction for clusters ($\approx$65\%). Even when considering only the (observable) \HI{} component in TNG100, \mbox{\citet{Watts_et_al_2020}} find statistically significant asymmetries that visually resemble observed \HI{} tails (e.g.,~\mbox{\citet{Chung_et_al_2007}}), with~a slightly higher occurrence (21 vs.~28\%) amongst group galaxies compared to those in lower-mass~halos.

While the stripping of gas is the clearest predicted effect of ram pressure, simulations have also begun revealing second-order effects due to the compressive effect on the leading edge of group satellites. In~RomulusC, for~instance, \mbox{\citet{Ricarte_et_al_2020}} demonstrated a correlation between ram pressure and black hole accretion rates as well as star formation rates, evidence for which has also been seen in recent observations \citep{Poggianti_et_al_2017b}. A~compression-induced enhancement of star formation has also been described in the EAGLE simulations by \mbox{\citet{Troncoso-Iribarren_et_al_2020}}. A~caveat applicable to both simulations, however, is that stars are formed directly from the tenuous ISM phase rather than from dense molecular gas that may have a different susceptibility to the effect of ram pressure, at~least in~detail.

The high resolution of contemporary galaxy group simulations has also enabled studies of their (stellar) morphology. \mbox{\citet{Feldmann_et_al_2011}} studied the transformation from disc to elliptical galaxy morphologies in one zoom-in simulation, and~identified major mergers prior to accretion as the key driver of this change. More recently, \mbox{\citet{Joshi_et_al_2020}} presented a detailed analysis of galaxy morphology in groups and low-mass clusters from the TNG50 and TNG100 simulations. They found that up to 95\% of (satellite) disc galaxies are transformed into non-discs by $z = 0$, with~redistribution of stars by tidal shocks during pericentric passages as the dominant mechanism behind the~transformation.

\textls[-15]{Even more fundamentally, simulations have investigated the tidal stripping of dark matter and stars from group satellites. With~the caveat that this stripping may be artificially enhanced by numerical artefacts (\mbox{\citet{vanDenBosch_Ogiya_2018}}, but~see \mbox{\citet{Bahe_et_al_2019}}), the~clear prediction is that dark matter stripping far outweighs that of stars: for example,~\mbox{\citet{Joshi_et_al_2019}}} found in an individual zoom-in simulation of a galaxy group that stellar stripping at a $>$10\% level typically only occurs after the loss of $\approx$80\% of their dark matter halo. This relatively minor role of stellar stripping, also borne out by EAGLE \citep{Bahe_et_al_2017a} and IllustrisTNG~\citep{Engler_et_al_2021}, leads to a stellar-to-halo mass relation with a much higher peak ratio ($\approx$0.15) for group satellites than centrals ($\approx$0.02). The~most extreme form of satellite stripping, their complete disruption, is even predicted to be somewhat more common in groups than clusters ($\approx$65 vs. 50\% of all accreted satellites with a total mass of $\sim$$10^{12}\,\msol$), due to the higher efficiency of dynamical friction driving satellites towards their dense centers~\citep{Bahe_et_al_2019}. 

Finally, we note that simulations are also increasingly demonstrating the importance of groups for the evolution of \emph{cluster} satellites: around 50\% of all $z = 0$ satellites in massive clusters of the TNG300 simulation were quenched in a group before being accreted onto their final host \citep{Donnari_et_al_2021}; \mbox{\citet{pallero20}} came to a similar conclusion with the Hydrangea/C-EAGLE simulations. Similarly, \mbox{\citet{Jung_et_al_2018}} found that almost half of all $M_\star \gtrsim 10^9\,\msol$ galaxies accreted onto clusters as (group) satellites are already gas poor at the time of cluster infall, compared to only 6\% of central galaxies. Even where groups and their galaxies have not (yet) joined a cluster, they therefore contribute to the large-scale environmental dependence of galaxy properties far beyond the ``edge'' of the cluster \citep{bahe13}.

\subsection{Simulating the Impact of Galaxy Group Astrophysics on Large-Scale Structure~Cosmology} \label{sec:cosmology}

Measurements of the growth of large-scale structure (LSS) can provide
powerful tests of our cosmological framework \citep{Peebles_80,Bond_80,Davis_85,Kaiser_87,Peacock_94}. Importantly, they are independent of, and~complementary to, constraints from analyses of fluctuations in the cosmic microwave background (CMB) and geometric probes, such as Type Ia SNe and baryon acoustic oscillations (BAOs). Generally speaking, the~different LSS tests (e.g., Sunyaev-Zel'dovich power spectrum, cosmic shear, group and cluster number counts, redshift space distortions, etc.) are just different ways of characterising the `lumpiness' of the matter distribution on different scales. On~very large scales, perturbation theory is sufficiently accurate to calculate this distribution reliably. However, most existing LSS tests probe well into the non-linear regime. The~standard approach is therefore either to calibrate the `halo model' (e.g., {\sc hm}code package; \mbox{\cite{Mead_16}}) using large N-body cosmological simulations, or~to use such simulations to correct linear theory empirically (e.g., {\sc halofit} package; \mbox{\cite{Takahashi_12}}).

If the matter in the universe were composed entirely of dark matter, these approaches would likely be sufficient. However, baryons contribute a significant fraction of the matter density and work by a number of different groups has shown, using cosmological hydrodynamical simulations, that feedback processes associated with galaxy formation can have a significant effect on the matter distribution on scales of up to a few tens of \mbox{megaparsecs \citep{vanDaalen_11,Schneider_15,Mummery_17,Springel_18}}. Therefore, while such effects are typically ignored or treated in a simple way as a first step when modelling LSS data, it is of critical importance to
understand their impact, as~they can introduce significant biases in
the inferred cosmological parameters in upcoming surveys if no action
is taken (e.g., {\cite{Semboloni_11,castro21,debackere21}}).

\textls[-5]{Galaxy groups (taken here to be bound systems with total masses of $\sim$$10^{13-14}$ M$_\odot$) play a particularly important role in LSS cosmology.  This is simply because a sizeable fraction of the galaxies, baryons, and~overall matter in the Universe resides in groups.  Consequently, LSS cosmology tests that probe more `typical' environments (such as cosmic shear, galaxy-galaxy lensing, galaxy clustering, redshift space distortions, CMB lensing, etc.), as~opposed to tests that sample only the most massive systems (such as the SZ effect power spectrum and current cluster count surveys), will be sensitive to the abundance of galaxy groups and the spatial and kinematical distributions of matter within and around~them.}

\begin{figure}[H]
\includegraphics[width=0.42\columnwidth]{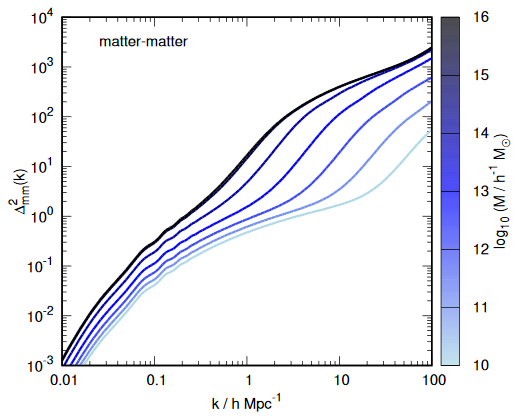}
\includegraphics[width=0.48\columnwidth]{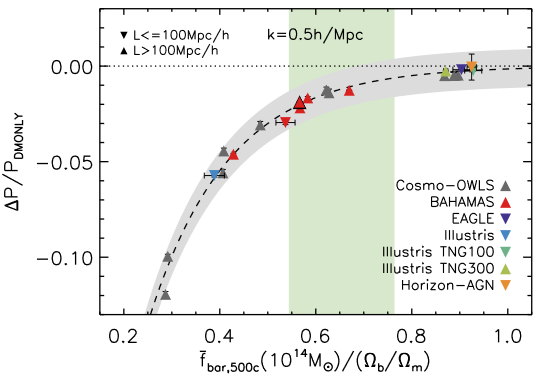}
\caption{(\textbf{Left}) The contribution to the dimensionless matter power spectrum for halos of different mass.  The~curves show the resulting power spectrum when integrated up to different choices of the maximum halo mass.
Here we can see that halos corresponding to galaxy groups ($10^{13-14}$ M$_\odot$) contribute a very large fraction of the power over the range of wavenumbers probed by LSS measurements (typically $k \la 10\ h$Mpc$^{-1}$).  This figure was reproduced with permission from \mbox{\citet{mead20}}.  (\textbf{Right}) The effect of galaxy formation on the matter power spectrum at the scale $k= 0.5\ h$Mpc$^{-1}$ as a function of the mean normalized baryon fraction in $\sim10^{14}$ M$_\odot$ halos.  Shown are simulations from cosmo-OWLS and BAHAMAS (grey and red), EAGLE (purple), Illustris (blue), TNG100 and TNG300 (cyan and green) and Horizon-AGN (orange).  Baryon fractions were calculated within $r_{500c}$ for halos in the mass range $M_{500c}= [6\times10^{13} , 2\times10^{14}]$ M$_\odot$. The~dashed curve shows that at this $k$, a~simple exponential function of the baryon fraction fits the predictions for the suppression of power of all simulations to within 1\% (grey band). The~vertical green band shows a range of mean group-scale baryon fractions roughly consistent with observations.  These results demonstrate that the differences in the simulation predictions for the impact of baryons on $P(k)$ can be understood to high accuracy based on differences in the baryon fraction on the group scale and that calibration to observed baryon fractions is a promising tool for constraining feedback processes in the simulations.  This figure was reproduced with permission from \mbox{\citet{vandaalen20}}.}
\label{fig:delta_k}
\end{figure}

\textls[-5]{A quantitative demonstration of the importance of the galaxy groups on LSS cosmology can be provided by examining the contribution by halo mass to the total matter power spectrum, $P(k)$.  Note that, at~present, virtually all current LSS tests probe cosmology through its effects on the matter power spectrum.  In~the left panel of Figure~\ref{fig:delta_k} we show the contribution to the dimensionless matter power spectrum by halos of different mass, as~calculated in \mbox{\citet{mead20}} using the \textsc{HMcode} halo model.  Note that the dimensionless matter power spectrum, $\Delta^2(k)$, is related to $P(k)$ via a multiplicative factor $4\pi (k/ 2\pi)^3$ and $k$ is the wavenumber related to the comoving size scale ($\lambda$) by $k = 2\pi/\lambda$.  The~curves show the resulting power spectrum when integrated up to different choices for the maximum halo mass.  The~results demonstrate that halos corresponding to galaxy groups contribute the majority of the power on the scales relevant for most LSS probes (typically $k \la 10\ h$Mpc$^{-1}$).  This conclusion is consistent with previous simulation-based findings presented in \mbox{\citet{vanDaalen_15}}.}

A clear ramification of groups contributing a large fraction of the signal to current LSS tests of cosmology is that theoretical models/simulations must be able to predict the abundance of groups and the matter distribution within them to a very high level of precision on average.  For~example, the~upcoming Rubin Observatory (formerly LSST), Euclid, and~Roman Space Telescope surveys are expected to measure the matter power spectrum to better than a few percent accuracy over a very wide range of scales, implying that the theoretical uncertainties in predicting $P(k)$ should be smaller than this to avoid biasing cosmological parameter constraints (e.g., \mbox{\citet{Huterer_05,Hearin_12}}).  As~already noted, because~baryons contribute a non-negligible fraction of the matter density, this means an accurate theoretical description of the baryons within groups (and their back reaction on the dark matter) is also~required.

Given the complexity of the physical processes involved in setting the thermodynamic properties of the IGrM and the difficulty in simulating the full range of scales at play (see Section~\ref{sec:modules}), the~prospects for accurately (to typically percent level) describing the impact of baryons and group astrophysics on LSS would at first sight seem daunting, if~not altogether hopeless at present.  Indeed, previous simulation work has shown that variations of the parameters associated with the efficiencies of feedback processes even within plausible bounds can lead to relatively large differences in the predicted properties of groups (e.g., \mbox{\cite{puchwein08,mccarthy10,lebrun14,planelles14}}).  Variations in resolution and method of solving the hydrodynamic equations may also produce important changes (e.g., \mbox{\cite{hahn17}}), though~they are arguably of secondary importance compared to changes in the subgrid modelling associated with feedback (e.g., \mbox{\cite{sembolini16}}).  A~consequence of these large simulation-to-simulation variations in the predicted properties of groups are relatively large study-to-study variations in the predicted impact of baryons on the matter power spectrum (see the simulation comparisons in \mbox{\cite{chisari18,vandaalen20}}). 

The study-to-study variation in the predicted properties of groups and the impact of baryons on $P(k)$ is not unexpected.  It is a consequence of not being able to derive the efficiencies for the relevant feedback processes from first principles (see discussion in \mbox{\cite{schaye15}}).  This problem is made particularly challenging in the context of simulations with finite resolution and approximations for other (coupled) physical phenomena.  As~we cannot derive the efficiencies from first principles, the~feedback in simulations must be \textit{calibrated} in order to ensure they reproduce particular quantities, after~which the realism of the simulations can be tested against other, independent quantities.  For~LSS cosmology, the~main problem we are trying to solve is to accurately model the impact of baryons on $P(k)$.  One way this could be achieved is to directly measure $P(k)$ from observations (e.g., via cosmic shear or galaxy-galaxy lensing and galaxy clustering) and compare this with the $P(k)$ predicted in the absence of baryon physics to measure the impact of baryons.  However, such an approach is generally a non-starter, as~to predict $P(k)$ in the absence of baryons requires that we assume a cosmology and therefore the process of deriving the impact of baryons on $P(k)$ becomes explicitly dependent on cosmology and we would have adopted a circular line of reasoning in our aim to constrain cosmology with LSS~measurements.  

An alternative approach is to modify the gravity-only predictions with simple,\linebreak physically-motivated prescriptions for baryon physics that have some number of associated free parameters and to {\textit{jointly}} 
 constrain the cosmological and feedback parameters through comparisons to LSS observables.  \textls[-15]{Examples of this approach include the halo model approach \textsc{HMcode} of \mbox{\citet{Mead_16,mead20}} and the `baryonification' approaches (which directly modifies the outputs of gravity-only simulations) of {\citet{Schneider_15,schneider20}}} and \mbox{\citet{arico20}}.  The~advantages of these approaches include: (i) they are considerably cheaper than running full cosmological hydrodynamical simulations; (ii) the baryon prescriptions are generally flexible and easy to adjust/improve; (iii) they can be straightforwardly incorporated within existing pipelines based on gravity-only simulations or the halo model.  The~disadvantage of these approaches are that the modeling of baryon physics and its back reaction on dark matter is simplistic and generally not self-consistent and that, unless~the impact of baryons is very different than the cosmological variations being explored, one expects there to be important degeneracies between the cosmological and feedback `nuisance' parameters and a degrading of cosmological constraining power (due to marginalization over uncertain baryon physics).

What about using cosmological hydrodynamical simulations directly for LSS cosmology predictions?  The approach of the \textsc{BAHAMAS} program \citep{mccarthy17,mccarthy18} (see also the recent FABLE simulations; \mbox{\citet{henden18}}), is to explicitly calibrate the feedback efficiencies so that they reproduce the observed baryon fractions of galaxy groups.  Except~for an explicit dependence on the universal fraction, $\Omega_b/\Omega_m$, the~baryon fractions of groups should be insensitive to changes in cosmology \citep{white93} and therefore they represent a highly useful metric on which to calibrate.  Furthermore, since the growth of matter fluctuations is fundamentally a gravitational process, by~ensuring the simulations have the correct baryon fractions on the scale of groups and gravitational forces are computed self-consistently, the~impact of baryons on $P(k)$ should be strongly constrained by this calibration approach.  Indeed, \mbox{\citet{vandaalen20}} have recently shown that one can understand the differences in the predicted impact of baryons on $P(k)$ from different simulations \textit{at the percent level} in terms of the differences in baryon fraction in the various simulations at a mass scale of $\sim10^{14}$ M$_\odot$ (see right panel of Figure~\ref{fig:delta_k}).  It is worth noting here that the simulations analysed in that study varied in resolution by more than a factor of 1000 in mass, used different hydro solvers, and~assumed different baseline~cosmologies.

The implication of this recent development is that, through calibration of feedback efficiencies on the observed baryon fractions of galaxy groups, we strongly limit the ways in which feedback can affect $P(k)$ (in other words we have a very strong \textit{prior} on the impact of baryons).  This, in~turn, means much stronger (and more robust) cosmological constraints from LSS.  However, before~claiming victory, a~number of important issues require further attention.  Firstly, as~the calibration is reliant on observations of galaxy groups, the~uncertainties in the observed baryon fractions need to be properly included in any cosmological analysis.  It goes without saying that the selection function of the group calibration data set must be also reasonably well understood and accounted for (otherwise we risk miscalibrating the feedback).  In~addition, while the results of \mbox{\citet{vandaalen20}} look very promising, we need to explicitly verify that different kinds of simulations (e.g., that vary how feedback, star formation, and so forth are implemented, resolution, how the hydro equations are solved, etc.) that are calibrated in the same way to the same precision actually produce the same $P(k)$.  In~other words, we need to check whether differences in the details and evolution of the simulations affect the end state (e.g., $P(k)$ at $z=0$) if some aspect of that end state has an imposed boundary condition (the baryon fractions at $z=0$).  We also have precious few observational constraints on the baryon fractions of groups beyond $z\sim0.3$ and it is therefore unclear to what extent the simulations calibrated at $z=0$ remain `well behaved' at significantly earlier times.  Finally, the~level of degeneracy between the various cosmological parameters and the parameters governing feedback efficiency needs to be fully quantified.  As~already noted, there is an explicit dependence on the universal baryon fraction, $\Omega_b/\Omega_m$, but~there may well be other less obvious degeneracies that require understanding in order to obtain percent level constraints on parameters such as the dark energy equation of state.  Thus, while a promising start has been made in addressing this complex challenge, a~number of important steps remain in order to fully account for the impact of group astrophysics on high-precision LSS~cosmology.

Finally, we note that, in this section, we have discussed the impact of group astrophysics on LSS cosmology, with~a focus on the non-linear matter power spectrum, $P(k)$, which is the basis of many precision LSS probes of cosmology including cosmic shear, CMB lensing, galaxy clustering, and~so on.  We have not specifically discussed the impact of group astrophysics on attempts to use the abundances (number counts) of galaxy group themselves to constrain cosmology.  Of~course, one advantage that groups have over clusters in this regard is that they are much more numerous, potentially allowing for stronger cosmological constraints than what might be obtained by clusters alone (e.g., \mbox{\cite{pillepich2012,pierre2016,bocquet16,pillepich18c}}).  The~challenge is that they are more difficult to detect and to model, as~already discussed.  These issues are particularly pertinent for group counts, as~they affect the cosmological observable in a much more direct way than, for~example, cosmic shear.  Nevertheless, as~our observational picture and ability to model galaxy groups improves, we expect galaxy group counts to play an increasingly important probe of cosmology and one that is complementary to constraints coming from cosmic shear, CMB lensing, and~other LSS~probes.




\section{Future~Directions} \label{sec:future}


\subsection{Using Simulations to Make Predictions for Existing and Future Missions and~Telescopes}\label{sec:5.1} 

Cosmological simulations are widely used to make predictions for future missions, often in such a capacity that a mission's approval or rejection may hinge on these predictions.  It is therefore incumbent upon simulators to generate mock observations that do not suffer from numerical effects, poorly implemented modules, or~insufficient resolution.  However, simulation predictions that are later refuted are not necessarily the result of numerical problems: instead, such discrepancies may also reveal new physical processes. This applies in particular to the complex interplay of non-gravitational physics and dynamics in~groups.  

\begin{figure}[H]
\includegraphics[width=0.95\textwidth]{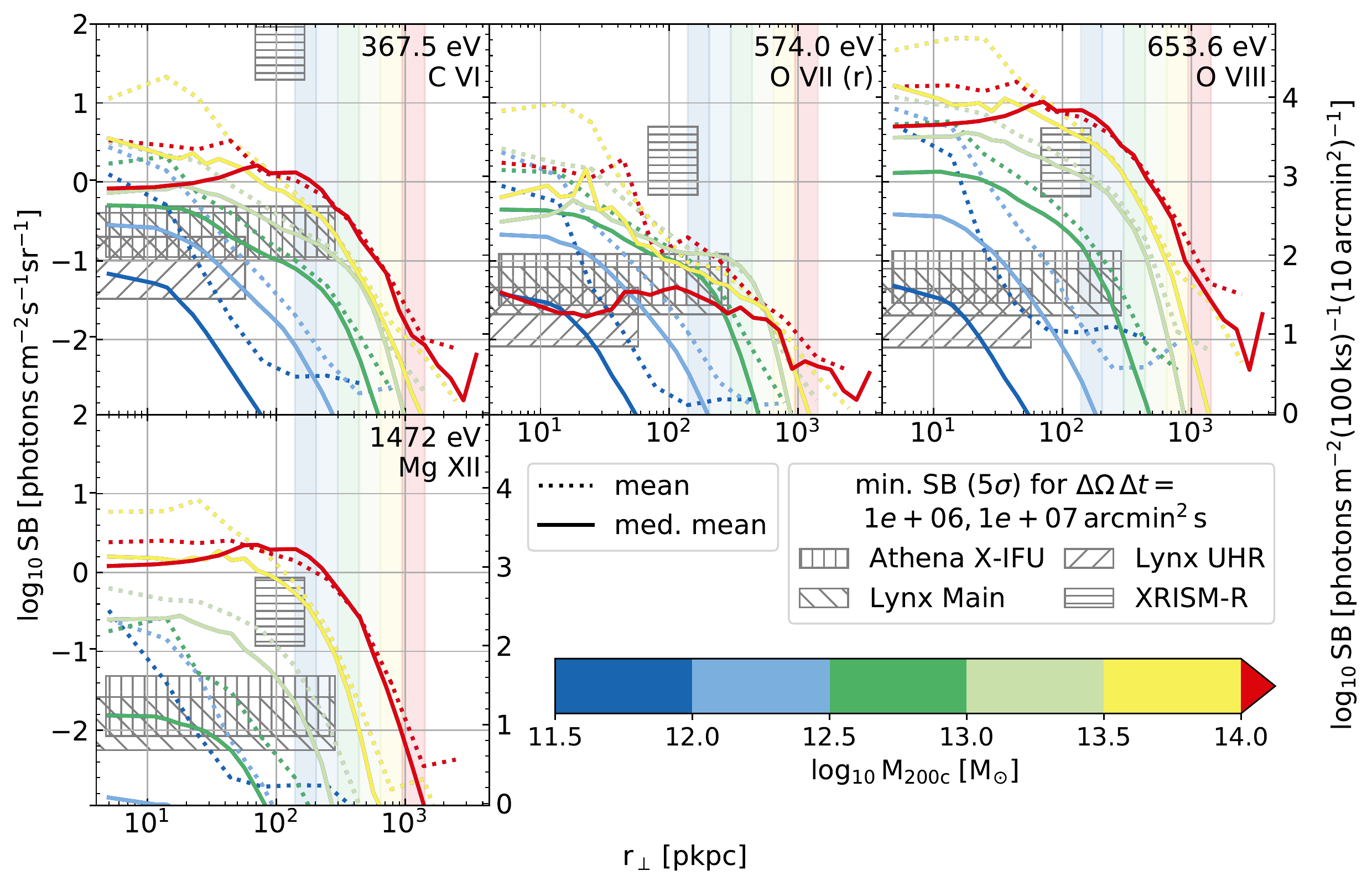}
\caption{{X-ray} 
 line emission predictions from the EAGLE simulation via Nastasha Wijers (in prep.) demonstrating the detectability by \XRISM, \Athena, \& \Lynx\ space-borne micro-calorimeters as a function of impact parameter from the central galaxy.  Lines are colored by halo mass with light green and yellow corresponding approximately to low-mass and high-mass groups.  Faded vertical stripes indicate $R_{200}$ for each halo mass.  High-mass groups are detectable by \XRISM\ in $\OVIII$ and $\MgXII$ at 100 kpc, while low-mass groups are detectable only in $\OVIII$.  \Athena\ and \Lynx\ should measure IGrM line emission for all group halos with high spatial~resolution.}
\label{fig:line_emission}
\end{figure}

A near-term example relevant for groups covers predictions of X-ray line emission from the EAGLE simulation that should be observed by several future missions, including \XRISM\ to be launched in 2022.  Figure~\ref{fig:line_emission} from Wijers~et~al. (in prep) demonstrates that \XRISM\ should be able to detect $\OVII$ and $\MgXII$ emission tracing the approximate temperatures of virialized IGrM gas.  This prediction follows on from the ``virial temperature thermometer'' model of \mbox{\citet{oppenheimer16}} and \mbox{\citet{wijers20}} that  specific metal ions should trace the volume-filling virialized halo gas corresponding to the temperature of the ion's peak collisional ionization fraction.  While (UV-band) $\OVI$ traces outer virialized galactic halo gas at $3\times 10^5$ K,  $\OVII$ and $\OVIII$ in the X-ray spectrum are predicted to trace $\geq 10^6$ K IGrM in groups.  Figure 8 of \cite{wijers20} predict that $\OVIII$ absorption is strongest in poor groups with $M_{500}\sim 10^{13}\ \msolar$ and $\FeXVII$ in intermediate groups with $M_{500} \sim 10^{13.5} \msolar$.

\XRISM\ should measure the significant metal and (for an assumed metallicity), baryon contents of the IGrM out to 100 kpc from the central galaxy as predicted by the EAGLE simulations in Figure~\ref{fig:line_emission}.  Micro-calorimeters on \Athena\ and \Lynx\ should be able to resolve the interior metal emission of the IGrM at superb ($<$10 eV) spectral resolution and signal to noise ratio. If, however, \XRISM\ does \emph{not} detect $\OVIII$ or $\MgXII$ at the levels predicted in this figure, this might point to one of the following scenarios. Firstly, the~IGrM might not be as metal-enriched as EAGLE predicts, for example,~because the nucleosynthetic yields assumed by the simulation are too high, or~due to a higher-than-predicted fraction of metals being retained within galaxies. Secondly, EAGLE might over-predict the IGrM baryon content, and~hence the gas density and emission line luminosity \footnote{As discussed in Section~\ref{sec:barcontent},  there is already evidence for this, although~at least the central regions of EAGLE groups appear to reproduce current observations (see Figure~\ref{fig:radial_profiles})}. 
A~third possibility is that metals in real groups are distributed more (or in principle also less) homogeneously than in EAGLE, which would affect the radially averaged cooling rates, and~hence line emission luminosities---metal-line emission is highly sensitive to the distribution of gas density and metallicity distributions on both the macro and micro IGrM scales. Finally, non-equilibrium ionization \citep{oppenheimer13a} and/or dual temperature electron-ion plasmas \citep{yoshida05} could alter simulation predictions that almost always assume ionization equilibrium and equipartition between electrons and~ions.

\begin{figure}[H]
\includegraphics[width=0.975\textwidth]{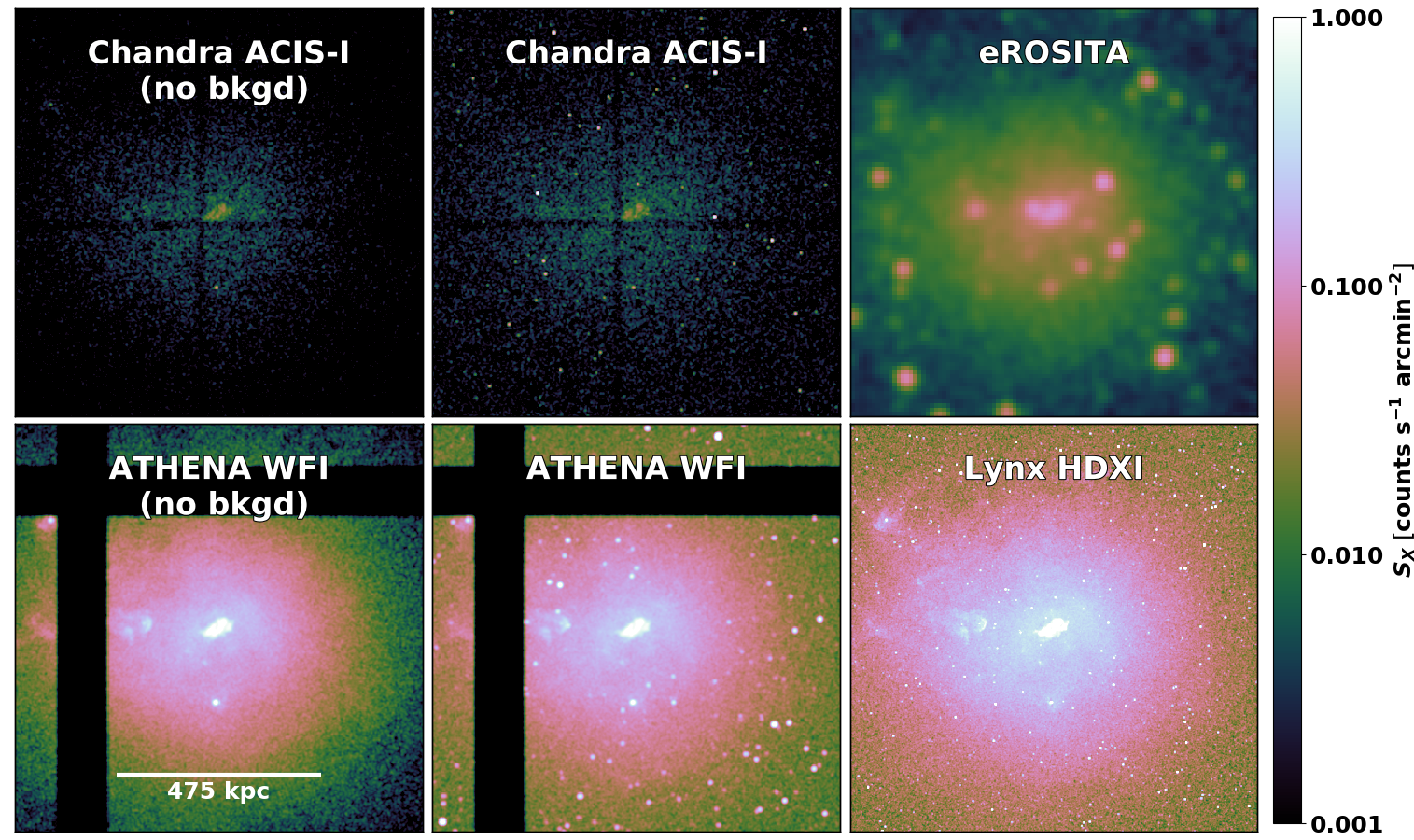}
\caption{Simulated $100$ {ksec} surface brightness maps of the same TNG100 $M_{500}=10^{13.6}\ \msolar$ group placed at $z=0.05$ by CCD detectors on \Chandra\ (Cycle 0 capability), \eROSITA, \Athena, \& \Lynx.  The~two left panels show \Chandra\ and \Athena\ mocks without noise, while all the other panels add noise from the instrument, the~Milky Way foreground, and~a randomly generated cosmic X-ray background (CXB).  The~field of view covers out to $R_{500}$.  All detectors use the color scale on the right.  A. Simionescu collaborated on the production of this figure.  Note: the CXB is differently randomly generated in each~panel.} 
\label{fig:pyxsim_mocks}
\end{figure}

When simulators generate mock observations, it is important to present their predictions in a manner that accurately showcases the capabilities of existing and proposed instruments.  Often it is advisable to work with instrumentalists and observers.  An~example of an attempt to fairly compare capabilities of existing and future missions appears in Figure~\ref{fig:pyxsim_mocks} for a TNG100 $M_{500}=10^{13.6}\ \msolar$ halo at $z=0.05$.  Each panel shows a mock {100ksec} 
 X-ray image based on the instrument capability at launch (Cycle 0 for \Chandra), but~the leftmost panels show mock observations without any noise.  We apply a forward modeling technique using the packages pyXSIM  \footnote{\url{http://hea-www.cfa.harvard.edu/~jzuhone/pyxsim/} {(accessed on 10 June 2021)} pyXSIM is an implementation of the PHOX algorithm \citep{biffi12,biffi13}} \citep{zuhone16} and SOXS  \footnote{\url{http://hea-www.cfa.harvard.edu/~jzuhone/soxs/} {(accessed on 10 June 2021)}}; 
 the SIXTE simulation software \citep{sixte} is used to create the \eROSITA~mock.  

We choose surface brightness units (counts s$^{-1}$ arcmin$^{-2}$) to show the relative throughput of the detectors. At~launch, \Chandra\ was able to detect groups out to $R_{500}$ in $\la$ 100 {ksec}, which appears consistent with the longest exposed groups of \mbox{\citet{sun09}}.  \eROSITA~can detect extended emission out to a good fraction of $R_{500}$, although~it will only approach such exposures at the ecliptic poles during its eRASS:8 4-year survey \citep{merloni12}, and~will require targeted follow-up on most groups in the sky to achieve such depth. \mbox{\citet{biffi2018c}} simulated \eROSITA~observations of clusters/groups between $z$ = 0.1--2.0 from the Magneticum Box2/hr run, inputting AGN to determine if the underlying ICM/IGrM emission can be separated from AGN contamination.  \mbox{\citet{oppenheimer20b}} simulated the stacking of galactic halos assuming 2 {ksec} exposures aimed at 50 $z=0.01$ galaxies with an average halo mass of $M_{500}=10^{12.5}\ \msolar$ to show that the eRASS:8 survey should be able to resolve the stacked profile out beyond 100 kpc.  Their forward modeling analysis used EAGLE and TNG100 galaxies as inputs, added in noise and attempted to subtract it, and~excised mock point sources from the cosmic X-ray background. However, \mbox{\citet{biffi2018c}} and \mbox{\citet{oppenheimer20b}} still did not attempt to mock the scanning mode of the 8 individual all-sky surveys, instead assuming single pointed exposures with the object placed at the center, which likely under-estimates sources of systematic errors.  Future data collected from \eROSITA~will need to be compared to simulations applying the same scanning exposures used by the eRASS:8~survey.  

\textls[-5]{Continuing on the lower panels of Figure~\ref{fig:pyxsim_mocks}, \Athena~will be able to distinguish IGrM structure associated with the central galaxy and satellites.  Finally, \Lynx\ should be able to clearly resolve azimuthal dependence, a~variety of satellite interactions, and~sharp shock fronts bow shocks associated with infalling satellite as seen to the left of the \mbox{simulated~central.}}

\subsection{Observations and Simulations in Support of Each~Other}\label{sec:5.2}  

It is clear that the increasing wealth of observational data on galaxy groups has led to continual improvement in cosmological simulations, as~we gain increased knowledge about the physical processes at play and constraints on their `efficiencies', both of which help inform the simulations.  However, an~important caveat to bear in mind is that observations themselves are subject to considerable uncertainties which should not be ignored when comparing with simulations, particularly if one intends to calibrate aspects of the simulations on said observations.  While good strides have been made in generating realistic synthetic X-ray, thermal Sunyaev-Zel'dovich (tSZ), optical, and so forth, `observations' of simulated groups to enable like-with-like comparisons (e.g., {\cite{mazzotta04,rasia06,lebrun14,zuhone16,barnes21}}), to~date much less attention has been devoted to ensuring a consistent method of \textit{selection}.  While the use of mock catalogs is standard practice in galaxy surveys to quantify the selection function (e.g., \mbox{\cite{avila18,korytov19,to21}}), the~use of realistic mocks to quantify the selection function of groups selected on the basis of their hot gas properties (particularly X-ray and tSZ) has lagged behind.  Instead very simplistic models (e.g., spherical beta models) are still regularly employed in characterizing the selection function.  There is also often a key difference in the way observers speak about the selection function (which they normally cast in terms of observable quantities such as flux, surface brightness, or~signal-to-noise ratio) and what a theorist or simulator would regard as the selection function (which is almost always with respect to halo mass).  Ultimately, what is required is an iterative process involving simulations and observations, whereby mock surveys of the simulations are used to inform a consistent definition of the selection function. A~like-with-like comparison is then made between the simulations and observations, shortcomings of the simulations are identified, new simulations are produced, and~the cycle repeats.  Each cycle yields not only an improved simulation and physical picture (hopefully), but~also informs our knowledge of how biased observational methods are in selecting groups and estimating their physical parameters.  Of~course, the~realism of the simulations must also be tested against independent observations which are not part of the calibration process (e.g., evolution of galaxy groups, environmental effects on galaxies, etc.).  

While the heterogeneous nature of most pointed X-ray observations with \Chandra~and \XMM~do not naturally lend themselves to the development of simple selection functions, upcoming \eROSITA~observations should be much more tractable in this regard, given the homogeneity of the survey.  Likewise, wide-field tSZ observations should benefit from mock surveys based on large hydrodynamical simulations, to~complement existing work based on simple spatial templates (e.g., \mbox{\cite{arnaud10}}), which are not expected to hold deep into the group regime \citep{lebrun15}.

\subsection{Timely Research Topics for Simulations of~Groups}  \label{sec:researchtopics}  

Arguably, simulations of $10^{13}-10^{14}\ \msolar$ halos have so far received less attention than neighboring halo mass ranges, which might be due to the current difficulty in the observational identification and characterization of groups, as~well as the complexity in their theoretical modeling. On~the observational side, breakthroughs are imminent on multiple fronts: deep and highly complete spectroscopic galaxy redshift surveys such as the 4MOST Wide-Field Vista Extragalactic Survey (WAVES) will deliver robust group catalogues;  IGrM probes with linear dependence on gas density---such as UV and X-ray absorption, kinetic Sunyaev-Zeldovich (kSZ), and~fast radio bursts \citep{mcquinn14,prochaska19}---instead of the $\rho^2$ scaling of X-ray emission will map the diffuse IGrM out to large radii. Below, we therefore list a selection of timely open research topics in the field of group~simulations.




\begin{itemize}[leftmargin=*,labelsep=5.5mm]
\item {\bf \textls[-5]{The relationship between the central SMBH and properties of the IGrM/ICM/CGM:}}  The strong correlations between the mass of the central SMBH in a group ($M_{\rm SMBH}$) and the temperature and X-ray luminosity ($T_{X}$ and $L_{X}$) of the IGrM/ICM \citep{gaspari19,bogdan18}, provide fundamental tests of AGN feedback in simulations (see also Section~4 of the companion review by {\citet{lovisari21}}).  
On~the one hand, the~$M_{\rm SMBH}$ scaling with $M_{\rm halo}$, assuming $T_{X}$ measures $M_{\rm halo}$, is a natural expectation of the SMBH growth being controlled by the binding energy of the halo \citep{booth10}.  The~EAGLE simulation prediction that $L_{X}$ scales inversely with $M_{\rm SMBH}$ {\it at fixed halo mass} and most strongly for the CGM (i.e., galactic halo masses below the group scale; \mbox{\citet{davies_jj19}} suggests the opposite, inverse trend, which is also seen in TNG100 \citep{davies_jj20}.  As~the latter paper explains, the~CGM $L_{X}$ is reduced in response to the integrated SMBH feedback lifting baryons out of galaxy halos, lowering the density, and~significantly increasing cooling times.  However, by~group masses, the~\mbox{\citet{gaspari19}} correlations appear reproduced by TNG100 with $T_{X}(<R_{500})$ showing surprisingly little scatter for quenched galaxies with $M_{\rm SMBH}>10^{8.2}\ \msolar$ \citep{truong21}.  This link between $M_{\rm BH}$ and halo-wide $T_X$, which is the best X-ray observational proxy for halo mass, indicates that simulations predict a fundamental relationship between $M_{\rm BH}$ and $M_{\rm halo}$ transmitted through the virialization of halo gas.  The~nature of this relationship contains both the virial temperature being set by hierarchical growth of group/cluster-scale halos (as discussed by \mbox{\cite{truong21}}), and~the mechanisms of gas accretion and AGN feedback determining SMBH growth, which span scales from the SMBH radius to $R_{\rm vir}$ (as discussed in Section~4.1 of the  companion review by {\cite{Eckert2021}. and in \mbox{\cite{gaspari20}})}.  
 \mbox{\citet{bassini19}} explored \gadget\ cluster zoom simulations, finding that they were able to reproduce the observed $T_{500}-M_{\rm SMBH}$ and other correlations in the group and cluster regime.  We emphasize the need for more simulations to explore the rich and diverse astrophysics contained in the relationship between group properties and their central~SMBHs.  
\vspace{0.10cm}

\item {\bf Cooling flows or cold rain?:}  \textls[-15]{One of the key phenomenon that links the IGrM/ICM to the BGG/BCG and ultimately, to~the SMBH hosted by the central galaxy is the flow of gas from the former to the latter two, particularly in CC clusters.  During~the cooling phase, the~conventional view is that the gas typically flows inwards subsonically and {\it en masse}, meaning that if gas is multiphased, then all phases move inwards in a comoving fashion. This is how cooling flows were originally conceived (\mbox{\cite{cowie77}}; see also reviews by \mbox{\cite{fabian94}} and \mbox{\cite{Peterson06}});}
this is what pre-AGN feedback simulations found (see Figures~17 and 19 of \cite{lewis00}); and recently, this is how the inflow is thought to behave during times when the central AGN is quiescent (or in quadrants about the cluster center where cooling is dominant). Recent very high resolution simulations of idealized galaxy groups and clusters \citep{Gaspari12,Gaspari13,Li14,Li15,Prasad15,Prasad17,Prasad18} 
find that when the ratio $t_{\rm cool}/t_{\rm ff}$ in a cooling group/cluster core drops below some threshold (nominally $\sim$10), local density perturbations can become thermally unstable \citep{mccourt12,Sharma12}, leading to the formation of cold dense clouds. 
These clouds then separate from the rest of the CGM and stochastically rain down upon central galaxy and its SMBH.  This ``cold rain'' fuels both star formation events as well as AGN~outbursts.  
\vspace{0.0cm}

There are number of problems with the conventional Bondi accretion model indicating that it is untenable (see, e.g., \cite{Prasad15,Prasad17,Nemmen2015,McNamara2011}), which the cold rain model appears to resolve.  We argue that this warrants further investigation of the cold rain model within the context cosmological hydrodynamics simulations of the formation/evolution of massive galaxies, groups, and~clusters. 
Two potential directions of study stand out: Firstly, current insights about the cold rain phenomenon come from idealized simulations that neither have satellite galaxies moving through the IGrM/ICM and inducing perturbations in their wakes, nor do they allow for interactions, like ram pressure stripping of these satellite galaxies. How these complications alter the thermal instability/cold rain picture remains unexplored. Secondly, it is not currently feasible to directly model the cold rain phenomenon in cosmological hydrodynamics simulations because that would require being able to resolve spatial scales approaching $\sim 1$ pc  see \cite{Hummels19} for further details).  This however means that there is an opportunity for developing innovative subgrid models that can capture the most important elements of the cold rain model.   There is precedence for the second option in that the torque-limited accretion subgrid model of \mbox{\citet{anglesalcazar17}} was created to encompass idealized simulations of gas-rich accretion discs from 10 kpc to 0.1 pc \citep{hopkins10}.
\vspace{0.10cm}

\item {\bf New models for AGN feedback:} 
Non-spherical, jet-like feedback appears necessary to impart energy to the IGrM/ICM while not over-evacuating the inner region, as~discussed in {Section~4 of the }{\citet{lovisari21}} 
companion review where they show in {their Figure~7} results from an idealized simulation by \mbox{\citet{gaspari14}} demonstrating a self-regulated jet capable of preserving the cool core (see also \citep{Li14,Li15,Prasad15,Prasad17,Prasad18}). Collimated jet feedback is currently inadequately modeled in cosmological simulations; however, thermal blast feedback should not be dismissed as a potential mode operating at late times until it is confirmed that cored NCC underluminous groups do not exist.  \mbox{\citet{meece17}} found a hybrid kinetic jet with thermal heating in idealized hydro simulations could best achieve self-regulation and produce a cool core, whereas a thermal-only jet results in a cored profile that rapidly radiates energy away leading to a cooling catastrophe.  Their kinetic-only model also achieves self-regulation, but~appears too steady compared to observed AGN duty cycles.  

\vspace{0.0cm}
The failure of cosmological simulations to reproduce the observed thermal structure of the IGrM (Section~\ref{sec:gasprofiles}) may be related to cosmological simulation's inability to model narrow, high momentum flux, jet outflows.  Narrow beams are better able to drill their way out through a highly pressurized IGrM/ICM that can easily stall an isotropic outflow, preventing the deposition of energy where it is needed.  Nonetheless, narrow outflows have their own challenges, which become apparent through running idealized simulations.  Firstly, as~shown by \mbox{\citet{Vernaleo06}} and confirmed by \mbox{\citet{cielo18}}, jets that fire in a fixed direction tend to deposit their energy at increasing larger distances and ultimately, end up doing so beyond the group/cluster core.  As~a result, such jets only delay the onset of catastrophic cooling, not prevent it.  The~second and equally vexing problem concerns the coupling between the narrow jets and the IGrM/ICM: how do narrow, bipolar jets manage to heat gas in the group/cluster cores in a near-isotropic fashion?  This motivated \mbox{\citet{babul13}} to argue for tilting jets, which change direction every so often as evidenced by observations detailed in this paper.  \mbox{\citet{cielo18}}---and most recently, \mbox{\citet{su21}}---found that not only is tilting necessary, but~the angle between jet events must also be reasonably large, and~furthermore heated jets work better than cold jets.  In~effect, the~desired outflows are those that have the appropriate energy/momentum flux, create near-spherical cocoons because these optimize energy transfer in transverse directions relatively to the jets, and~in a time-averaged sense, distribute their energy in a near-isotropic fashion within the group/cluster~core.  

\vspace{0.10cm}

\item {\bf X-ray detectability of new classes of groups:} The largest number of diffuse object detections by \eROSITA~will be groups (e.g., \cite{pillepich18c}).  The~complete eRASS:8 survey should detect groups with $M_{500}>10^{13}\ \msolar$ out to  $z=0.05$ \cite{kaefer20}.  While simulations of EAGLE and TNG100 eRASS:8 stacking show galactic-scale halos at $M_{500} < 10^{13}\ \msolar$ will not be individually detected \citep{oppenheimer20b}, \eROSITA~should observe groups in the local volume covered by CLoGS.  Simulations will provide necessary guidance in the interpretation of any prospective cored NCC under-luminous groups and/or coalescing groups that have yet to virialize.  It may well be that significantly under-luminous groups potentially exist as \mbox{\citet{pearson17}} cannot detect with \Chandra\ two of their 10 optically selected groups, which do not show signs of being unvirialized.  The~near future holds promise to detect new potential classes of groups---poor, under-luminous, and~coalescing---in~X-rays.  

\vspace{0.10cm}

\item  \textbf{Multi-phase gas stripping from group satellites:} \textls[-15]{A shortcoming in common to all simulations that we have discussed in this review is the lack of a cold ($T \ll 10^4\, \mathrm{K}$) and dense molecular phase in the ISM of galaxies. Both observations (e.g.,~\mbox{\cite{Moretti_et_al_2020}})} and idealized hydrodynamic simulations (e.g.,~\mbox{\cite{Tonnesen_Bryan_2009}}) clearly indicate that ram pressure has a different effect on the dense molecular phase from which stars are formed than the more tenuous, warmer components traced by \HI{} and \HII{}. Although~post-processing can be used to estimate the molecular content of group satellites (albeit with strong assumptions; \mbox{\cite{Stevens_et_al_2021}}), it cannot capture the different dynamical evolution of the two phases. Simulations with direct modeling of molecular gas---as is now often done in high-resolution zooms of individual galaxies (e.g.,~\mbox{\cite{Hopkins_et_al_2018, Applebaum_et_al_2021}})---would therefore reveal a fundamentally new aspect of the interaction between the IGrM and satellite galaxies. Recent advances in subgrid cooling models \citep{Ploeckinger_Schaye_2020} make such large-scale cold ISM simulations possible, but~the high resolution required to resolve giant molecular clouds at least marginally ($\lesssim\!10^4\,\msol$) makes them unfeasible on cluster scales for the foreseeable future. Galaxy groups, on~the other hand, would be perfectly suited to exploring this additional facet of the baryon cycle in a full cosmological~setting.

\vspace{0.10cm}

\item {\bf The Sunyaev-Zel'dovich Effect:} SZ stacking is already measuring the pressure and density profiles of groups from large radii inward.  Cross-correlating large spectroscopic surveys (e.g., BOSS \cite{ahn14}) with high-resolution maps of the CMB from the Atacama Cosmology Telescope (ACT) has measured the extended pressure and density profiles of groups via the tSZ and kSZ effect respectively.  \mbox{\citet{amodeo21}} detected elevated gas pressure profiles outside $R_{200}$ of $z=0.55$ $M_{200}=10^{13.5}\ \msolar$ groups indicating that feedback energy equivalent to double the gaseous halo binding energy needs to coupled directly to the IGrM, which is significantly above predictions from TNG100 simulations and even more so the EAGLE simulations~\citep{davies_jj20}.  \mbox{\citet{schaan21}} showed that $z=0.31$ $M_{200}=10^{13.7}\ \msolar$ groups are far more devoid of baryons in kSZ measurements than a \citet{navarro97} (NFW) profile.  \mbox{\citet{lim21}} tested groups in Illustris, EAGLE, TNG300, and~Magneticum simulations against \mbox{\citet{Planck13}} stacks, finding that the $M_{500}\sim 10^{13.0-13.5}\ \msolar$ scale provides a very promising scale to constrain the nature of AGN feedback.  The~measurements of pressure, density, and, through division, temperature profiles of groups will dramatically increase in the 2020's as the Rubin Telescope comes on line and the Roman and Euclid Telescopes are launched, providing spectroscopic surveys to cross-correlate further CMB observations from the ACT, the~Simons Observatory, the~Large Millimeter Telescope, and~CMB-S4.  These future SZ surveys will provide standard calibrations against which simulated groups are~compared.  
\end{itemize}

\section{Final~Statement} \label{sec:final}

Cosmological hydrodynamic simulations provide an aggregate picture of groups that is not yet available observationally.  Some of the latest state-of-the-art simulation projects can reproduce key stellar observations of galaxies, while others are able to match essential gaseous properties of clusters and large scale structure.  We approach the breakthrough when a single high-resolution cosmological simulation suite can match gaseous and stellar properties of both galaxies and clusters.  The~simulated groups in the intermediate range provide true testable predictions for upcoming observational datasets that include comprehensive galaxy surveys, all-sky X-ray maps, and~deep radio surveys.  Future observed group datasets synthesizing galaxy catalogues down to dwarf galaxies, X-ray emission extending beyond $R_{500}$ and to masses below $M_{500}< 10^{13.5}\ \msolar$, thermal and kinetic Sunyaev-Zel'dovich measurements, and~UV absorption compilations plus 21-cm maps accounting for warm and cool gas component will provide creative new stress tests of the non-gravitational, baryonic physics in cosmological~simulations.

\dataaccess{{Much} 
 of the simulation data used here are publicly available, including for the EAGLE~\citep{EAGLE_data}, IllustrisTNG \citep{TNG_data}, and~SIMBA (\url{http://simba.roe.ac.uk/} {(accessed on 10 June 2021))} 
 datasets.  Figures uniquely generated for this review by the authors include Figures~\ref{fig:mhalo_fgas}--\ref{fig:radial_profiles}, \ref{fig:entropy_profiles} (left panel), Figures~\ref{fig:temperature_profiles}--\ref{fig:pyxsim_mocks}.  }



\authorcontributions{B.D.O.: lead author Sections~\ref{sec:1}, \ref{sec:overview}, \ref{sec:hydro}--\ref{sec:AGNfeedback}, \ref{sec:baryons}, \ref{sec:5.1}, \ref{sec:researchtopics} and \ref{sec:final}; 
A.B.: lead author \mbox{Section~\ref{sec:BGG}}, major contributions to Sections~\ref{sec:baryons} and \ref{sec:researchtopics}; 
Y.B.: lead author  \mbox{Section~\ref{sec:satellite}},  major contributions to \mbox{Sections~\ref{sec:1}--\ref{sec:modules}} and \ref{sec:baryons}; 
I.S.B.: lead author Sections~\ref{sec:3.10} and \ref{sec:multiphase}; 
I.G.M.: lead author Sections~\ref{sec:3.9}, \ref{sec:cosmology} and \ref{sec:5.2}. 
All authors have read and agreed to the published version of the manuscript. }

\funding{This research received no external~funding.}

\acknowledgments{We are grateful to the three anonymous referees who provided thorough reports that substantially improved the scope and clarity of this review.  We thank the following astrophysicists for thought-provoking conversations, observational and simulation data and figures, and~essential guidance in the production of this review:  
Sarah Appleby, Michelle Cluver, Weiguang Cui, Romeel Dav\'e, Dominique Eckert, Seoyoung Lyla Jung, Amandine Le Brun, Ilani Loubser, Lorenzo Lovisari, Daisuke Nagai, Ewan O'Sullivan, Douglas Rennehan, Vida Saeedzadeh, Zhiwei Shao, Prateek Sharma, Aurora Simionescu, Ming Sun, Michael Tremmel, Nastasha Wijers, and~Mark Voit.  This research was supported in part by the KITP National Science Foundation under Grant No. NSF PHY-1748958, and by the Netherlands Organization for Scientific Research (NWO) through Veni grant number 639.041.751.  This review is reliant on the public data release and accessibility generously provided by the EAGLE, IllustrisTNG, and~SIMBA teams.  A~number of calculations and figures presented in this review were done using (i) high performance computing facilities at Liverpool John Moores University, partly funded by the Royal Society and LJMU’s Faculty of Engineering and Technology, and~(ii) advanced research computing resources provided by Compute/Calcul Canada.  I.G.M. acknowledges support by the European Research Council (ERC)
under the European Union’s Horizon 2020 research and innovation
programme (grant agreement No 769130). A.B. acknowledges research support from Natural Sciences and Engineering Research Council of Canada (NSERC) and  Compute Canada.}

\conflictsofinterest{{The authors declare no conflict of interest.}}

\reftitle{References}

\externalbibliography{yes}
\bibliographystyle{Definitions/aa}
\bibliography{groups_simulations}



\end{document}